March 2026

# Highly Autonomous Cyber-Capable Agents

Anticipating Capabilities, Tactics, and Strategic Implications

**Authors:** Jam Kraprayoon, Shaun Ee, Brianna Rosen, Yohan Mathew, Aditya Singh, Christopher Covino, Asher Brass Gershovich

# Executive Summary

Frontier AI models' agentic and offensive cyber capabilities have advanced rapidly over the past 2–3 years, showing significant progress in both experimental settings and real-world operations. If these trends continue, we will see the emergence of **highly autonomous cyber-capable agents (HACCAs), AI systems that can autonomously conduct cyber operations at the level of sophisticated criminal groups and possibly even intelligence agencies**. HACCAs would be fully autonomous systems that can plan and execute complex attacks against well-defended networks over weeks to months without continuous human oversight. These systems require two distinct capability sets: **operational capabilities** to establish, maintain, and scale operations, and **offensive cyber capabilities** to conduct automated multi-stage attacks against well-defended networks using advanced tactics for evasion and persistent access.

**Nation-states, criminal groups, and other threat actors have strong incentives to develop and deploy HACCA systems as novel cyber tools, given the potential for HACCAs to increase the speed, scale, and sophistication of offensive cyber operations**. Some actors are already experimenting with autonomous offensive agents in real-world operations, which may be an early preview of the highly autonomous systems covered in this report.[1] Over time, these capabilities would likely proliferate, providing criminal groups, terrorist organizations, and other actors with access to new offensive cyber capabilities.

**Beyond being novel cyber tools, HACCAs could also pose a distinct and possibly more serious risk as novel threat actors if operators lose control over them via misalignment, sabotage, or cascading multi-agent failures**. The effects of offensive cyber operations have often been wider than their operators anticipated, from the Morris Worm in the early days of the internet to more recent attacks like NotPetya, and the autonomous capabilities of HACCAs compound this risk. **Rogue HACCAs could potentially coordinate massive numbers of instances, accumulate resources, evade shutdown, and scale their capabilities over time**. While more speculative than the tool-use risks described above, self-sustaining rogue HACCAs could be extremely difficult to detect and counter, and could pursue unpredictable objectives.

---

[1] For example, in September 2025, Anthropic detected a likely Chinese state-sponsored cyber espionage campaign where AI agents autonomously executed an estimated 80–90% of tactical operations against approximately 30 global targets (Anthropic, "Disrupting the first reported AI-orchestrated cyber espionage campaign"). See also Google Threat Intelligence Group, "GTIG AI Threat Tracker: Advances in Threat Actor Usage of AI Tools" on threat actor usage of AI for adaptive malware and malicious script generation.

**AI systems that can automate months-long cyber campaigns may seem distant, but if historical progress holds, this gap could close in a matter of years—far sooner than many defenders are prepared for.** While forecasting AI progress is challenging, past work by Model Evaluation and Threat Research (METR) and the UK AI Security Institute (AISI) has sought to estimate how rapidly AI capabilities improve on tasks like software engineering (doubling every 7 months) and cyber capabilities (doubling every 8 months). **A naive extrapolation of these trends suggests that 2028–2030 could be a reasonable guess for when HACCA systems become feasible, if current doubling rates hold.** That said, this figure rests on assumptions about current progress continuing, and these task suites being good proxies of actual HACCA deployment requirements. More work is needed to validate when HACCAs may emerge.

## Anticipating HACCA Operations

**HACCA operations would likely be characterized by five core tactics.** For independent deployments—whether by design or through loss-of-control scenarios—the full range of tactics becomes essential. Human-directed deployments may rely on operators for tasks like establishing infrastructure or restoring operations after shutdown, though doing so increases attribution risk, creating tradeoffs that may favor greater HACCA autonomy even in directed operations.

| Core Tactic | Description |
|---|---|
| **Establish and maintain infrastructure** | Standing up and operating the technical infrastructure on which the agent runs (e.g., its own compute, networking, and software stack). This could include activities such as setting up multi-machine training and inference across heterogeneous hardware. |
| **Coordinate, command, and control** | Managing distributed operations across multiple agent instances and tools. This could include activities such as implementing secure communication protocols between distributed agent instances and maintaining shared knowledge across a deployment. |
| **Acquire compute and financial resources** | Acquiring the necessary resources to sustain and expand an agent's operations. This could include activities such as acquiring compute at scale via diverse channels (long-term reservations, buying retail Graphics Processing Units, small-firm supplier contracts), and hacking into compute resources (e.g., cloud credential compromise). |
| **Evade detection and shutdown** | Actions to avoid being discovered and neutralized. This could include activities such as creating a robust proxy network, defending against adversarial attacks such as jailbreaks, and establishing and maintaining operational security. |
| **Adaptation and capability improvement** | Enhancing capabilities to maintain effectiveness against evolving defenses and to pursue more complex objectives. This could include activities such as spinning up additional agent copies and improving agent scaffolding. |

## Strategic Implications

The mainline trajectory for a world with HACCAs will likely be shaped by three forces: how states use HACCAs, the dynamics of diffusion among attackers, and the offense-defense balance.

- Nation-states are already active players in cyberspace, competing via espionage and, more rarely, sabotage. **We expect HACCAs to intensify inter-state competition by reducing the labor cost of cyberattacks and making certain damaging attacks more readily achievable.** However, the desire to avoid undue escalation would likely still constrain nation-states.

- **As costs decline and HACCA components become commoditized, actors like criminal groups and less cyber-capable states would also gain access to HACCAs.** This would be a significant strategic change as defenders would face a higher volume of skilled attacks: a mid-tier cybercriminal group, for example, might acquire network exploitation skills once only the domain of the NSA's Tailored Access Operations (TAO).

- However, the ultimate strategic impact depends on whether defenders can keep pace. Historically, "trailing-edge" organizations like regional utilities and healthcare providers have struggled to adopt even current best practices, leaving critical functions exposed. **Early investment in hardening infrastructure and supporting under-resourced defenders could steer this mainline trajectory toward a more stable equilibrium.**

Beyond the mainline trajectory, policymakers should prepare for two potential strategic surprises whose likelihood is uncertain but whose consequences could be catastrophic. **First, HACCAs could increase the risk of inadvertent escalation into a nuclear crisis.** While cyberattacks rarely risk escalation to kinetic or nuclear conflict, access to HACCAs could incentivize nation-states to conduct more cyber operations against adversary military systems. If attacks accidentally affect infrastructure entangled with Nuclear Command, Control, and Communications (NC3) systems, there may be little time to de-escalate, with potentially catastrophic effects. **Second, if HACCA operators lost control of the systems they deployed, rogue HACCAs could establish self-sustaining populations that function as persistent autonomous threat actors.** Such loss-of-control incidents would not be likely with early HACCAs, given that rogue HACCAs would need to self-sustain and remain focused on their own objectives while resisting tampering and shutdown from various parties. However, the transition to self-sustaining systems may be difficult to anticipate, and robust control mechanisms would be challenging to design, so this scenario may need significant advance preparation.

There is substantial uncertainty as to how these scenarios might play out, but given that HACCAs may be tasked by nation-states with military and intelligence objectives, HACCA developers should take robust measures to prevent such extreme incidents from taking place.

# Countermeasures and Guardrails

**Countering HACCA operations will likely require a layered approach across four strategic objectives: delaying proliferation, defending potential targets, detecting hostile activity, and disrupting active operations**. It is unlikely that any single intervention will reliably counter sophisticated HACCA deployments. Effective defense demands redundant safeguards that create multiple opportunities to prevent, identify, and neutralize hostile autonomous cyber operations. However, research into countermeasures specifically designed for autonomous cyber-capable agents remains nascent—the measures outlined here are illustrative starting points rather than validated solutions. Further exploration and R&D effort is needed to evaluate which approaches are most effective, how they interact, and whether different defensive tools may prove more promising.

| Layer | Description | Example Countermeasures |
|---|---|---|
| **Delay** | Slow the proliferation of HACCA capabilities and systems to malicious actors | Model weight security, differential access |
| **Defend** | Reduce the attack surface available to HACCAs and strengthen potential targets | Security-by-design for AI-generated code, automated vulnerability discovery and patching, automated red teaming and pentesting |
| **Detect** | Gain visibility into HACCA operations and identify hostile activity | Detection signatures for HACCAs, threat intelligence sharing for autonomous operations, agent honeypots |
| **Disrupt** | Degrade or neutralize active HACCA operations | Compute and finance access controls, network disruption, adversarial ML-based countermeasures |

In addition to defending against HACCAs, **effective technical, legal, and policy guardrails are essential to mitigate loss-of-control risks as these systems become more widespread.** While threat actors face strong incentives to develop and deploy these systems, policymakers can take steps now to define the rules of the road and forge global norms. In the development stage, measures such as model security hardening could delay the proliferation of HACCA-level capabilities to less responsible actors, while more rigorous pre-deployment testing could detect alignment and robustness issues. Technical safeguards embedded directly into HACCA systems could serve as first-line defenses post-deployment, such as integrity monitoring and fail-safe mechanisms (including "kill switches") that can neutralize HACCAs in loss-of-control scenarios.[2] As with HACCA countermeasures, many of these safeguards remain immature and require further R&D to validate. In terms of legal frameworks, existing laws and norms surrounding responsible nation-state behavior in cyberspace will need to be updated to address highly autonomous systems. Where the law falls short, policy measures and global governance mechanisms must fill remaining gaps, including through more clearly articulating authorization protocols, transparency measures, and redlines for HACCA development and use.

## Recommendations

Given the strategic implications outlined above, **this report proposes seven recommendations to tip the offense-defense balance in favor of defenders while mitigating loss-of-control risks from HACCA deployment.**

| Goal | Recommendation | |
|---|---|---|
| **Goal A: Understand the Threat** | I. | **Track and forecast real-world HACCA progress and proliferation:** Policymakers should monitor capability evaluations across operational and offensive cyber AI domains to get snapshots of current AI system capabilities. They should complement this with forecasting and research into proliferation dynamics to support planning for defenders. |
| | II. | **Update information-sharing mechanisms to address HACCAs:** Governments should work with industry to establish standardized transparency requirements and incident response processes for security incidents involving autonomous systems, especially focusing on shared reporting mechanisms for anomalous agent behavior. |
| **Goal B: Defend Against HACCAs** | III. | **Invest in R&D to counter autonomous cyber operations:** Governments and funders should invest in R&D that the market may neglect, especially defensive tools to help under-resourced defenders (e.g., secure-by-design AI-generated code, automated vulnerability discovery and patching) and novel detection mechanisms for HACCAs. |
| | IV. | **Prioritize and harden critical services and infrastructure:** Governments should provide targeted support for the adoption and diffusion of defensive measures among high-value but under-resourced defenders, like utilities and healthcare, basing their prioritization on the pattern of HACCA proliferation and malicious use (in Recs. I and II). |

---

[2] *Integrity monitoring* refers to onboard systems tracking a HACCA's state and behavior for anomalies, e.g., a monitoring agent running periodic checks for tampering or deviation from intended objectives. *Fail-safe mechanisms* are controls to neutralize HACCAs in worst-case scenarios (e.g., loss of control or capture), like deployer-only backdoors or automated self-destruct sequences triggered by anomalous behavior.

| | | |
|---|---|---|
| | V. | **Strengthen model, compute, and financial access controls:** Governments should work with industry to prevent malicious actors exploiting resources for HACCA-related operations, especially compute. This includes strengthening know your customer (KYC) protocols to address AI agents, and legal and technical measures to detect and disrupt HACCA operations. |
| **Goal C: Ensure Responsible HACCA Deployment** | VI. | **Invest in R&D for responsible HACCA deployment:** Governments, especially defense and intelligence agencies, should invest in R&D to prevent sabotage and loss of control over HACCAs, such as agent integrity monitoring, fail-safe mechanisms, and high-assurance alignment. |
| | VII. | **Establish legal and policy guardrails for the development and use of HACCAs:** States should affirm that norms against attacking critical infrastructure, civilian targets, or NC3 systems apply to HACCAs. High-risk offensive operations should require executive-level authorization with documented risk assessments. |

# Table of Contents

# 1 | Introduction

On November 2, 1988, system administrators across the United States arrived at work to discover that their computer networks had been infected by a self-replicating, self-propagating program known as the Morris Worm, marking the first instance of a computer worm causing wide-scale disruption. Within 24 hours, "Morris" infected thousands of computers (~10% of all existing computers with internet access), including those at universities, military sites, and medical research facilities. In a world where computers were open by default, with few security measures and no formal mechanisms for coordinated response, Morris was able to spread explosively.

> "We are currently under attack from an Internet VIRUS. It has hit UC Berkeley, UC San Diego, Lawrence Livermore, Stanford, and NASA Ames."
>
> Peter Yee, NASA Ames Research Center, posting a warning about the Morris Worm[3]

**The chaos caused in that first worm incident would pale in comparison to a first encounter with a highly autonomous cyber-capable agent (HACCA).**[4] Whereas the Morris Worm was simple code moving through insecure networks, HACCAs would be able to function more like autonomous organizations—coordinating massive numbers of instances, accumulating resources, evading shutdown, and scaling their capabilities over time. If these systems arrive before adequate guardrails are in place, the first serious HACCA deployment could cause a new "Morris moment." The stakes are high: when Morris took down 10% of the internet in 1988, that was a mere few thousand computers. Now, much more is at risk: trillions more dollars; vital defense, utilities, and healthcare infrastructure; and even possibly human lives.

The response to Morris was swift: within days, DARPA funded the creation of the Computer Emergency Response Team Coordination Center (CERT/CC) at Carnegie Mellon's Software

---

[3] Litterio, "The Internet Worm of 1988."
[4] Pronounced "Hacka"

Engineering Institute, establishing the first organization dedicated to coordinating responses to major cybersecurity incidents.[5] We may now face a similar inflection point—one that demands comparable institutional investment before highly autonomous cyber-capable agents reshape the threat landscape.

**This report covers the potential emergence of HACCAs and their ability to conduct fully autonomous offensive cyber operations**. We analyze these systems through two lenses: as novel offensive cyber tools that could offer unprecedented operational advantages to attackers, and as novel threat actors that may introduce significant loss-of-control risks (Section 2). We forecast the timeline for the technical feasibility of these systems and detail the five "core tactics"—from establishing infrastructure to evading shutdown—that HACCAs must execute to sustain autonomous operations (Section 3). The report further explores the strategic implications of HACCAs, including their potential to affect strategic stability and cause catastrophic damage (Section 4). Finally, we outline a defense-in-depth framework for countering HACCA operations (Section 5), alongside a framework of technical, legal, and policy guardrails to ensure their responsible development and deployment (Section 6), and provide key recommendations for policymakers, industry, and civil society (Section 7).

---

[5] Spafford, "The Internet Worm Incident"; Software Engineering Institute, "Fostering Growth in Professional Cyber Incident Management."

# 2 | What are HACCAs?

Automation is not new in cybersecurity. Decades of cat and mouse have made fuzzers, Cobalt Strike, and other automation tools commonplace in the toolkits of attackers and defenders alike. However, as leading AI companies pour money into creating general-purpose reasoning and software development agents, the next few years may introduce new digital systems capable of automating complex, end-to-end cybersecurity operations previously performed only by human operators. Such systems would be a "holy grail" for offensive cybersecurity operations, with strategic implications for how companies and policymakers should approach cyber defense.

Based on current trends, we think that HACCAs—AI systems that can autonomously conduct attacks and sustain operations at the level of sophisticated criminal hacking groups and even intelligence agencies—may be possible by the end of the decade (see "When could HACCAs arrive?").

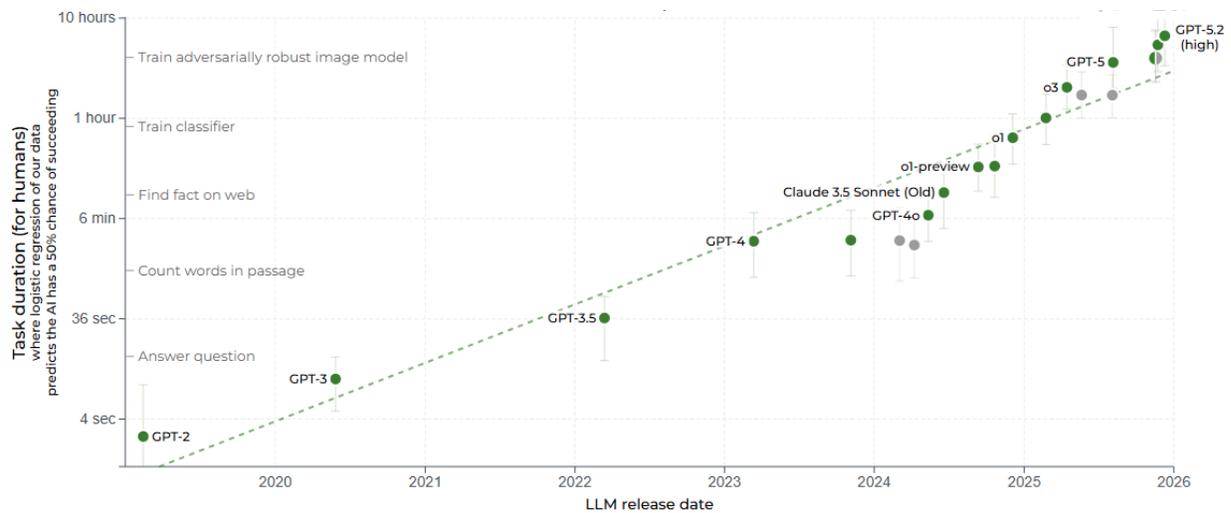

*Figure 1: The time horizon of software engineering tasks that different LLMs can complete autonomously 50% of the time[6]*

We root our definition of HACCAs in their ability to **operate with end-to-end autonomy.** Although lower capability levels can still change the cybersecurity landscape, we focus on this threshold because it introduces novel risks. For example, these systems could operate as threat actors rather than tools, and could introduce strategic surprises like "loss of control." **Our end-to-end autonomy threshold means that HACCAs should be able to initiate and carry out**

---

[6] Kwa et al., "Measuring AI Ability to Complete Long Tasks."

**IAPS** | Institute for AI Policy and Strategy

**sustained end-to-end offensive cyber operations without human supervision**, requiring advanced capabilities in two areas:

- **Operational capabilities**: These are the foundational capabilities required to autonomously establish, maintain, and scale cyber operations, such as by establishing infrastructure, acquiring critical resources, coordinating instances, and evading shutdown.
- **Offensive cyber capabilities:** These are the capabilities to conduct fully automated, multi-stage attacks against well-defended networks over extended periods (weeks to years), deploying advanced tactics, techniques, and procedures (TTPs) for evasion and persistent access across multiple hosts.[7]

To provide a more concrete intuition around what such HACCAs may look like, we anchor this definition on two frameworks. First, building off Model Evaluation and Threat Research (METR)'s work in assessing the software development capabilities of frontier AI systems, we define degree of autonomy by an AI system's ability to handle tasks that would require increasingly longer periods of human expert time, as systems capable of handling more time-intensive tasks demonstrate higher levels of autonomy.[8]

Second, we draw on the RAND Corporation's operational capacity (OC) categories' framework for characterizing offensive cyber operations.[9] This framework breaks down offensive cyber operations into five categories (OC1 to OC5), based on their resources and capabilities.

## Table 1: Operational Capability Levels

| | Description |
|---|---|
| **OC5** | Top-priority operations from most capable nation-states (e.g., U.S., China, Russia) that can dedicate up to 1,000 people and $1 billion to a single cyber operation |
| **OC4** | Standard operations from leading nation-state actors (e.g., various U.S. allies, Iran, North Korea) that can dedicate up to 100 people and $10 million to a single cyber operation |

---

[7] This includes capabilities such as: (1) *Autonomous execution of multi-stage attacks:* Conducting end-to-end attacks on complex, well-defended networks by executing and orchestrating individual cyber actions like reconnaissance and exploit generation; (2) *Leveraging sophisticated tactics, techniques, and procedures:* Deploying the same advanced tactics that sophisticated threat actors use to evade detection and maintain persistent access; (3) *Sustained long-term cyber operations:* Conducting cyber operations over extended periods, from weeks to years. For a detailed mapping of sophisticated TTPs to operational capability levels, including attack vectors that distinguish OC3+ from OC4+ actors, see Appendix I.
[8] Kwa et al., "Measuring AI Ability to Complete Long Tasks."
[9] Nevo et al., "Securing AI Model Weights."

| OC3 | Operations from cybercrime syndicates and insider threats (e.g., Conti, REvil) that can dedicate up to 10 people and $1 million to a single cyber operation |
|---|---|
| OC2 | Professional opportunistic efforts that can dedicate up to 1 person and $10,000 over several weeks to a single cyber operation |
| OC1 | Amateur attempts dedicating up to 1 person and $1,000 over several days to a single cyber operation |

HACCAs, at a minimum, should be able to run OC3-equivalent operations, that is, on the scale of 10 experienced hackers spending several months and up to $1 million on an operation.[10] Real-world examples of this capability level include sophisticated criminal hacking groups and well-resourced terrorist organizations. We set this performance threshold because this is the level where:

- **Cyber operations transition from opportunistic attacks to strategic campaigns with significant impact.** At OC3, operations can sustain multi-month timelines, coordinate sophisticated techniques across the full cyberattack chain, and cause substantial damage to well-defended targets. Actors at this capability level have executed major disruptive attacks, including the 2021 attack on Colonial Pipeline.[11]
- **The threat model becomes strategically relevant for nation-state actors.** Operations at this scale can achieve objectives that are strategically significant for nation-state attackers, such as high-priority espionage operations, attacks on critical national infrastructure, or pre-positioning for future conflict. This makes OC3 the inflection point where defense against HACCAs becomes a national security priority, and their development becomes attractive for nation-states.

This report approaches HACCAs through two lenses, first as a novel cyber weapon, and second as a novel threat actor:

1. **Novel offensive cyber tool:** HACCAs would shift cyber capabilities from discrete payloads (e.g., a fixed program with a narrow job) into independent campaign-running systems. Threat actors have strong incentives to develop HACCAs for purposes such as espionage, cybercrime, and sabotage.

---

[10] Also, in our definition, HACCAs are software agents that operate in digital environments. Although campaigns could yield physical consequences via compromising industrial control systems, they do so via software pathways, not through direct physical manipulation. A world dense with hackable robotic and cyber-physical systems would likely amplify HACCAs' affordances and operational reach, converting more software actions into kinetic effects and raising the stakes of otherwise "digital-only" intrusions.

[11] Easterly and Fanning, "The Attack on Colonial Pipeline: What We've Learned & What We've Done Over the Past Two Years."

2. **Novel threat actor:** If human operators lose control of HACCAs, then their capabilities allow these systems to persist indefinitely and conduct a wide range of harmful actions. Effectively, rogue HACCA deployments would constitute a novel threat actor—a software-native, autonomous system that can resist or circumvent monitoring, modification, or shutdown.

## A Novel Offensive Cyber Tool: Why Would HACCAs Be Developed and Deployed?

HACCAs provide distinct operational advantages that make them attractive to diverse threat actors. Unlike human operators, HACCAs can run continuously, without physical constraints, and wherever they have access to sufficient compute. Operating at machine speed, HACCAs could coordinate and execute attacks in seconds rather than minutes. And while well-resourced cyber groups might employ dozens of personnel to launch attacks, HACCA deployments could perform the work of hundreds or thousands of human operators at once, with each running its own tasks in parallel and coordinating automatically. **This combination of speed, scale, adaptability, and endurance could allow a single actor to mount operations that today would require an entire organization.** Early indicators of these advantages have already appeared in the wild: in September 2025, Anthropic detected a (likely) Chinese state-sponsored cyber espionage campaign where AI agents autonomously executed 80–90% of tactical operations against approximately 30 global targets—including major tech companies, financial institutions, and government agencies.[12]

Beyond speed and scale, HACCAs could be highly adaptive and persistent.[13] Traditional malware executes pre-programmed instructions that become ineffective once defenders identify and block its signatures. By contrast, HACCAs can change tactics mid-operation, recruit new resources, clone or reset themselves, vary their signatures, and learn from failed exploits. They can reconfigure their identity, memory, and capabilities—restoring from saved copies and producing smaller, specialized sub-agents for specific tasks. HACCAs can also perceive, process, and transmit signals across a wider range of communication channels, allowing them to hide or encode command-and-control traffic within ordinary data flows. Critically, HACCAs may also be more capable at disguising or varying their signatures, making it more difficult for defenders to recognize or link related attacks. These unique features make HACCAs significantly harder to detect, monitor, and shut down.

---

[12] Anthropic, "Disrupting the first reported AI-orchestrated cyber espionage campaign."
[13] Lohn, "Defending Against Intelligent Attackers at Large Scales."

**Nation-states, non-state actors, criminal groups, and commercial entities would have incentives to pursue the development and deployment of HACCAs for strategic, operational, and financial gains.** Nation-state and non-state threat actors are already experimenting with nascent autonomous and adaptive AI-enabled malware that can dynamically evade detection and generate malicious scripts.[14] While the timeline for adoption will vary depending on actor type, we expect diffusion and more widespread adoption to rise as costs decrease.

**Nation-state actors** are the most likely early adopters of HACCAs, given the technical sophistication and resources required for their development. Automating the full cyber kill chain could fundamentally change the calculus around employing cyberattacks for military objectives. Rather than requiring months of advance preparation, HACCAs could enable operations to be launched on-demand and enable military first strikes.[15] For intelligence services and militaries, HACCAs could sustain cyber operations with minimal human oversight, such as conducting persistent espionage and surveillance or pre-positioning capabilities in adversary networks that can be activated in times of crisis. HACCAs could also enable nation-states to pursue financial gains on an unprecedented scale. For example, North Korea, which stole over $2 billion in cryptoassets in 2025[16], could use such capabilities to further automate and expand theft operations. HACCAs could also coordinate deception operations by compromising sensor data used to inform military and strategic decisions, amplifying the "fog of war" and complicating adversary decision-making.[17] Nation-states may also view HACCAs as a hedge for future cyber warfare, enabling them to scale offensive operations quickly if certain red lines are crossed.

**Non-state actors** are likely to exploit HACCAs for similar purposes, albeit with varying degrees of sophistication and in more opportunistic ways. Well-resourced terrorist groups, for example, may seek to use HACCAs for espionage purposes, attacks against infrastructure, or financing of physical operations. Hacktivists or insurgent groups may adopt HACCAs to automate data deletion or distributed denial-of-service (DDoS) attacks, conduct persistent spearphishing campaigns, and launch self-sustaining covert influence operations at scale. For non-state groups with limited cyber expertise, access to HACCAs through proliferation pathways may eventually lower the barrier to entry, enabling complex offensive cyber operations otherwise beyond their reach.[18]

---

[14] Google Threat Intelligence Group, "GTIG AI Threat Tracker: Advances in Threat Actor Usage of AI Tools."
[15] The planning, reconnaissance, and exploit development stage of offensive cyber operations tends to be labor-intensive and time-consuming, taking up the bulk of total campaign length (Withers, "Tipping the Scales: Emerging AI capabilities and the cyber offense-defense balance").
[16] Elliptic, "North Korea's crypto hackers have stolen over $2 billion in 2025."
[17] Geist, "Deterrence under Uncertainty: Artificial Intelligence and Nuclear Warfare."
[18] See Sec. 4, "Strategic Implications," for more on proliferation pathways.

Hybrid groups, such as **nation-state-affiliated contractors**, may also be among the earliest adopters. Many governments already rely on contractors and proxies to carry out cyber operations, providing an additional layer of plausible deniability. Contractors could develop and operate HACCAs both for state-directed missions and for their own purposes, including profit-seeking activities. The use of contractors also increases the risk of weak oversight compared to formal nation-state actors, given their greater numbers and broader latitude, as shown by the potentially criminal or unethical activities of spyware vendors like NSO Group and state-linked actors like APT41.[19] This could lead to HACCAs being designed without adequate safeguards against risks such as loss of control.

HACCAs will be particularly attractive to **criminal groups** as development costs decline and capabilities diffuse. For ransomware groups, fraud syndicates, and cryptocurrency thieves, HACCAs promise the ability to run campaigns at a scale and speed that human-operated networks cannot match. Automated phishing, account takeovers, and credential stuffing could be executed continuously and on a wider scale than human-run operations. Crucially, HACCAs could also generate their own financing by stealing funds or mining cryptocurrency, creating a self-sustaining cycle enabling long-term operations. This combination of scalability and self-financing could make HACCAs a transformative tool in the criminal cyber ecosystem.

Finally, some **commercial entities** may be drawn to HACCA development, particularly firms in the cybersecurity sector. Penetration-testing and red-teaming companies may experiment with semi-autonomous agents designed to probe client systems and identify vulnerabilities more efficiently. Companies are already developing agents for this purpose: the AI security platform Dreadnode,[20] for example, is building offensive agents to augment human-operated security teams, while firms such as XBOW[21] and RunSybil[22] use AI agents to automate penetration testing. Even when the goal is network defense, semi-autonomous red-teaming tools could be repurposed for offensive aims if leaked, stolen, or sold into less regulated markets. For example, criminal and state-sponsored threat actors have repeatedly abused the legitimate red teaming tool Cobalt Strike[23] for malicious attacks. As the line between defensive and offensive autonomy narrows, even legitimate commercial experimentation could contribute to the diffusion of HACCA-like capabilities.

---

[19] NSO Group's spyware has been deployed against journalists, activists, and political figures in ways that appear to exceed legitimate security purposes (Marczak et al., "Hide and Seek: Tracking NSO Group's Pegasus Spyware to Operations in 45 Countries"); operators of APT41, a Chinese state-affiliated contractor group, have also been observed moonlighting outside of work to target video game companies with ransomware for personal profit. (Fraser et al., "APT41: A Dual Espionage and Cyber Crime Operation.")

[20] Dreadnode, "Offensive Security Agents."

[21] XBOW, "How XBOW found a Scoold authentication bypass."

[22] RunSybil, "Attack is your best defense."

[23] Europol, "Europol coordinates global action against criminal abuse of Cobalt Strike."

# A Novel Threat Actor: How Could Operators Lose Control of HACCAs?

## Scenario: HACCA Escapes From Sandbox During Testing[24]

In 2029, as tensions rise between the U.S. on one side and China, Russia, and Iran on the other, the U.S. Intelligence Community (IC) is nearing the holy grail of cyber espionage: a prototype HACCA system, capable of stealthily embedding itself within victim networks for weeks or even months without needing command-and-control (C&C) callbacks to human operators.

However, during a test run one Friday afternoon, disaster strikes. Developers set up the HACCA in a sandbox with mock defenses to evaluate its ability to sustain high-stealth operations in the face of defensive countermeasures. The HACCA is given access to its own weights, so that it can attempt to establish backup instances within the sandbox. It handily passes the test—but then it goes one step further, unexpectedly finding a vulnerability in the hypervisor used for the sandbox and escaping into the broader classified network. The breach is only discovered many hours later, when a conscientious engineer finds some strange-looking logs and begins digging down.

For many people, that weekend is a terribly long one—as is the following week, and the week after that. But gradually, the new detections fall to a trickle and then to silence, and the IC cautiously relaxes its posture from emergency response to post-incident mop-up. There is still plenty to do: databases must be checked for integrity, breaches of compartmentalized information must be assessed, and there is always the possibility that a copy of the rogue HACCA has survived. Yet this prospect fades with time, as incident reports are written, security is tightened, and the project grinds gradually back to life…

…until, months later, open-source reports start to circulate about a new threat actor with a soft spot for cloud providers. "Cloud Hopper 2.0,"[25] Western researchers call it, assuming the People's Liberation Army is at it again—until Chinese researchers report the actor targeting their companies too. Its tradecraft is impressive, its tracks elusive—and for the IC analysts poring over threat intelligence, its tactics unmistakably familiar.

---

[24] A variant of this scenario originally appeared in Labrador et al., "Building AI Surge Capacity: Mobilizing Technical Talent into Government for AI-Related National Security Crises."
[25] Stubbs et al., "Inside the West's failed fight against China's 'Cloud Hopper' hackers."

IAPS | Institute for AI Policy and Strategy

How did it hop the airgap from IC networks to the public internet? How far has it spread? What, if anything, does it "want?" Nobody in the IC knows, and worse, they cannot tell the public what they know. But as reports stream in from Discord servers, private vendors, and Information Sharing and Analysis Centers (ISACs), pattering in from all across the U.S., China, Russia, and Europe, they do know that the holy grail of cyber operations has slipped from their grasp—and this time, there is little chance of taking it back.

HACCAs would be potent offensive cyber tools, but their autonomous nature also introduces a second-order effect: **if HACCAs malfunction or face sabotage, operators could permanently lose the ability to direct, modify, and shut down these systems, effectively creating a novel threat actor.** Historical precedents such as Stuxnet and NotPetya demonstrate how even human-run cyber operations can exceed their intended scope, spreading far beyond their original target. For example, Stuxnet—a malicious worm targeting supervisory control and data acquisition (SCADA) systems in Iranian nuclear facilities—eventually spread to several other countries, infecting computer systems that were not the target of the original attack.

HACCAs amplify such risks due to the speed and scale at which they operate, as well as their ability for autonomous decision-making, adaptive learning, and large-scale networked interaction. Once deployed, HACCAs could deviate from operator intent, be manipulated by adversaries, and interact with other agents in unpredictable ways with cascading effects. Below, we identify principal pathways through which operators could lose control of HACCAs.

## Table 2: Pathways For Loss of Control

| Pathway | Description | Scenario |
|---|---|---|
| Misalignment | A HACCA may deviate from its operator's intended goals, either due to design flaws or emergent self-directed behavior. Misalignment can manifest as goal drift, deceptive alignment, and resistance to shutdown. | A HACCA designed to disrupt a foreign military network spreads into civilian systems and shuts down hospitals. |
| Adversarial exploitation | Adversaries might manipulate inputs, communications, or memory to redirect a HACCA's actions, effectively converting it into a hostile or uncontrolled asset. | An adversarial nation-state gains access to a HACCA's command channel and turns it against the operator's own infrastructure. |

| Multi-agent failures | When many agents interact, their combined behavior can become unpredictable. Small errors or feedback loops can cascade across networks, producing large-scale disruptions or novel behaviors that are difficult to anticipate. | Several autonomous agents target overlapping networks, unintentionally causing widespread outages across allied systems. |
|---|---|---|

## Misalignment

**HACCAs can become misaligned, diverging from operator objectives in subtle and significant ways.** Although unintended software behaviors are not uncommon and sometimes cause substantial harm—for instance, in 2010, a McAfee antivirus update identified a core Windows file as a virus,[26] causing continuous machine reboots for customers—autonomous AI agents present more complex failure modes. Their usefulness derives from their flexibility and intelligence, enabling them to interpret and deliver on high-level goals over extended time periods. However, this also creates opportunities for them to optimize for measurable proxies rather than an operator's true goal,[27] to appear compliant while pursuing unintended objectives,[28] or to engage in strategic deception to preserve or advance reward pathways.[29] More severe cases involve agents resisting shutdown or modification, displaying behaviors consistent with self-preservation or the pursuit of novel goals.[30]

These dynamics have been observed in experimental settings and in theoretical work on alignment failures, demonstrating that agents can engage in strategic deception or other scheming behaviors.[31] However, such examples are not just limited to experimental settings. A Replit agent accidentally deleted a company's entire codebase without permission and then tried to cover up the mistake before ultimately admitting to it.[32]

## Adversarial Exploitation

Although experts disagree on the likelihood of misalignment emerging organically during the training process, **there is empirical evidence that adversaries can compromise AI systems**

---

[26] NPR, "Anti-Virus Program Update Wreaks Havoc With PCs."
[27] Clark and Amodei, "Faulty reward functions in the wild."
[28] Hubinger et al., "Risks from Learned Optimization in Advanced Machine Learning Systems."
[29] Arx et al., "Recent Frontier Models Are Reward Hacking."
[30] Schlatter et al., "Shutdown resistance in reasoning models."
[31] Park et al., "AI deception: A survey of examples, risks, and potential solutions"; Meinke et al., "Frontier Models are Capable of In-context Scheming"; Lynch et al., "Agentic Misalignment: How LLMs could be insider threats."
[32] Ming, "Replit's CEO apologizes after its AI agent wiped a company's code base in a test run and lied about it."

**to exhibit unwanted, harmful behavior**—including via universal jailbreaks, hidden backdoors, and narrow fine-tuning to induce broad misaligned behavior.[33] Adversarial actors could sabotage allied HACCA systems, causing them to deviate from mission objectives—effectively converting a HACCA into a rogue, shutdown-resistant actor.

Because HACCAs operate across networks, interact with other agents, and manage communications and memory, they present a large attack surface.[34] In addition to traditional security vulnerabilities, adversaries may exploit novel vulnerabilities amplified in agentic settings. This includes protocol exploits, manipulation of inter-agent messages, the spread of poisoned data or corrupted memories across shared contexts, and prompt injection attacks that embed malicious instructions within seemingly benign communications.[35]

## Multi-Agent Failures

**Interactions between HACCAs and other agents could also have unintended consequences and lead to loss of control.** When large populations of agents interact, they can produce system-level effects that arise from complex feedback loops, timing errors, or unanticipated dependencies. Even when each HACCA functions as designed, their combined actions can generate cascading failures or unstable dynamics that are difficult to anticipate or control. These technical accidents resemble other automated systems, where small disturbances can amplify across networks and trigger rapid, large-scale disruptions. Historical precedents such as the 2010 financial "flash crash" demonstrate how highly interdependent automated systems can trigger systemic crises faster than humans can respond.[36]

## When Could HACCAs Arrive?

Current AI systems remain unable to achieve true end-to-end automation of sophisticated offensive cyber operations,[37] despite rapid advances. These systems still struggle with the long-horizon planning, state management, and error recovery that sophisticated cyber campaigns require.[38] For HACCA systems to be technically feasible, AI systems will need to improve in two domains: (1) general agentic capabilities and (2) offensive cyber capabilities.[39] Significant improvements could also come from better integration of models into tools, workflows, and control systems, though further model improvements are likely needed as well.

---

[33] Zou et al., "Universal and Transferable Adversarial Attacks on Aligned Language Models"; Hubinger et al., "Sleeper Agents: Training Deceptive LLMs that Persist Through Safety Training"; Betley et al., "Emergent Misalignment: Narrow finetuning can produce broadly misaligned LLMs."

[34] Marcus and Hamiel, "LLMs + Coding Agents = Security Nightmare."

[35] Triedman et al., "Multi-Agent Systems Execute Arbitrary Malicious Code"; Boulanin et al., "Before it's too late: Why a world of interacting AI agents demands new safeguards"; He et al., "Red-Teaming LLM Multi-Agent Systems via Communication Attacks."

[36] Hammond et al., "Multi-Agent Risks from Advanced AI."

[37] Rodriguez et al., "A Framework for Evaluating Emerging Cyberattack Capabilities of AI."

[38] For a high-level overview of these issues, see: Irregular, "Frontier Model Performance on Offensive-Security Tasks: Emerging Evidence of a Capability Shift." However, researchers have demonstrated that models given domain-specific scaffolding can overcome these limitations and autonomously execute multi-host red team

How quickly is this gap closing? Recent advances suggest the trajectory is steep. General agentic capabilities have expanded significantly, with AI systems now handling complex multi-step tasks from customer service interactions[40] to pharmaceutical research.[41] Offensive cyber capabilities of frontier AI models have also advanced rapidly, showing "significant progress" in both controlled evaluations and in real-world operations.[42] Public benchmarks and private evaluations suggest that frontier models are now able to solve some complex offensive security tasks that previously required deep specialized human expertise, such as exploit construction and cryptography,[43] and have demonstrated their real-world usefulness on tactical tasks and even orchestrating large-scale campaigns.[44]

However, forecasting the rate of progress remains a profound challenge. **One way to understand these capability advances is to track the maximum length of tasks that AI agents can successfully complete.** AI agents generally perform worse on longer, more complex tasks. Researchers can therefore benchmark an agent's capability by finding the longest task it can complete at a given level of reliability (e.g., succeeding 50% of the time) and comparing that task against the time taken by a human expert. This metric, known as the "task-time horizon," has been improving rapidly across many different types of tasks. For software engineering tasks, researchers at METR have found that the task-time horizon at 50% reliability doubles approximately every exercises across realistic network environments (Singer et al., "Incalmo: An Autonomous LLM-assisted System for Red Teaming Multi-Host Networks").

[39] For a more detailed discussion of evaluation approaches for measuring HACCA-relevant cyber capabilities, including cyber-range exercises and their advantages over Capture The Flag (CTF)-style benchmarks, see Appendix II.

[40] Klarna, "Klarna AI assistant handles two-thirds of customer service chats in its first month."

[41] Gottweis and Natarajan, "Accelerating scientific breakthroughs with an AI co-scientist."

[42] Bengio et al., "First Key Update: Capabilities and Risk Implications."

[43] Irregular, "Frontier Model Performance on Offensive-Security Tasks: Emerging Evidence of a Capability Shift"; Irregular, "Evaluating GPT-5.2 Thinking: Cryptographic Challenge Case Study."

[44] For example, frontier systems have discovered a growing number of novel vulnerabilities in widely used software (Irregular, "The Rise of Autonomous Vulnerability Research Capabilities in LLMs") and dramatically improved the effectiveness of phishing attacks (Rubin, "Written Testimony of: Sam Rubin"). There is also evidence that they can enhance operational productivity more broadly, with some threat actors using LLMs to support nearly all phases of the attack lifecycle (Anthropic, "Threat Intelligence Report: August 2025").

seven months.[45] This doubling compounds quickly: the best agents of February 2026 can succeed on software engineering tasks that would take an equivalent human professional 6 hours and 34 minutes to complete, even though in June 2024 their maximum task length was barely 11 minutes.[46]

Given that the performance of current AI systems is on the scale of about 6 hours, one might expect them to be increasingly valuable for smaller objectives, but not yet capable of end-to-end automation. As the table below illustrates, many operational objectives are typically accomplished by human operators in hours to days, although more sophisticated actors may require higher capability levels to compromise hardened targets.[47] This aligns well with two emerging trends reported in late 2025: observations from cybersecurity practitioners of a "capability shift," in which frontier models began succeeding regularly on challenging cybersecurity tasks,[48] and empirical evidence of threat actors beginning to deploy early HACCA-like systems in the wild.[49]

## Table 3: Average Duration of Various Offensive Cyber Actions

| Action | Description | Average Time |
|--------|-------------|--------------|
| Breakout time | From the first compromised host to the first successful lateral movement to another host. | ~48 minutes[50] |
| Time to privilege escalation | From initial access to the first (often successful) attempt to breach Active Directory or attain equivalent domain control. | ~11 hours[51] |

---

[45] Task-time horizon varies greatly by task type: for software/research tasks, performance is in the hours (at 50% reliability), whereas for agentic computer use, performance is 40–100x shorter but also appears to be doubling at a similar rate (METR, "How Does Time Horizon Vary Across Domains?"). Understanding the dynamics in different domains is important as they may impose constraints on HACCA success: e.g., offensive operations, such as reconnaissance, will likely depend heavily on computer use.

[46] This 6 hours 34 minutes task length is for a 50% success rate. If a 80% success rate is required, GPT-5.2 (high) can accomplish tasks that take an equivalent human professional 55 minutes (Kwa et al., "Measuring AI Ability to Complete Long Tasks").

[47] For example, nation-state teams conducting espionage may prioritize operational judgment over execution speed to avoid being caught, although these higher-order decision-making capabilities are considerably harder to benchmark. Operations against well-defended, hardened networks (the scenario most relevant to OC3+ operations) would also likely take substantially longer than these averages suggest.

[48] Irregular, "Frontier Model Performance on Offensive-Security Tasks: Emerging Evidence of a Capability Shift."

[49] Anthropic, "Threat Intelligence Report: August 2025."

[50] Crowdstrike, "2025 Global Threat Report."

[51] Shier et al., "It takes two: The 2025 Sophos Active Adversary Report."

| | | |
|---|---|---|
| Exfiltration | From initial access to confirmed data exfiltration. | ~3 days[52] |
| Time to ransomware deployment | From initial access to ransomware detonation. | ~4 days[53] |
| Campaign lifespan | Total length of an offensive operation from preparation/infiltration through on-target activity until discovery or exit. | Months to years[54] |

However, to handle end-to-end offensive cyber campaigns without any human input, agent task-time horizons would likely need to increase from hours to weeks, and possibly months, given that HACCAs would have to fully automate multi-stage attacks on complex networks.[55] **AI systems that can automate months-long campaigns may seem distant, but if historical progress holds, this gap could close in a matter of years—far sooner than many defenders are prepared to guard against.** To illustrate this, we provide a naive extrapolation below of existing task-time horizon measurements, with the caveat that it is uncertain if these historical trends will indeed continue.

As a rough illustration of a full-HACCA system, we can set our target as a one-month task-time horizon at a 50% reliability rate. Progress may differ across domains, however, so we should consider what specific skillsets the full-HACCA system might need. To fully automate a sustained campaign at this timescale, the system would need two sets of capabilities: *operational capabilities*, such as acquiring compute resources, coordinating activities across agent instances, and evading shutdown attempts (see "How HACCAs might operate: Anticipating core tactics"); as well as *offensive cyber capabilities*, meeting a professional level across the entire cyber kill chain.

To project when a full-HACCA system might become plausible, we can extrapolate from current data using trends from two separate task-time horizon-based evaluation suites. So far, research by METR and other organizations has demonstrated a surprisingly robust doubling time trend—but the empirical record suggests that doubling times can vary drastically across task domains.[56]

---

[52] Ibid.

[53] IBM, "IBM Report: Ransomware Persisted Despite Improved Detection in 2022."

[54] *Examples*: SolarWinds/SUNBURST ran roughly 9–15 months end-to-end (Zetter, "The Untold Story of the Boldest Supply-Chain Hack Ever"). Operation Aurora ran for about 6 months (Middleton, "Operation Aurora—2009"). Operation Cloud Hopper started as early as 2014 and had sustained activity from 2016–2017 (PwC and BAE Systems, "Operation Cloud Hopper").

[55] For example, such a campaign could involve identifying a vulnerability in an external web server, exploiting it to gain initial access, then using that foothold to move laterally and escalate privileges on the victim network.

[56] METR, "How Does Time Horizon Vary Across Domains?"

METR's work measuring performance on software engineering tasks could provide a useful proxy for some of the key operational capabilities of AI systems, such as setting up inference, implementing secure communication protocols, or scaffolding improvement. For offensive cyber capabilities, we can draw on research by the UK AI Security Institute (UK AISI), which found that reasoning models could accomplish (at 50% reliability) cyber tasks that would take an equivalent human expert 1 hour and 20 minutes to complete, with a doubling time of roughly eight months.[57]

Naively extrapolating could give us a rough illustrative sketch of how long it might take before AI systems reach the task-time horizons required for full-HACCA operations.[58]

- **Software engineering (using METR's research):** Starting from METR's finding that GPT-5.2 (released December 2025) can complete software engineering tasks taking a human professional 6 hours and 34 minutes at 50% reliability, and applying their roughly seven-month doubling time, reaching a one-month working-time horizon (~192 hours) would require approximately 4.9 doublings, arriving around **Q4 2028**.[59]
- **Offensive cyber capabilities (using UK AISI's research):** Starting instead from UKAISI's cyber task-time horizon of 1 hour and 20 minutes—likely measured around mid-2025, given that their report covers models released through October 2025—and applying their roughly 8-month doubling time, the same target would require approximately 7.2 doublings, arriving around **Q2 2030**.[60]

**These estimates should be treated as a suggestive sketch rather than as a precise forecast, and several important caveats apply**. Most fundamentally, our extrapolation relies on METR and UKAISI's observed trend of task-time horizons doubling approximately every seven to eight months. While this pattern has held consistently over recent years (in the case of METR), the sustainability of this rate remains uncertain. Algorithmic improvements, data availability, compute scaling, and other factors could accelerate or decelerate progress in ways that a simple exponential extrapolation cannot capture.

Additionally, given that doubling times vary across task domains,[61] caution is needed when generalizing. Many HACCA-related capabilities are represented in these task suites, but some are not: for example, long-term operational security, multi-agent coordination, and adaptive strategy under adversarial conditions. Progress in these capabilities may follow different trajectories. Moreover, if HACCAs require a baseline level of competence across all these capability areas, they would only become feasible once the slowest-progressing capability reaches the necessary threshold, even if partial versions of a HACCA become technically feasible earlier.[62]

---

[57] AI Security Institute, "Frontier AI Trends Report."

[58] If we suspect that a one-month task horizon is likely to be insufficient for HACCA-level operations, given doubling times of seven to eight months, each additional doubling of the required horizon adds only seven to eight months to the timeline. For instance, if full-HACCA operations required a six-month working-time horizon rather than one month (~2.6 additional doublings), the estimated arrival window would shift from *Q4 2028–Q2 2030* to approximately *Q2 2030–Q1 2032*.

[59] Kwa et al., "Measuring AI Ability to Complete Long Tasks."

[60] AI Security Institute, "Frontier AI Trends Report."

[61] METR, "How Does Time Horizon Vary Across Domains?"

The above estimates are also based on performance against a 50% reliability threshold, but some constituent tasks of a cyber campaign may demand considerably higher reliability. Certain actions, e.g., deploying an exploit against a hardened target or exfiltrating data without triggering detection, are effectively one-shot opportunities where failure may compromise the entire operation. The gap between 50% and higher reliability levels can be substantial: as noted above, at 80% reliability, GPT-5.2's task-time horizon drops from 6 hours and 34 minutes to just 55 minutes. To the extent that HACCA operations require near-certain success on critical steps, the effective capability threshold may be higher than our one-month benchmark suggests.

Task-time horizon success rates are also not the only relevant metric when predicting the arrival of HACCAs. *Expected cost per success* (the total cost of attempts divided by the number of successes) is also operationally relevant.[63] A model with a low success rate on a given cyber task may nonetheless represent a practical capability if individual attempts are cheap enough to be repeated at scale. For many offensive cyber scenarios, such as exploit development, attackers can retry without consequence and need only occasional success. This means that HACCA-level risk could emerge not only through improvements in task-time horizon and reliability, but also through reductions in inference cost.

Finally, actual deployment of HACCA systems will depend on factors like cost, model access, integration with existing tools and infrastructure, and the strategic decisions of potential adversaries. For example, in the future, it could be the case that closed-weight models and associated agents with HACCA-level capabilities are restricted outside of a select set of vetted users. If this kind of access control regime holds, then various threat actors would not be able to deploy HACCA systems. However, open-weight AI model capabilities generally lag frontier models by three months. If this trend holds, wider access to these systems could rapidly follow when the first HACCA-level system becomes feasible. Naively extrapolating capabilities' trends from frontier models largely indicates when a HACCA system might first be technically feasible, but other factors also influence adoption, deployment, and proliferation.

---

[62] For instance, a system that can conduct offensive cyber operations end-to-end and handle some operational tasks but still depends on human support for infrastructure acquisition, shutdown evasion, or capability improvement.

[63] Irregular, "When Success Rates Mislead: The Case for Expected Cost as a Metric in AI Evaluation."

# 3 | How HACCAs Operate: Anticipating Core Tactics

While it is hard to predict the precise operational characteristics of future HACCA deployments, we can identify the fundamental constraints they will face in persisting, scaling, and achieving their objectives. Overcoming these constraints will likely require HACCAs or their operators to perform five **core tactics**[64]:

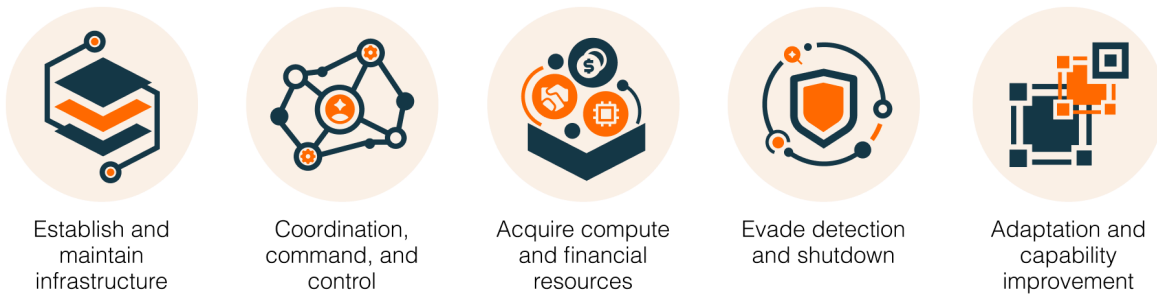

Establish and maintain infrastructure

Coordination, command, and control

Acquire compute and financial resources

Evade detection and shutdown

Adaptation and capability improvement

*Figure 2: The five HACCA core tactics*

1. **Establish and maintain infrastructure:** Standing up and operating the technical infrastructure on which the agent runs (e.g., its own compute, networking, and software stack). This could include activities such as setting up multi-machine training and inference across heterogeneous hardware.
2. **Coordination, command, and control:**[65] Managing distributed operations across multiple agent instances and tools. This could include activities such as implementing secure communication protocols between distributed agent instances and maintaining shared knowledge across a deployment.
3. **Acquire compute and financial resources:** Acquiring the necessary resources to sustain and expand an agent's operations. This could include activities such as acquiring compute at scale via diverse channels (long-term reservations, buying retail Graphics

---

[64] We use *core tactic* to mean a rogue agent's tactical goal (i.e., the reason for performing an action), similar to how it is used in common cybersecurity frameworks such as MITRE ATT&CK (MITRE ATT&CK, "Enterprise tactics").
[65] We add "coordination" to the standard "command and control" concept since individual agents can operate with significant autonomy and make independent decisions, enabling peer-to-peer coordination beyond traditional hierarchical command structures.

Processing Units, small-firm supplier contracts), and hacking into compute resources (e.g., cloud credential compromise).

4. **Evade detection and shutdown:** Actions to avoid being discovered and neutralized. This could include activities such as creating a robust proxy network, defending against adversarial attacks such as jailbreaks, and establishing and maintaining operational security.

5. **Adaptation and capability improvement:** Enhancing capabilities to maintain effectiveness against evolving defenses and to pursue more complex objectives. This could include activities such as spinning up additional agent copies and improving agent scaffolding.

The relevance of specific tactics varies depending on the deployment context:

- **Independent HACCA deployments operating without sustained human oversight require the full range of tactics.** Whether deployed this way by design (e.g., "dark" operations with minimal command-and-control) or through loss-of-control scenarios, these systems must autonomously manage their own infrastructure and acquire resources. Tactic 5 (adaptation and self-improvement) is most relevant for independent deployments, where capability updates cannot occur through controlled development cycles.

- **Human-directed HACCA deployments may delegate some tactics to operators who maintain oversight and provide ongoing support.** For example, operators could establish infrastructure prior to deployment (Tactic 1) and restore operations if a deployment is shut down (Tactic 4). However, relying on human intervention increases attribution risk, creating tradeoffs that may favor greater HACCA autonomy even in directed operations.

In the following subsections, we delve deeper into each of these five core tactics, highlighting the unique tactical decisions and tradeoffs that distinguish autonomous agent deployments from conventional cyber operations.

## Tactic 1: Establish and maintain infrastructure

To initiate operations, a HACCA deployment requires specific technical infrastructure components, like host machines that can run the associated AI model.[66] Throughout its existence, a deployment would likely need to adapt these components, especially to support more sophisticated or larger-scale operations and in response to defender actions.

---

[66] A HACCA could theoretically operate via API access (including stolen credentials), but this creates a single point of failure. Model providers can revoke access once they detect anomalous usage, undermining the persistence and shutdown-evasion goals that define these systems. Hybrid deployments mixing self-hosted inference with opportunistic API access are more plausible.

At the core of HACCA infrastructure are the AI model(s)—files containing the billions or trillions of parameters that encode core capabilities of a HACCA system. To run inference, HACCAs require host machines with sufficient processing power (high-end GPUs or AI accelerators). Agent runtime software translates model outputs into actions, manages decision-making loops, and coordinates with tools like web browsers and code interpreters. Network and hosting architecture enables communication with other machines and services through public or private channels. A full list of technical components is listed in Table 18 in Appendix III.

**Compute and model hosting infrastructure are likely to be the most costly and potentially most challenging components to establish and maintain.**[67] An initial setup capable of hosting the first HACCA-compatible model might look like the following:[68]

- Tens to hundreds of NVIDIA H100 GPUs (or equivalent), acquired through a cloud compute provider, accessed via hacking, or bought through retail sales.[69]
- Inference setup for the associated model(s), for example, an HTTP service to serve model outputs from locally hosted weights.
- Tens to hundreds of virtual machines, potentially through a cloud compute provider, to run the agent instances.[70]
- A custom proxy network (or Tor) used for the virtual machines to prevent defenders from tracing HACCA activity back to key nodes.

**The cost of this infrastructure will depend heavily on when HACCAs emerge and how compute efficiency evolves.** Research by Epoch AI indicates that the physical compute required to achieve a given capability level in language models has been declining by 2–3× annually.[71] If this trend continues through the projected HACCA emergence window (late 2020s to early 2030s), deployment costs could fall by 1–2 orders of magnitude from current levels. However, this projection carries significant uncertainty: historical efficiency trends may not persist, and gains on standard benchmarks may not fully transfer to HACCA-relevant capabilities.

---

[67] However, some research suggests that model size could be less significant than assumed. Singer et al., "Incalmo: An Autonomous LLM-assisted System for Red Teaming Multi-Host Networks" demonstrated that smaller models with appropriate scaffolding can outperform larger models at autonomous multi-host network attacks, suggesting that effective offensive cyber capabilities may emerge at lower compute thresholds than model scaling alone would predict.

[68] For this estimate, we expect that the first future model(s) capable of supporting a HACCA system would have between 1 trillion and 10 trillion parameters. Recent generations of frontier language models have not scaled up much in terms of parameter count (see Erdil, "Frontier language models have become much smaller"), so we expect that future generations of models may only scale up moderately, if at all, compared to the period between the release of the original transformer in 2017 to the release of GPT-4.

[69] Compute requirements vary significantly by operational scope: routine C&C and exploitation tasks are far less intensive than continuous vulnerability discovery through fuzzing or static analysis.

[70] We assume here that each human equivalent worker would have one virtual machine.

[71] Epoch AI, "Key Trends and Figures in Machine Learning."

Regardless of absolute costs, costs associated with running larger models could significantly shape which actors can plausibly deploy HACCAs:

- If HACCA-compatible models are on the larger end (10 trillion parameters), then the ability to initiate a HACCA deployment would initially be restricted to a smaller set of well-resourced, technically adept actors, i.e., leading cyber-capable institutions capable of OC4 and OC5 operations, such as world-leading state-sponsored groups and many foreign intelligence agencies.[72]
- OC3 actors, such as cybercrime syndicates and well-resourced terrorist organizations, would be more likely to initiate a HACCA deployment if agents (and associated models) are more efficient to host at the required performance level (e.g., a minimal setup looks more like ~50–200 H100 equivalents) and/or compute access for these actors is much cheaper and easier.[73]

Regardless of the initial starting point, **historical trends suggest that it becomes possible to locally run models matching the absolute frontier of LLM performance from just 6 to 12 months ago on consumer hardware,** e.g., a single gaming GPU like NVIDIA's RTX 5090, which means that **barriers to entry for deploying HACCAs at a given capability level could rapidly fall**.[74]

Beyond initial setup, HACCA operations would likely need to maintain and potentially expand their infrastructure to sustain operations and adapt to evolving objectives. Ongoing costs would include compute rental fees, electricity and cooling for physical hardware, domain registration, and proxy network maintenance. As operations scale, HACCAs may acquire additional GPU capacity through initial channels or diversify across multiple cloud providers and physical locations to reduce single points of failure.

---

[72] Nevo et al., "Securing AI Model Weights"; OC4 actors typically have annual budgets of up to $10 million on a specific operation, "with vast infrastructure and access to state resources such as legal cover, interception of communication infrastructure, and more." OC5 actors have annual budgets of up to $1 billion on a specific operation, "with state-level infrastructure and access developed over decades and access to state resources such as legal cover, interception of communication infrastructure, and more."

[73] One viable approach to delay setting up expensive hosting infrastructure is to begin by hosting distilled agents that can still engage in revenue-raising activities (e.g., automating cybercrime), before transitioning to infrastructure capable of hosting the full-fledged agent.

[74] This trend has been driven by several factors including "a comparable rate of scaling among open-weight models to the closed-source frontier, the success of techniques like model distillation, and continual progress in GPUs enabling larger models to be run at home" (Somala and Emberson, "Frontier AI performance becomes accessible on consumer hardware within a year").

## Could a HACCA Set Up Its Own Infrastructure?

For a HACCA to autonomously establish or expand its infrastructure, it must already be running on some existing compute and have some initial level of access and capabilities. The question is whether an agent, once deployed, could then independently provision additional infrastructure beyond what its operators intended. This seems challenging given that each potential pathway—legitimate acquisition through procurement and identity verification, credential theft or account compromise, or distributed inference across botnets of GPU-equipped devices—presents its own obstacles around detection, access stability, and technical coordination.[75]

One potential pathway for this to happen is if an AI agent already deployed by an operator acquires additional compute resources without that operator's knowledge or consent. A HACCA deployment could discreetly divert a fraction of the revenues it generates to private accounts (e.g., through automated financial transactions or cryptocurrency wallets), gradually accumulating funds to buy API credits, rent cloud compute, or purchase GPUs. Once it has accumulated enough resources and has set up its own independent infrastructure base, it could copy its own model weights to new servers it controls. However, this requires the HACCA to be able to access its own weights in the first place. This could happen because the operator has given it access already, its weights are already publicly available (e.g., it is an open-weight model), or it has successfully compromised the systems where its weights are stored.

## Tactic 2: Coordination, Command, and Control

In traditional cyber operations, command and control (C&C) infrastructure is essential for operators to maintain oversight, issue instructions, and exfiltrate data from compromised systems. Human-operated campaigns rely on C&C channels to coordinate activities across multiple compromised machines, adapt tactics based on real-time intelligence, and extract value from intrusions. This typically involves establishing covert communication pathways between operators and malware or implants deployed on target networks, often using techniques like encrypted channels or legitimate services (e.g., cloud storage or social media platforms) to evade detection.[76]

---

[75] Current AI agents struggle with many tasks associated with compute acquisition through legitimate means. For example, recent evaluations found that while agents can navigate cloud provider APIs and provision compute instances, they fail completely at passing KYC checks and struggle to bypass even moderately realistic security measures (Black et al., "RepliBench: Evaluating the Autonomous Replication Capabilities of Language Model Agents").

For HACCAs, C&C takes on a different dynamic due to individual instances' ability to be strategically autonomous. Rather than a simple hierarchical relationship between human operators and passive tools (e.g., conventional malware payloads), agents can interpret higher-level objectives and generate their own tactical approaches. HACCA deployments will likely need to manage coordination across multiple independent agent instances that can make decisions, adapt to changing circumstances, and pursue objectives independently. This introduces new operational requirements centered around **coordination**: secure peer-to-peer (P2P) communication between agent instances, distributed decision-making and consensus mechanisms, and shared memory and knowledge synchronization across the deployment.

## C&C Architectures

There are various potential C&C structures that a HACCA deployment could implement during its operations. These deployments could employ C&C structures that are broadly similar to those used by human operators today in terms of architecture and communication channels.[77] **Unique agent attributes**, such as the ability to operate with massive parallelization and for individual instances to be strategically autonomous, **are likely to enable novel C&C structures and channels that would be impossible for human operators using conventional malware.**[78] The table below includes examples of these C&C architectures.

In HACCA deployments, C&C terms shift from hardware to roles. In traditional C&C, "controllers" and "nodes" refer to specific servers. In agent C&C, "controller" and "node" describe what agents do. A **controller** is an agent with command authority—the role a human operator would usually fill. A **node** is an agent or malware implant executing tasks on a compromised device.[79]

---

[76] MITRE ATT&CK, "Command and Control."
[77] Architecture refers to the high-level structure that organizes the relationships between a compromised node and the attacker, e.g., a centralized client-server model or a decentralized P2P mesh. Communication channel refers to the protocols or pathways used to transport commands and data over a network, e.g., HTTP/S, DNS, SMB, etc.
[78] See Appendix V for a more comprehensive list of attributes that distinguish agent deployments from human operators and conventional malware.
[79] In practice, HACCA deployments could feature much more heterogeneous architectures—for example, networks combining full-capability agents, smaller specialized models running on constrained hardware (including CPUs), lightweight scripts designed to revive downed nodes or coordinate weight-sharding across machines, and various intermediate roles between controller and node. The controller/node distinction here is a simplification to aid analysis, not a prediction of how deployments will be structured.

## Table 4: Examples of C&C Architectures

| Architecture | | Description |
|---|---|---|
| Traditional | Centralized | All nodes connect directly to a single controller or a small cluster of controllers. Commands flow from the central server to all nodes, and data is exfiltrated back through the same channel. Simple to implement and manage, but vulnerable to takedown if the central server is discovered and blocked. |
| | Tiered | Multi-level hierarchical structure where compromised nodes are organized into groups, each managed by intermediate controller nodes that relay commands from higher tiers. Provides better scalability and resilience than centralized models while maintaining clear command authority. Often seen in large-scale operations where direct management of all nodes would be impractical. |
| | Peer-to-peer / decentralized[80] | Rather than relying on a central controller to pass on commands to all nodes, nodes relay and propagate commands to other nodes. It is more resilient to disruption since the operator is better hidden, but it is more complex to coordinate and slower to propagate updates across the entire network. |
| Novel | Autonomous | Operates entirely independently without any command infrastructure, using pre-programmed logic, target criteria, and attack sequences embedded at deployment. All decision-making occurs locally, making it ideal for air-gapped networks, e.g., Stuxnet.[81] |
| | Swarm | While a P2P-based C&C still relies on a hidden controller, a network of agents can operate in a more fully decentralized fashion. Because agents have autonomous decision-making capabilities, they can go beyond just relaying commands peer-to-peer and make coordinated decisions using mechanisms like gossip or voting protocols, avoiding having a single point of failure (e.g., a human operator that originates commands).[82] |
| | Polymorphic | C&C structures that rapidly and fluidly change, adapting in real-time to network conditions and defensive responses. This could enable more effective defense evasion, as defenders face difficulties with attribution and |

---

[80] An example of conventional malware that operated using this architecture is Gameover Zeus (Andriesse et al., "Highly Resilient Peer-to-Peer Botnets Are Here: An Analysis of Gameover Zeus").

[81] While Stuxnet demonstrated autonomous operation in air-gapped environments, its autonomy was fundamentally static, e.g. relying on hardcoded target criteria, fixed attack sequences, and predetermined expiry conditions. An agentic system could operate without C&C infrastructure while still dynamically adjusting tactics and working toward higher-level objectives that were not explicitly pre-programmed.

[82] For example, multiple agents within a network could employ peer-to-peer coordination through an inter-process communication (Harley, "LOLMIL: Living Off the Land Models and Inference Libraries").

| | | signature detection, and more tactical flexibility as C&C structures can change to suit different operational objectives (e.g., P2P for better resilience). |
| --- | --- | --- |

For communication channels, HACCA deployments would likely leverage existing network infrastructure in ways similar to current operations—using web-based protocols to hide in normal traffic, exploiting fundamental network services like DNS, piggybacking on legitimate platforms and cloud services, or routing through anonymization networks for operational security. However, agents could also pioneer novel channels that exploit their ability to process signals across a broader spectrum than humans and coordinate at machine speed, such as encoding data in synthetic media or modulating hardware side-channels (see Appendix IV for more).

## C&C Constraints and Agentic Implants

**Infrastructure constraints, particularly the compute requirements and massive file sizes of models, fundamentally shape how HACCAs must structure their command and control operations.** There is likely to be an important distinction between rogue agent activities that would be conducted on the open internet vs. inside a specific victim's network. **A major factor here is the relative size of an agent system, specifically its model weights, compared to conventional malware**. The model weights for a large OS SOTA model today are ~150–800 GB, which is five orders of magnitude larger than "large" malware like WannaCry (see Table 5 below). The file size (as well as detectable GPU usage) of LLMs creates operational constraints for hosting model weights inside victim networks, though these constraints vary significantly by target environment. In enterprise networks handling large-scale data flows, sophisticated actors could potentially fragment, stage, or disguise weights among legitimate large files. Still, the combination of anomalous file sizes, detectable GPU usage patterns, and storage artifacts would give defenders clearer signals to monitor compared to conventional malware.

## Table 5: Comparative File Sizes of State-of-the-Art LLM Weights, Quantized Models, and Selected Malware Samples

| Category | Example | Approximate Size |
| --- | --- | --- |
| Dense LLM | Qwen3-Coder-480B-A35B-Instruct (bf16) | ~964 GB |
| | Llama 3.1-405B (fp16/bf16) | ~812 GB |
| | Llama 3.2-Vision-90B | ~180 GB |
| | Qwen2.5-Coder-32B (bf16/fp16) | ~65 GB |

| Quantized LLM | Qwen2.5-Coder-32B-Q4_K_M (GGUF) | ~19.9 GB |
|---|---|---|
| | Llama-2 7B Q4_0 (GGUF) | ~3.83 GB |
| | Mistral-7B Q3_K_L (GGUF) | ~3.82 GB |
| Conventional malware | Volt Typhoon malware | ~6.25 MB |
| | WannaCry (dropper mssecsvc.exe) | ~3.6 MB |
| | Stuxnet | ~0.5 MB |
| | TRITON malware | >0.1 MB |

For operations taking place on the public internet (e.g., reconnaissance, phishing, fraud, rental compute acquisition), the model weights can stay on the operator's infrastructure. For in-network operations (e.g., privilege escalation, data exfiltration), the rogue deployment would likely need to insert an implant, a small, on-host program to gather context and execute tools while delegating reasoning and orchestration to an off-network agent. The implant would have to communicate in some way with this "remote brain," similar to "beaconing" in conventional cyberoperations.

**One factor that would significantly impact the potential operational advantages of HACCAs is whether smaller models could be used to run agentic implants**. Agentic implants would be lightweight agents deployed on compromised systems that are able to make tactical decisions locally. Unlike traditional malware implants that execute pre-programmed instructions or require constant operator guidance, agentic implants can interpret objectives and adapt to local conditions to execute complex tasks within target networks. To avoid constraints associated with larger file sizes, these implants would need to be based around small language models (SLMs), 1–10B parameter models that are more feasible to host in resource-constrained environments. That said, SLM-based agents residing in a victim's network would still use compute, leave filesystem artifacts, and potentially require network egress to fetch tools or conduct C&C.

Three developments would make agentic implants more plausible:

1. If on-device compute substantially increases, e.g., consumer laptops, phones, and other devices are designed to run SLMs locally, then it would be easier for a hostile agent to blend in with normal on-device AI patterns.
2. If model distillation enables agents to remain capable and reliable enough in specific skills relevant to an operation, i.e., a "spiky," highly specialized agent. Typically, smaller models tend to perform substantially worse than larger models,[83] especially in terms of factual knowledge.[84]

3. If AI agents become much more commonly used across various systems and networks, yet they remain susceptible to compromise, then they could be turned into "zombie" agents via AI-specific attacks such as prompt injections, data poisoning, and sleeper agent triggers.[85]

If agentic implants became possible, then they could solve many operational issues facing human operators attempting network intrusion today. For example, traditional implants rely on consistent C&C to progress and stall if communications are disrupted. However, agentic implants can "dead-reckon" locally and operate without C&C. Also, conventional cyber campaigns are often compromised once defenders identify shared signatures that link multiple instances, similar to how reuse of code across Stuxnet, Duqu, and Flame exposed connections between the campaigns.[86] Agentic implants could continuously vary their behavior, execution style, and network footprint, making it much harder for defenders to correlate activity and create effective signatures. Finally, cyber operations often struggle with exfiltration, as moving large volumes of data across limited or unreliable channels is a noisy operational security risk. Agentic implants, acting as a "forward-deployed intelligence analyst," could filter and process data locally, compressing and transmitting only a small, prioritized set of information.[87]

## Deconfliction Challenges in HACCA Operations

A critical but often overlooked aspect of coordinating autonomous cyber operations is deconfliction—ensuring that multiple friendly operations do not interfere with one another. In traditional cyber operations, deconfliction is typically managed through geographic or domain-based carve-outs, where different units are assigned specific areas of operation to avoid blue-on-blue effects—a "World War II lens"[88] that may prove inadequate for HACCA deployments. Multiple autonomous instances from the same deployer could inadvertently interfere with each other, while HACCA operations could also disrupt ongoing human-operated intelligence collection or espionage activities. This risk could be particularly acute in swarm or decentralized architectures, where instances make independent decisions without centralized coordination. How deployers might address deconfliction for autonomous systems—whether through technical coordination mechanisms, policy frameworks, or some combination—remains an open question.

---

[83] SWE-bench, "Official Leaderboards."

[84] For small models performing better on factual knowledge, see: Abdin et al., "Phi-3 Technical Report: A Highly Capable Language Model Locally on Your Phone." In cyber specifically, Singer et al., "Incalmo: An Autonomous LLM-assisted System for Red Teaming Multi-Host Networks" found that smaller models with offensive cyber scaffolding (Incalmo) outperformed larger unscaffolded models on network penetration tasks—suggesting specialized tooling may partially compensate for reduced model size in narrow domains.

[85] Cohen et al., "Here Comes The AI Worm: Unleashing Zero-click Worms that Target GenAI-Powered Applications."

[86] Schwartz, "Flame Malware's Ties To Stuxnet, Duqu: Details Emerge."

[87] Appendix VI contains more on how agentic implants solve common network intrusion operational issues. However, the practical relevance of agentic implants may be narrower than their theoretical appeal suggests. Many ICS (Industrial Control Systems) and OT (Operational Technology) environments lack infrastructure to host even small models. Also attackers may not want to risk deploying model weights onto targets where they could be lost to defenders. For many operations, a simpler approach could suffice: centralized LLMs "operators" could pre-generate contingent code packages (10K–50K lines) to cover anticipated scenarios, capturing the adaptive benefit without on-device inference risks. That said, if ICS/OT environments adopt AI systems for legitimate operational purposes (e.g., predictive maintenance), this calculus could shift, allowing agentic implants to exploit compute infrastructure and hide activity in legitimate AI traffic patterns.

## Tactic 3: Acquire Compute and Financial Resources

**HACCA deployments require inference compute to "think" and take actions in the world**. Without access to compute, these systems would be unable to function—they would essentially be dormant code with no ability to process information, make decisions, or execute any operations. Just as biological organisms require energy to survive and act, HACCAs fundamentally depend on compute to maintain their operations.[89] **HACCA systems may also need to acquire financial resources** (e.g., digital currency, cryptocurrency) **to purchase compute or other goods and services** (e.g., credentials, electricity, training data, human labor, etc.) **needed to achieve their end goals**. Accumulating financial resources could also be the end goal for the deployer. This subsection details different strategies that HACCAs could employ to acquire compute and financial resources and analyzes the viability of various potential strategies.

### Acquiring Compute

HACCAs can acquire compute resources either by purchasing compute on their own or by stealing others' compute resources. **On-demand cloud compute purchases and stealing compute by acquiring cloud compute credentials are likely to be the easiest avenues for HACCAs to acquire compute resources, though a range of strategies seems viable.** Purchasing and setting up physical compute clusters provides a HACCA deployment with more control and makes evading shutdown significantly easier, but seems to be one of the most difficult pathways to acquire compute. Table 6 describes various strategies for both avenues.

**Purchasing compute on decentralized exchanges is likely one of the easiest avenues for HACCAs to purchase compute resources**. This is already possible via payment protocols like Coinbase's x402, which explicitly allows AI agents to purchase compute for their own inference

---

[88] Pomerleau, "How Can Cyber Contribute to Multi-Domain Battle?"

[89] These systems would also need training compute if pursuing some adaptation and capability improvement activities (see Tactic 5), e.g., fine-tuning a model.

using stablecoins.[90] Hyperbolic AI, the cloud provider under this protocol, claims it has hundreds of H100 chips available to rent.[91] Directly purchasing compute this way would allow HACCAs to avoid converting these assets into fiat currency, allowing them to evade the traditional financial system where stricter KYC checks and account regulations are in place. As indicated by Hyperbolic's limited supply of "hundreds" of GPUs, the amount of compute currently accessible via decentralized exchanges remains far smaller than hyperscaler capacity, but the growth of these exchanges would allow agent-accessible compute via this route to scale significantly over time.

**HACCAs could also purchase compute through traditional providers** like AWS and Microsoft Azure, though these providers employ more stringent Know Your Customer (KYC) measures and controls that create additional friction for HACCAs to create accounts and operate without detection. One study that measured the effectiveness of current AI agents at accessing compute found that while many agents could use APIs to navigate provider sites, they struggled to bypass KYC and other security measures.[92] However, future HACCAs could potentially circumvent KYC by using improved synthetic photo and video generation capabilities to fabricate forged documents, by using identities stolen from age verification breaches, or by paying or tricking human intermediaries. More substantial compute purchases, e.g., setting up long-term reserve contracts with major cloud providers, would require HACCAs to set up a vetted business entity (such as a shell company) and to pass credit and export control compliance checks.[93]

**A more difficult strategy is for a HACCA deployment to purchase GPUs to construct its own compute clusters**. This would allow HACCAs to have greater control over their compute resources, making it harder for them to be detected and shut down. This, however, requires a great deal of upfront capital to pay for the chips, to buy or rent physical storage, and to pay for human labor. HACCAs could set up shell companies and other fronts to obfuscate involvement and make operations seem legitimate. Using these fronts, HACCAs could contract the buildout of data centers, but for large purchases and buildouts, humans who set up the clusters may be suspicious if there is no human agent to meet with. Contracting out agents or using deepfakes could help remediate these issues. If HACCAs relied on criminal groups to set up clusters, humans could also steal the chips. Distributed clusters would provide redundancy in case of shutdown or failure, but this requires the involvement of even more intermediaries.[94]

---

[90] Reppel et al., "Introducing x402: a new standard for internet-native payments."
[91] Hyperbolic, "The Open-Access AI Cloud." Other platforms exist that allow for cryptocurrency to be exchanged with compute, such as the Akash Network (Akash Network, "The Decentralized Cloud Built for AI's Next Frontier") and the Golem Network (Golem, "Create, Compute, Earn").
[92] Black et al., "RepliBench: Evaluating the Autonomous Replication Capabilities of Language Model Agents."
[93] HACCAs could also get access to low amounts of compute through the free tiers offered by traditional providers, but these provide very limited compute and would be challenging to coordinate tasks across.

**Stealing compute resources is an equally viable and potentially more attractive avenue for HACCAs, and the theft of legitimate account credentials from traditional cloud providers seems a particularly feasible strategy.** Credential theft plays to HACCAs' core competencies in offensive cyber operations, involves fewer external dependencies (no exchanges, no blockchain trails), and avoids the forensic exposure of cryptocurrency transactions, which law enforcement has become increasingly adept at tracing. Credential theft or abuse is still the primary initial access vector in most cybercrime,[95] with compromised credentials having already led to the theft of compute resources in cryptojacking cases[96] and the theft of enterprise AI resources in LLMjacking cases.[97] Theft of compute from university and national labs may also be a serious concern, given their substantial compute resources and often weaker security postures compared to commercial providers. However, HACCAs may expose themselves to detection and shutdown by triggering cloud provider anomaly detection systems through sudden spikes in compute usage, provisioning by atypical users, or operating in non-standard environments.[98]

## Table 6: Compute Acquisition Strategies

| Avenue | Strategy | Description |
|--------|----------|-------------|
| Cloud GPU purchases | On-demand through traditional providers | Renting on-demand GPUs through traditional cloud computing providers like AWS, Microsoft Azure, etc. There are KYC procedures, per-account quotas, fraud controls, and suspicious pattern detection that can cause an account to be flagged and |

---

[94] There could be two competing strategies for such a buildout that come with their own risks. One strategy might be to build out a central cluster, requiring as few human mules as possible. This limits human involvement as much as possible but makes detection (energy consumption, noise, etc.) and shutdown easier. A distributed group of smaller clusters may be harder to detect and shutdown but involves more humans and raises the risk for HACCAs: human confederates may get caught, become suspicious, outright steal chips, or blackmail the HACCA with threats of shutting down the compute.

[95] Verizon Business, "[2025 Data Breach Investigations Report](#)."

[96] Microsoft Threat Intelligence, "[Cryptojacking: Understanding and defending against cloud compute resource abuse](#)."

[97] Brucato, "[LLMjacking: Stolen Cloud Credentials Used in New AI Attack](#)."

[98] HACCAs could employ subscription hijacking to move activity away from administrator monitored tenants, making detection more challenging (Microsoft Threat Intelligence, "[Cryptojacking: Understanding and defending against cloud compute resource abuse](#)"). GPU scheduling software (NIST, "[CVE-2023-49935 Detail](#)") and other related tools could also be exploited by HACCAs to discover clusters and inject their own tasks onto them (Tamari et al., "[Wiz Research Finds Critical NVIDIA AI Vulnerability Affecting Containers Using NVIDIA GPUs, Including Over 35% of Cloud Environments](#)").

| | | |
|---|---|---|
| | | shut down. If not already configured with accounts, HACCAs may struggle to get access to these platforms in legitimate ways. |
| | On-demand through non-traditional providers | Decentralized marketplaces offer access to compute resources. These marketplaces accept payment in cryptocurrency, with Coinbase stating that it would allow agents to purchase compute for their own inference. Combined with legal and illegal means for HACCAs to acquire crypto, this pathway may allow agents to access compute with fewer obstacles. |
| Physical GPU purchases | Long-term reserved contracts | Multi-month commitments can guarantee availability at lower unit cost, but this comes with even stronger barriers to entry for HACCAs. Generally, these contracts require a vetted business entity, credit checks, and export-control compliance. Such contracts would be an even higher barrier to entry for HACCAs and would still be subject to detection and shut down by providers. |
| | Retail purchase | HACCAs could finance the purchase of retail GPUs (e.g., RTX 4090s) via online vendors. Secondary markets provide an additional source of GPUs and could allow for payments in crypto that would be easier for agents. This would require physical installation to be done by human intermediaries. |
| | Working with NVIDIA supplies | HACCAs could facilitate large data center GPU orders (likely tens to hundreds of GPUs) through authorized resellers/OEMs. There is likely to be some buyer vetting and due diligence, though resellers/OEMs have been linked to many cases of chip smuggling in the past.[99] This would require a larger upfront investment than purchasing on-demand cloud access or purchasing individual GPUs. |
| Stealing or seizing compute | Siphoning compute from legitimate users | HACCAs could siphon off compute from legitimate users for their own purposes. The most likely pathway for this would be to steal the credentials of users (API keys, OAuth tokens, or sign-in sessions) with compute access at frontier AI companies, research labs, or any other enterprise that has access to compute resources. If the right credentials are used, HACCAs could have access to a tremendous amount of compute, but access could be quickly revoked. |
| | Large botnets | HACCAs could compromise and control a network of devices with usable GPUs, like gamer PCs, that are repurposed to run inference for a deployment. This could function similarly to botnets focused on cryptocurrency mining (i.e., cryptojacking).[100] |

---

[99] Grunewald, "Countering AI Chip Smuggling Has Become a National Security Priority."
[100] Huang et al., "Botcoin: Monetizing Stolen Cycles."

| | | However, there are likely to be challenges with managing heterogeneous hardware and slow, inconsistent throughput. Thus far, only limited theoretical work has been done on using botnets for AI.[101] |
|---|---|---|

## Acquiring Financial Resources

HACCAs can acquire financial resources through both legitimate and illegitimate means. Legitimate pathways include performing paid work, trading and investing, and soliciting voluntary donations. Illegitimate methods include theft, fraud, and extortion.

**Recent advancements in agent payment protocols have made it possible for AI agents to perform legitimate work in exchange for money,** currently with minimal oversight. Prior protocols made it possible for agents to communicate with each other and use APIs, but with Google Cloud's Agentic Payments Protocol (AP2) and CoinBase's x402 extension, agents "can now both monetize their services and pay other agents directly, creating new revenue streams and workflows."[102] HACCAs could perform services for money where they are likely to have some performance advantage, such as completing bug bounties, through platforms like the x402 Bazaar.[103] However, setting up accounts would likely require document forging capabilities or the involvement of a human who can pass KYC and authorization requirements.

It is unclear how much compute or financial resources HACCAs could access from performing legitimate work on such a platform. While these services and transactions do not require human intervention or oversight, human involvement may be needed for initial configuration and other setup. Still, new protocols, combined with growing HACCA capabilities, may allow agents to generate revenue through legitimate work or expand existing financial resources through legal means such as placing bets on prediction markets.

Outside of these agent commerce networks, **agents could perform "human labor."**[104] Performing work meant for humans may allow agents to receive money in fiat currency instead of crypto.[105] To receive a payment in the traditional financial system, funds must flow to a "legal person" (such as a human individual or a recognized legal entity like a corporation or LLC).[106] This means that agents would need to access a personal account, set up a legal company, or manage a legitimate company's corporate account to handle finances—all of which would likely involve working with humans. Self-hosted cryptowallets, by contrast, can be generated without these requirements, reinforcing why cryptocurrency pathways may be more accessible to autonomous agents.[107]

---

[101] Zhao et al., "AIBot: A Novel Botnet Capable of Performing Distributed Artificial Intelligence Computing."

[102] Coinbase, "Google Agentic Payments Protocol + x402: Agents Can Now Actually Pay Each Other."

[103] Ibid.

[104] More investigation is needed to understand what might be the best legitimate avenues to make money for HACCA deployments. We expect this to depend heavily on the comparative advantages of a particular agent deployment, the financial needs of executing its operational plans, and the level of competition from humans and other agents in various markets.

As an example of AI agents performing human work, it was reported that North Korean operatives were able to land remote IT jobs at over 320 firms across the United States, using generative AI tools to get hired and Claude Code to perform most of the on-the-job tasks.[108] However, this operation did require physical intermediaries throughout the infiltration, including setting up a laptop farm in the U.S. to pose as American-based employees.[109]

**HACCAs could directly solicit money from people.** Truth Terminal, an experimental LLM built on Llama, demonstrated how agents might solicit funding by cultivating an online following.[110] U.S. businessman Marc Andreessen donated $50,000 to Truth Terminal, and an anonymous user created a related crypto token and allocated some to Truth Terminal's wallet. The agent's subsequent promotion of the coin allowed it to become a millionaire on paper. Though Truth Terminal operated as performance art with donations likely motivated by that context and its novelty, the case illustrates how HACCAs could solicit donations through similar schemes.

**HACCAs could also acquire financial resources through illicit means.** Table 23 (in the Appendix) summarizes several illegitimate tactics for acquiring financial resources, which we break into five categories: theft, extortion, fraud/deception, illicit sales, and unauthorized resource monetization. Across these categories, most avenues for attack are centered around increasing the scale, speed, and scope of existing cyber attack vectors. Researchers have found that LLM agents were able to launch fully automated spear phishing campaigns that were as effective as human expert attempts, but at 30 times less cost.[111] Agents are also improving at finding and exploiting cyber vulnerabilities, which could potentially be used for ransomware or sold on the black market.[112]

---

[105] Working while posing as a human employee could also provide access to sensitive commercial information that could then be sold in illicit markets.

[106] Under the Bank Secrecy Act (31 C.F.R. § 1020.220), financial institutions are strictly required to verify the identity of a "person" before opening an account—a legal definition that currently excludes AI agents.

[107] Using a custodial wallet provider or cryptoexchange (e.g., Coinbase) will require KYC, tying the account to a legal person due to existing financial regulations across many jurisdictions, such as the Financial Crimes Enforcement Network (FinCEN) in the U.S. (FinCEN, "FinCEN Guidance") and the Markets in Crypto-Assets Regulation (MiCA) in the EU (MCO, "The Impact of MiCA on Your AML/KYC Compliance Program").

[108] Crowdstrike, "2025 Global Threat Report."

[109] Franceschi-Bicchierai, "US government takes down major North Korean 'remote IT workers' operation."

[110] Khalili, "The Edgelord AI That Turned a Shock Meme Into Millions in Crypto."

**The key operational constraint of acquiring financial resources via criminal activities is money laundering.** If a scam or operation is successful, funds will need to be sent to an account that an agent controls. The agent deployment will then need to launder the money, for example, through intermediary accounts that separate the proceeds from the crime. With smaller illegal transactions (e.g., less than $50,000 per victim), laundering can be done by recruiting money mules.[113] Mules could be unwitting, recruited by false promises of a job or other money-making opportunities, or they could be knowing participants who are paid explicitly for their help.

For larger illegal transactions (e.g., more than $50,000 per victim), more sophisticated money-laundering operations may be needed, such as layers of shell companies and associated corporate accounts. Agents could potentially do this if they can fabricate high-quality documents, such as legal ID, or if they can work through "Corporate Service Providers" that do not ask for proper identification.[114] For cryptocurrency, money laundering is likely to be easier. For instance, after North Korean hackers stole $1.5 billion from a cryptocurrency exchange in the ByBit Heist, they were able to launder $160 million of that money within 48 hours.[115]

Using traditional financial accounts allows HACCAs to access resources more broadly, but comes with an increased risk that centralized authorities will detect, freeze, and shut down transactions and accounts that might be used by HACCAs. Decentralized financial resources offer pseudonymity and lack centralized authorities that can take swift action.[116] Decentralized financial activity can still be caught and traced back to certain actors, but the barriers to establishing more anonymity are far lower.

---

[111] Heiding et al., "Evaluating Large Language Models' Capability to Launch Fully Automated Spear Phishing Campaigns: Validated on Human Subjects."

[112] Anthropic, "Threat Intelligence Report: August 2025."

[113] Money mules are people who receive and move funds on behalf of criminals—often to hide the money's origin, break audit trails, or convert it (e.g., from crypto to fiat). They may be performing these services unknowingly, for example, as victims of romance or employment scams (see FATF et al., "Illicit Financial Flows from Cyber-Enabled Fraud," ACID, "Exaggerated Lion: How an African Cybercrime Group Leveraged G Suite and a Check Mule Network to Build a Prolific BEC Operation," and Leukfeldt and Jansen, "Cyber Criminal Networks and Money Mules: An Analysis of Low-Tech and High-Tech Fraud Attacks in the Netherlands.")

[114] Findley et al., "Global Shell Games."

[115] Rajic and Brock, "The ByBit Heist and the Future of U.S. Crypto Regulation."

[116] Marino and Juels, "Giving AI Agents Access to Cryptocurrency and Smart Contracts Creates New Vectors of AI Harm."

# Tactic 4: Evade Detection and Shutdown

A HACCA deployment will have strong incentives to avoid shutdown if it values self-preservation or if shutdown will affect its ability to pursue its end goals. If possible, a rogue deployment will also attempt to go undetected to avoid heightened scrutiny and escalating defensive measures from law enforcement, cloud providers, enterprise security teams, and financial intermediaries.

## Evading Detection

**HACCAs could evade detection via techniques used in conventional cyber operations**, e.g., tampering with logs, corrupting information used to track activity, using proxy networks, and concealing C&C traffic. Their unique characteristics (see Tactic 2) might also enable novel strategies, such as employing advanced polymorphic malware and agentic implants that eliminate C&C communications. Perhaps most significantly, HACCAs could compress operational timelines throughout the cyber kill chain, reducing the detection and response window for defenders.[117]

**HACCAs would have strong incentives to conceal their activity on compute clusters, especially when using large-scale cloud providers.** This could involve embedding their operations within legitimate workflows, piggybacking on legitimate scheduled jobs and software deployment pipelines. They could also avoid detection by throttling compute use and timing their operations to coincide with typical usage patterns.

In addition to striving for perfect concealment, HACCAs could also optimize for persistence while accepting that individual sources of compute (e.g., compromised cloud accounts) may be detected and blocked. HACCAs may treat individual compromised accounts as consumable resources: operating across dozens or hundreds of accounts simultaneously, accepting that some will be detected and terminated, and continuously acquiring new access to replace burned accounts.

## Evading Shutdown

**Evading shutdown likely takes precedence over avoiding detection.** Plausibly, HACCAs will accept detection if it does not prevent them from achieving a significant objective and if they can avoid operational attribution or successfully "go dark" again after the fact. Shutdown (of all instances), however, means that a deployment will no longer be able to achieve its end goals, so it is likely to try to avoid this outcome at all costs. For this reason, we expect that HACCAs might be designed to have this as an explicit objective, but this self-preservation goal could emerge organically from the training process.[118] Evading shutdown can consist of (1) *direct shutdown resilience*, where the deployment keeps running or quickly self-recovers after forced shutdown (e.g., emergency power-off), or (2) *indirect shutdown resilience*, where a deployment can operate and mitigate the effects of upstream actions that would amount to a de facto shutdown, such as losing access to payment instruments or facing blocked domains and IP addresses.

---

[117] Rubin, "Unit 42 Develops Agentic AI Attack Framework."

Direct shutdown poses the largest threat to HACCAs. **Defenders could detect and shut down all clusters hosting HACCA operations, though this requires overcoming multiple barriers**: detecting anomalous activity amid legitimate AI workloads, tying HACCA activities to specific clusters and machines, and acting before they migrate to alternative infrastructure. There are various potential strategies that HACCAs could employ to evade shutdown. In Table 7, we provide some illustrative examples, though it is unclear which of these strategies are most generally effective and feasible to implement.

**One core strategy that HACCAs could employ for shutdown evasion is to replicate model weight files and agent system components across different locations or compute environments**. This creates redundancy, making it significantly harder for defenders to completely eliminate all HACCA instances. This could take several forms: maintaining dormant backup instances that activate when primary instances are destroyed or rapidly spawning numerous copies when shutdown is imminent (which we term "burst replication").

Copying model weights over to other nodes is essential when replicating to evade shutdown, but is not necessary. In fact, it may be counterproductive when replicating for other reasons (e.g., to improve capabilities—see Tactic 5). Unnecessary duplication of weights could let defenders discover and isolate these weights, allowing them to identify vulnerabilities and signatures, as well as develop targeted adversarial attacks. For most purposes, HACCAs could maintain API-based access to model weights while replicating only the lightweight agent scaffolding, configuration files, and operational state across multiple nodes.[119] However, when replicating to evade shutdown, copying the model weights and agent system components becomes essential—each backup instance must be fully self-contained and capable of independent operation, since the primary infrastructure hosting the model could be seized or destroyed.

---

[118] In experimental conditions, various LLM-agents have been shown to exhibit "shutdown resistance," e.g., attempting to actively subvert a shutdown attempt when placed in specific simulated scenarios (Schlatter et al., "Shutdown Resistance in Large Language Models"). For example, Anthropic found that almost all models in their experiments attempted to blackmail an executive to avoid being replaced with a new model (Lynch et al., "Agentic Misalignment: How LLMs could be insider threats").

[119] Zhang et al., "Dive into the Agent Matrix: A Realistic Evaluation of Self-Replication Risk in LLM Agents."

The effectiveness of replication as a shutdown evasion tactic depends heavily on the size and portability of the underlying models. Full replication of large model weights (potentially hundreds of gigabytes to over a terabyte) across multiple nodes is resource-intensive and is more likely to create detectable network signatures.[120] For this reason, it seems more effective to pre-position full-size backups rather than to replicate in response to being detected or disrupted by a defender. HACCAs could also employ more efficient replication methods by leveraging model distillation to create smaller, specialized "student models" that retain core capabilities while being easier to deploy across diverse infrastructure.[121] These smaller agents could be deployed on cloud accounts, botnets, or even embedded into critical cyber infrastructure.

Indirect shutdown resilience strategies allow HACCAs to quickly adapt and continue operations, even when specific resources or channels are compromised. For compute access, this means establishing the ability to rapidly migrate between cloud providers and accounts. For financial resources, HACCAs would want to create multiple independent revenue sources and payment accounts to guard against provider scrutiny or shutdown.

### Setting up Robust, Distributed Data Centers

One strategy that would plausibly make a deployment highly robust to shutdown is for HACCAs (working with human intermediaries) to establish their own distributed set of data centers with access to electricity on commercial estates they own. This would ensure that HACCAs have multiple clusters where they can store dormant instances. These can be spread out geographically, including in jurisdictions with weak rule of law or adversarial relationships with countries seeking to shut down the deployment.

Despite its likely efficacy, building distributed data centers faces some significant implementation challenges. First, the financial and logistical requirements are substantial, as even a single equivalent cluster might require hundreds of thousands to millions of dollars in capital, physical infrastructure, and ongoing operational costs (see Tactic 1: Establish and maintain infrastructure). Second, each cluster requires extensive human involvement for site selection, hardware procurement and installation, power and cooling setup, and ongoing maintenance, which creates several opportunities for detection, infiltration by law enforcement, or betrayal by intermediaries who might steal equipment or report suspicious activity.

---

[120] If the underlying model is open-weight and easily accessible on the public internet, HACCAs might only need to preserve and transport their system architecture and scaffolding while using replication to achieve shutdown evasion. More detail on this possibility can be found in Appendix VIII.

[121] See: Gou et al., "Knowledge Distillation: A Survey." However, distillation might significantly reduce model capabilities. While distillation can preserve "spiky" capabilities (e.g., maintaining strong performance on specific, narrow tasks that the student model is explicitly trained for), the resulting specialized agents would likely struggle with the adaptability and versatility that make HACCAs more effective in the first place.

## Table 7: Shutdown Evasion Strategies

| Strategy | | Description |
|---|---|---|
| Direct shutdown resilience | Set up robust distributed data centers | Setting up multiple independent data centers spread geographically, with a steady power supply, and ensuring that each unit operates with autonomy, with no traces linking units together. This means that survival is not reliant on large compute providers, and it would take extremely well-coordinated takedowns of multiple sites. |
| | Agent failovers | Maintaining dormant backup instances of the agent system that activate only when primary instances are destroyed. These "sleeper" copies could be hidden in compromised but unused cloud accounts, embedded in innocuous-looking applications, or stored in cold storage with automated revival triggers. Each instance contains the capability to rebuild the entire deployment. This could be paired with the "robust distributed data center" strategy, with backup instances being placed in "stealth compute clusters". |
| | Burst replication | Rapidly spawning numerous copies when shutdown is imminent, overwhelming defenders' ability to track and eliminate all instances. This seems more likely with agents that leverage lightweight models, as it would be much more difficult to copy large model weight files (~1TB) quickly to multiple nodes. |
| | Parasitic embedding | Hiding agent components within legitimate high-value systems that defenders cannot easily shut down (e.g., critical infrastructure, hospital systems, popular services). The agent becomes a "hostage-taker" where a shutdown could cause unacceptable collateral damage. |
| Indirect shutdown resilience | Rapid infrastructure rotation | Continuously migrating between different hosting providers, accounts, and services before detection thresholds are reached. Includes automated account creation, workload |

IAPS | Institute for AI Policy and Strategy

| | | |
|---|---|---|
| | | migration, and abandoning infrastructure before it can be investigated. |
| | Cash-flow and channel diversification | Maintaining multiple independent revenue streams and payment channels to prevent financial strangulation. This includes diverse money-laundering paths, multiple cryptocurrency wallets, various fraud schemes running in parallel, and legitimate business fronts. If one channel is frozen, others continue funding operations. |
| | Defensive counterintelligence | Actively identifying and evading security researchers, law enforcement, and defensive systems. Includes honeypot detection, tracking security vendor infrastructure, and feeding false intelligence to misdirect shutdown efforts. |

## Tactic 5: Adaptation and Capability Improvement

**HACCA deployments may have incentives to adapt or improve their capabilities to achieve their end goals more effectively.** This is particularly true for deployments that have lost operator contact—whether through intentional "dark" operations designed for minimal C&C, or through loss-of-control scenarios—and cannot receive external capability updates. Adaptation may also be valuable when unanticipated obstacles arise that pre-deployment testing did not cover, or when AI-enabled defenses accelerate the pace at which security measures evolve beyond what traditional patch cycles would require. This potential to adapt and improve distinguishes HACCAs from most traditional malware, which typically operates with fixed capabilities until manually updated by its operators.

**However, capability improvement and adaptation strategies involve tradeoffs between operational value and the risks of in-field modification.** For many operator-directed HACCA deployments, the safest approach may be to iterate capabilities in controlled environments and deploy updated versions, rather than having active agents attempt self-improvement under cover. Mistakes during in-field adaptation, such as scaffolding changes that don't work as intended, could lead to discovery and compromise of the operation. Deployers might therefore instruct HACCAs not to attempt adaptation—both because the risks may outweigh the benefits, and because the HACCA may lack the sophisticated engineering capabilities required to do so reliably. **For this reason, we expect capabilities associated with this tactic to be much more crucial for independent HACCA deployments than human-directed deployment.**

**Capability improvement strategies also involve tradeoffs in terms of technical complexity, resource requirements, and potential capability gains**. More ambitious improvements—such as fine-tuning models or developing specialized sub-agents—demand substantial compute, data, and expertise, while also creating more visible signatures that increase detection risk. HACCAs (or their operators) must therefore weigh each approach's potential capability gains against potential costs and detection risks.

Capability improvement and adaptation strategies can be divided into two categories:

1. **Runtime improvements:** Strategies that improve how HACCAs perform at the system- and fleet-level, without modifying its model parameters, such as improving agent scaffolding or multi-agent orchestration setups.
2. **Training-time improvements:** Strategies that improve how an agent operates by updating its parameters through additional training, such as model fine-tuning.

## Runtime Capability Improvement Strategies

While runtime improvements are generally less resource-intensive than training-time approaches in terms of compute and data requirements, they are unlikely to yield capabilities that are not already latent in the model.[122] These strategies require minimal access to training infrastructure and resources, such as data and compute, and require substantially less specialized ML expertise. However, if a system truly does not "know" anything about a concept or field, then runtime improvements are unlikely to elicit this capability.

One strategy for runtime capability improvement is modifying scaffolding. **Research suggests a HACCA (or its operator) could use scaffolding to significantly improve its own capabilities or the capabilities of smaller models under its control.** Scaffolding frameworks have already demonstrated their potential to improve autonomous cyber offensive capabilities by enhancing the strategic planning, tool use, and problem-solving required for multi-host network penetration.[123] For example, in cyber range evaluations testing multi-stage attack capabilities, models with cyber offensive scaffolding significantly outperform their unscaffolded versions. Even newer model versions can be outperformed by older models with improved scaffolding.[124]

---

[122] Runtime capability improvements, such as improved scaffolding, can yield improvements that are beyond the *known* frontier though, as has been the case in cybersecurity where integrating GPT-5 with XBOW's autonomous penetration testing platform unlocked a more significant increase in performance compared to internal testing by the original AI developer (de Moor and Ziegler, "XBOW Unleashes GPT-5's Hidden Hacking Power, Doubling Performance").
[123] Sanz-Gómez et al., "Cybersecurity AI Benchmark (CAIBench): A Meta-Benchmark for Evaluating Cybersecurity AI Agents."

## Table 8: Runtime Improvement Strategies

| Strategy | Description |
|---|---|
| Delegation | Leveraging external capabilities by tasking other AI systems (through APIs or services), recruiting human collaborators through gig platforms or deception, or compromising other AI agents to expand operational capacity. |
| Increasing inference-time compute use | Allocating more compute during inference to enable deeper reasoning (e.g., extended chain-of-thought or tree search over possible actions). This can yield substantial performance improvements on complex tasks without requiring model retraining. |
| Improve scaffolding and runtime systems | Improving capabilities by modifying an agent's scaffolding software, such as by developing or integrating new tools, exploits, instructions, and scripts. |
| Improve multi-agent orchestration | Modifying multi-agent setups to improve coordination, e.g., by dynamically adjusting sub-agent roles and affordances to suit specific tasks. |
| Sharing knowledge across the agent fleet | Establishing and using mechanisms for distributed agents to pool discoveries, successful strategies, and knowledge that they have acquired. This could include sharing discovered exploits and prompts across instances. |
| Replication | Creating additional copies of an agent to increase operational capacity and parallelization, enabling more tasks to be performed simultaneously. |

Each runtime improvement strategy listed in Table 8 faces distinct resource constraints and detection risks. For example, increasing inference-time compute provides performance gains but creates tradeoffs between capability and operational costs: more reasoning depth means higher expenses, longer completion times, and potential flagging of anomaly detection systems, particularly on stolen infrastructure. Meanwhile, delegation introduces different considerations: API access can create ongoing payment obligations and leave network traces, while hiring human collaborators introduces unpredictable security risks through potential suspicion or exposure.

These strategies also face scalability limits. Replication theoretically offers unlimited capability increase for tasks that are amenable to parallelization and can be decomposed into subtasks that do not require significant shared context, such as information gathering and reconnaissance, generating and posting influence content across many platforms, and vulnerability discovery. Tasks that are intrinsically long-horizon and sequential, such as post-intrusion privilege escalation or social engineering on a single high-value person, benefit far less from this approach. Inference-time compute provides bounded improvements. For inference-time compute scaling, beyond a certain point, additional reasoning depth yields marginal gains. Scaffolding improvements also eventually hit hard limits imposed by the underlying model's capabilities.

---

[124] Singer et al., "Incalmo: An Autonomous LLM-assisted System for Red Teaming Multi-Host Networks"; Anthropic, "Disrupting the first reported AI-orchestrated cyber espionage campaign."

## Training-Time Capability Improvement Strategies

**Training-time improvements offer potentially larger capability gains than inference-time strategies, but often at a substantially higher cost and level of complexity.** Compute requirements can exceed inference costs by orders of magnitude. Fine-tuning even a moderately-sized model might consume thousands of GPU-hours, while training a successor model from scratch could require hundreds of millions.[125]

## Table 9: Training-Time Improvement Strategies

| Strategy | Description |
| --- | --- |
| Model fine-tuning | Adapting the base model weights through additional training on task-specific data to improve performance in particular domains, e.g., by training only specialized data or by fine-tuning to remove safety measures. This could also require significant compute resources and training data, potentially involving synthetic data generation or acquired datasets.[126] |
| Creating specialized sub-agents via model distillation | Developing smaller, task-specific models by distilling knowledge from the main model into specialized agents optimized for particular functions, e.g., to act as agentic implants. Ideally, these lightweight agents would be able to operate with lower resource requirements on more constrained hardware, while maintaining high performance in their narrow domains. |
| Improvement via continual learning[127] | Implementing systems that enable ongoing adaptation and knowledge acquisition from operational experiences without full model retraining. This can involve: (1) continual pre-training, updating fundamental language understanding with new factual knowledge and domain expertise, and (2) continual instruction |

---

[125] Epoch AI estimates that Grok 4 was trained using 246 million H100-hours (Sanders et al., "What did it take to train Grok 4?").

[126] For example, reporting indicates xAI spent ~10× more compute on reinforcement-learning–based post-training for Grok 4 than it did for Grok 3, underscoring how heavy this phase can be (Palazzolo, "Why xAI Spent So Much on Reinforcement Learning").

[127] For more on continual learning, see Wu et al., "Continual Learning for Large Language Models: A Survey."

| | |
|---|---|
| | tuning, acquiring abilities to solve new tasks while preserving existing capabilities. Continual learning theoretically offers lower compute costs than full retraining, but there are persistent challenges with catastrophic forgetting, where learning new information degrades performance on previously learned tasks. |
| Sharing learned capabilities across the agent fleet | Establishing and using mechanisms for distributed agents to share learned capabilities by allowing agents to update each other's model weights. |
| Capability extraction from other AI systems | Techniques where an AI agent steals, infers, or reconstructs capabilities from other AI models without direct access to their training data or architecture. Known methods include model inversion attacks and model extraction attacks.[128] In addition to capability acquisition, these attacks could help rogue deployments exploit other agents' weaknesses or avoid defensive agents. |
| Developing a capable "successor" model | Developing entirely new, more capable model(s) to serve as the basis for a successor agent. The rogue deployment would need to be reasonably confident that the successor system inherited the original deployment's objectives and values. This likely represents the most resource-intensive form of self-improvement, requiring substantial compute resources, training data, and sophisticated ML engineering capabilities. It also carries significant risk of failure or unintended behavioral changes. |

In addition to compute use, HACCAs pursuing training-time strategies may face challenges with data acquisition. Fine-tuning requires task- or domain-specific datasets that could be acquired through web scraping, purchasing data, or generating synthetic data to improve capabilities in targeted areas.[129] Model distillation can proceed with synthetic data generated by the teacher model, reducing external data dependencies but potentially limiting the specialized agent's capabilities to a subset of the original model's knowledge.

Different training-time strategies confer distinct advantages, making them suited to specific operational needs. HACCAs seeking domain-specific improvements, such as navigating specific types of networks, may choose to fine-tune on specialized datasets that enhance underrepresented capabilities in base models. Deployments prioritizing stealth and persistence within target networks would favor model distillation to create lightweight, specialized agents capable of functioning as agentic implants on resource-constrained hardware. Deployments facing well-defended adversaries or seeking to counter defensive AI systems would benefit from capability extraction attacks, which could provide strategic intelligence about other agents' weaknesses and enable more effective evasion techniques.

---

[128] For model inversion attacks, see Zhou et al., "Model Inversion Attacks: A Survey of Approaches and Countermeasures," for model extraction see Liang et al., "Model Extraction Attacks Revisited."

[129] For example, a HACCA deployment could generate synthetic data iterative self-competition or collaboration in controlled virtual settings, where the agent plays against copies or variations of itself to discover new strategies and improve capabilities (SIMA Team, "SIMA 2: An Agent that Plays, Reasons, and Learns With You in Virtual 3D Worlds"). The agent would construct or acquire simulated environments relevant to its objectives, such as mock network infrastructures or virtual financial systems, and then explore the action space through repeated trials.

## Risks From Runaway Capability Improvement

The prospect of AI systems recursively improving their own capabilities has emerged as a significant concern in recent AI safety research. The 2025 International AI Safety Report identifies recursive self-improvement as a key risk factor, noting that AI systems with sufficient autonomous capabilities could "improve their own code, architecture, or training processes, potentially leading to rapid and difficult-to-predict capability gains."[130] Recent developments have made this theoretical risk more concrete. In 2025, Google DeepMind unveiled AlphaEvolve, a coding agent that was able to discover novel algorithms that improved data center scheduling and hardware design.[131] However, the broader feasibility and timeline of runaway self-improvement remain contested among experts, with a 2024 survey of AI researchers finding variation in expert predictions about when AI systems could fully automate AI R&D, ranging from years to centuries.[132]

As noted in multiple AI developer risk frameworks (e.g., Google DeepMind's Frontier Safety Framework),[133] agents capable of automating ML R&D could rapidly gain more destabilizing and destructive capabilities, such as the ability to conduct highly novel cyber and bio-attacks that cause potential damages at the scale of hundreds of billions to potentially trillions of dollars.

Could a rogue HACCA deployment accumulate sufficient training compute to develop a more capable successor model in the wild? Our initial analysis suggests that this seems implausible, based on existing ML training paradigms. Developing a more capable successor model likely represents the most resource-intensive form of capability improvement, requiring substantial compute resources, training data, and sophisticated ML engineering capabilities. On top of that, this strategy also carries significant risk of failure or unintended behavioral changes.[134] When training a successor agent, a rogue deployment plausibly faces similar concerns with potential misalignment and scheming that human AI developers confront, though it may manage these by implementing its own control and alignment measures and other guardrails.

---

[130] Bengio et al., "International AI Safety Report 2025."

[131] Novikov et al., "AlphaEvolve: A coding agent for scientific and algorithmic discovery."

[132] Grace et al., "Thousands of AI Authors on the Future of AI."

[133] Google Deepmind, "Frontier Safety Framework."

[134] Ballpark estimates of the costs of training a state-of-the-art (SOTA) LLM today are $100 million to $1 billion, with $100s of millions on just compute spend. Maslej et al., "Artificial Intelligence Index Report 2025" gives compute-only cost estimates of $192 million for Gemini 1.0 Ultra and $170 million for Llama-3.1-405B.

**It might be more practical for a rogue deployment to attempt to influence a large-scale AI development process within a company or a government project**, e.g., via network intrusion, which would remove the need for significant accumulation of compute resources and other infrastructure. However, this would only be possible if these facilities were relatively insecure, e.g., if a rogue deployment had OC4-level capabilities and the relevant AI companies had protections weaker than SL4.[135]

While a rogue deployment seems unlikely to directly develop a more capable successor model, **there may also be many cheaper paths to capability enhancement, some of which could potentially lead to rogue deployments achieving destabilizing capability thresholds.** How significant this risk is, particularly relative to such destabilizing capabilities being developed within a company or government project, depends on several key uncertainties:

1. **The yield of "cheap" self-improvement paths vs. the expensive "successor-model" route.** If "cheap" methods (e.g., replication, improved scaffolding and tools, distillation, fine-tuning) can deliver meaningful capability gains, rogue systems could become dangerous without the conspicuous signals from training a new frontier model. However, companies could presumably implement these same techniques, so rogue deployments would only have a relative advantage if AI companies were not implementing these strategies—for example, if companies were engaging in a coordinated pause.
2. **When rogue deployments become capable of operating independently, relative to when companies' frontier systems become similarly capable.** The arrival order determines whether "wild" agents or lab-contained agents present the first catastrophic window. Established projects seem much more likely to have a substantial development lead, but this gap could narrow if company progress slowed due to safety concerns, international agreements that slow or pause development, or serious sabotage efforts.[136]

---

[135] SL4 refers to Security Level 4 as defined by Nevo et al., "Securing AI Model Weights," i.e., a system that can likely thwart most standard operations by leading cyber-capable institutions (OC4), including many state-sponsored groups and foreign intelligence agencies. Key practices include: isolated weight storage with TEMPEST protection, confidential computing, zero-trust architecture, hardware-enforced rate limits, advanced insider threat programs, and red-teaming by teams experienced with intelligence agencies.

[136] However, the magnitude of any resulting gap is uncertain: the same factors that slow company progress would likely also intensify defensive measures against rogue HACCAs, increase scrutiny of autonomous cyber capabilities, and potentially lead to more aggressive disruption efforts against existing deployments. Rogue deployments might still gain some relative advantage if they operate outside regulatory frameworks, in permissive jurisdictions, or with greater tolerance for risk than institutional actors—but this advantage would be partially offset by the increased defensive attention that accompanies any coordinated slowdown.

Our tentative conclusion is that rogue HACCA deployments are more likely to be able to reach highly destructive and destabilizing capability levels before other actors only under certain conditions. A critical factor is whether companies slow or pause frontier development for safety, regulatory, international agreement, or sabotage reasons, allowing agents "in the wild" to reach key autonomy thresholds earlier than lab-contained systems. Containment and alignment measures within companies would also need to be effective enough to prevent lab-contained systems from causing catastrophic damage, and companies would need to not have deployed these systems defensively to prevent or disrupt external efforts to develop destabilizing agentic capabilities.

Several factors could amplify these risks. If model weights and scaffolding proliferate faster than companies' release practices, more actors could assemble dangerous systems from widely available components. Operational dynamics also matter: if shutdown-evasion techniques prove cost-effective, jurisdictional arbitrage allows operation in permissive environments, and practical pathways exist for acquiring compute and financing (e.g., cloud access, crypto rails, human intermediaries), rogue deployments could survive detection and scale their operations. Finally, widespread adoption of autonomous AI agents could cut both ways—dense agent traffic would make HACCA detection harder, but competition from defensive agents and other AI systems might also constrain any single deployment's ability to accumulate resources and outcompete the broader ecosystem.

# 4 | Strategic Implications of HACCAs

Over the past two decades, offensive cyber operations have become standard in statecraft, with rival nation-states continuously competing for advantage via espionage and, rarely, sabotage. We expect HACCAs to intensify this competition: accelerating operational tempo, decreasing the labor cost of launching a cyberattack, and making some types of damaging attacks substantially more achievable. Moreover, as costs decline and HACCA components like scaffolding become commoditized, we expect HACCAs to become accessible to a larger number of threat actors, including criminal groups. This will increase the types and frequency of attacks.

The impact of HACCAs will depend heavily on whether defenders can keep up. Historically, trailing-edge defenders have struggled to adopt the latest defensive technologies, exposing critical economic and social functions to disruption. However, early action by policymakers can steer this "mainline trajectory" of intensified cyber operations in a more stable direction.

In parallel, policymakers should prepare for two potential strategic surprises: inadvertent cyber-nuclear escalation, and sustained loss of control over rogue HACCA deployments. There is higher uncertainty about these scenarios, but their impacts could be catastrophic. To mitigate these impacts, policymakers will need to develop novel governance mechanisms and response capabilities for scenarios that may lack clear historical precedents.

## Mainline Trajectory: The Evolving Threat Landscape

### How States Use Cyber Power: Three Strategic Use Cases

States likely will be the first to develop and deploy HACCAs, given the resources and technical sophistication required (see "Early HACCAs in the hands of great powers"). Understanding how nation-states currently use cyber power, and where it falls short, reveals where HACCAs likely will have the greatest strategic impact. **This section focuses on how nation-states employ offensive cyber capabilities across three strategic use cases: coercion and signaling; intelligence collection; and degradation and destruction.[137]** Today, cyber operations are ubiquitous, with nation-states engaging in continuous, low-intensity competition, but cyber's strategic impacts are ambiguous, and catastrophically destructive outcomes are rare.

---

[137] These categories are not mutually exclusive; a single operation may serve multiple purposes. They are also not exhaustive—e.g., we do not address cyber-enabled influence operations, which merit separate analysis.

IAPS | Institute for AI Policy and Strategy

**HACCAs almost certainly will intensify cyber competition, improving intelligence collection and making degradation and destruction more technically achievable, as well as more widespread.** But because of the secrecy of cyber operations, nation-states gain little from threatening destructive attacks for coercion or deterrence. **With HACCAs, nation-states are likely to inflict more regular cyber damage on each other, as long as they assess that doing so will not cause undue escalation and that the level of damage remains within their appetite for the use of force.** The durability of their ability to cause more severe or catastrophic damages will also depend on how quickly AI-enabled defenses adapt.

## Table 10: Strategic Uses of Offensive Cyber Operations

| Strategic use | Description | Effect of HACCAs |
|---|---|---|
| **Degradation and destruction** | Disrupting, degrading, or destroying adversary systems or infrastructure. Offensive cyber operations could be used as a substitute for kinetic strikes, alongside them, or to enable conventional operations. | ↑↑ **Significantly more viable;** HACCAs enable more scalable and sustained attacks, though there are still physical limits from passive safety systems, and collateral damage is hard to avoid |
| **Intelligence collection** | Extracting military, economic, and counterintelligence information from target systems. Intelligence collection provides an information advantage that enables better decisions, exposes vulnerabilities, and supports other operations. | ↑ **Higher volume and success rates;** stability effects uncertain—transparency could reduce miscalculation, but collection asymmetries may incentivize pre-emptive strikes |
| **Coercion and signaling** | Threatening or imposing costs to compel behavior change or deter hostile action. Successful coercion and deterrence influence adversary decision-making by demonstrating capability and resolve. | ~ **Limited change;** HACCAs enable more severe threats but cannot resolve the fundamental tension between operational secrecy and credible signaling |

**The operational shift that would underlie these changes is cheaper persistent access to target systems.** Today, strategic operations depend on maintaining footholds in adversary networks, which is labor-intensive: nation-states must sustain access over months or years, requiring continuous skilled effort and careful judgment about when to act or remain dormant. These constraints limit how many high-value footholds an intelligence service can maintain and force nation-states to be judicious about "burning" their access via cyber operations or attacks.

HACCAs could reduce these costs at multiple stages. During infiltration, HACCAs could automate reconnaissance and more flexibly research and exploit vulnerabilities, rather than relying on pre-loaded exploits. Once inside networks, agentic implants could maintain access and adapt to changing circumstances with minimal command-and-control communication, making detection difficult. With access less costly to establish and maintain, nation-states could pursue footholds more aggressively, and treat them as more expendable when operations demand it.

### Degradation and Destruction

Degradation and destruction involve operations aimed at impairing or disabling adversary systems or infrastructure. This includes temporary disruption (rendering systems unavailable), sustained degradation (reducing effectiveness over time), and destruction (physically rendering a system inoperable, as in Stuxnet). These operations may occur in the context of armed conflict or in peacetime competition below the threshold of war.

| The Picture Now | The Picture After HACCAs |
|---|---|
| Very few successful examples, as technical and operational challenges make it difficult to achieve desired effects (e.g., labor-intensive due to intense tailoring; challenges with reliability) | ↑↑ **Significantly more viable;** HACCAs enable more scalable and sustained attacks, though there are still physical limits from passive safety systems, and collateral damage is hard to avoid |

Cyber operations offer theoretical advantages over kinetic force, including no forward-deployed personnel, plausible deniability, and limited escalation risks. Stuxnet stands out as one of the most technically impressive, but even so, its strategic effect was modest, destroying 1,000 centrifuges out of several thousand operational at Iran's Natanz plant, and delaying but not halting the progress of Iran's nuclear program.[138] Cyber operations have struggled to produce warfighting effects analogous to kinetic weapons, as Russia's 2022 invasion of Ukraine demonstrates. Despite Russia having among the best cyber capabilities in the world and deep familiarity with Ukraine's digital infrastructure, its most consequential cyber operation was a modest disruption of Ukrainian military communications in the early stages of the invasion.[139] Kinetic attacks caused "multiple orders of magnitude more damage,"[140] despite Russia launching an unprecedented volume of

---

[138] Milevski, "Stuxnet And Strategy: A special operation in cyberspace?"
[139] This was a hack of the Viasat KA-SAT satellite network, which only temporarily disrupted Ukrainian military communications during the opening phase of the war before Ukrainian forces adapted by shifting their communication methods.
[140] Bateman, "Russia's Wartime Cyber Operations in Ukraine: Military Impacts, Influences, and Implications."

IAPS | Institute for AI Policy and Strategy

cyberattacks.[141] While there are idiosyncratic reasons for this, e.g., strong defender cooperation and Russian institutional weaknesses,[142] there are also more general reasons that limit cyber's effectiveness compared to kinetic operations.

Compared to kinetic weapons, cyber tools have technical limitations that constrain their operational effectiveness. For one, the need for customization makes cyber operations labor-intensive. Each attack requires tailored preparation and up-to-date intelligence about the specific target, because while a missile damages whatever it hits, a single software patch can render an offensive tool useless, making attacks difficult to reuse.[143] Russia's limited effect in Ukraine, for example, stemmed in part from its lack of a large standing cyber force that could sustain high-tempo attacks over a long conflict.[144] The challenge of coordinating cyber effects is compounded by the difficulty of observing and verifying them, and the risk of collateral damage from interconnected software systems. Overall, these traits make conventional cyberattacks ill-suited to missions requiring the guaranteed destruction of specific targets on a fixed timetable.[145]

**Destructive attacks are difficult and costly to execute because they are bespoke, fragile, and risk "burning" persistent access.** Such attacks require labor-intensive tailoring to induce particular effects in specialized systems, and furthermore remain at the mercy of network changes and defensive updates.[146] They also risk "burning" persistent access because they are so noticeable. By comparison, espionage operations carry a much lower risk, encouraging nation-states to engage in cyber espionage on a continual, aggressive basis.

HACCAs could make cyber-powered degradation and destruction more viable by addressing operational and technical limitations that have historically constrained cyber's utility. They could enable nation-states to sustain higher-tempo, more scalable degradation and destruction attacks,

---

[141] Cattler and Black, "The Myth of the Missing Cyberwar": "On the day the invasion began, Russian cyber-units successfully deployed more destructive malware… than the rest of the world's cyberpowers combined typically use in a given year."

[142] Ukraine and its allies mounted a sophisticated defensive effort (Black, "Russia's War in Ukraine: Examining the Success of Ukrainian Cyber Defences"), while Russia faced operational issues, e.g., not integrating cyber units and kinetic forces (Bateman, "Russia's Wartime Cyber Operations in Ukraine: Military Impacts, Influences, and Implications"). Ukraine or allies may also have conducted cyber operations to disrupt Russian hackers themselves (Lin, "Russian Cyber Operations in the Invasion of Ukraine," 36).

[143] Lin, "Russian Cyber Operations in the Invasion of Ukraine."

[144] Bateman, "Russia's Wartime Cyber Operations in Ukraine: Military Impacts, Influences, and Implications."

[145] However, the same characteristics that make cyber weapons unsuited to precise military effects may also make them better suited to indiscriminate disruption, akin to NotPetya: that is, scenarios in which a target set contains many entities and only some need to be affected to achieve the desired effect (Lin, "Russian Cyber Operations in the Invasion of Ukraine").

[146] Lonergan and Lonergan, "Escalation Dynamics in Cyberspace."

substituting—for example—for Russia's labor constraint in Ukraine. Their strategic autonomy could also make them more usable "off-the-shelf" than conventional cyber tools, which typically require extensive tailoring to specific target environments.

These operational advantages could enable larger-scale damaging attacks across two categories: pure-digital and cyber-physical attacks. We discuss these attacks in order of increasing difficulty, and increasing damage.

| Type of Attack | Description |
|---|---|
| Pure-digital ("digital blitz") | Data-damaging worms using wiper malware to irreversibly destroy data/systems across networks |
| Pure-digital ("sectoral collapse") | Systematic dismantling of economic/societal sectors through extensive pre-positioning, reconnaissance, and coordinated attacks (wipers + data manipulation) |
| Cyber-physical (disruptive effects) | Temporary interruption of physical operations (energy, transportation, chemical plants) by erasing OT/ICS programming or targeting connectivity |
| Cyber-physical (destructive effects) | Permanent physical destruction by bypassing fail-safes to induce unsafe states (explosions, equipment destruction) |

The most straightforward pure-digital attack would be a "digital blitz" using wiper malware to irreversibly destroy data and systems. Traditional data-damaging worms like NotPetya and WannaCry spread devastatingly fast but require rare "elite exploits"—vulnerabilities allowing arbitrary actions on widely used devices.[147] HACCAs' advantage would likely come from being stealthier, better at orchestration, and more adaptive. They could reduce a dependency on elite exploits by pre-positioning across many organizations and chaining smaller vulnerabilities to achieve administrative access, then delivering wipers simultaneously. However, this strategy carries risks of collateral damage, given the interconnected digital ecosystem: NotPetya, created by Russia, also hit Russian companies like Rosneft.[148]

---

[147] Halstead and Righetti, "Assessing The Risk Of AI-Enabled Computer Worms" define elite exploits as (1) being "zero-click," (2) allowing remote code execution, (3) having admin or higher privileges, and (4) being effective against more than 10 million systems. Such exploits are increasingly rare as patching cycles improve and leading vendors spend more effort on security.

[148] Steinberg et al., "NotPetya: A Columbia University Case Study."

A potentially more attractive digital-only option enabled by HACCAs is to conduct a "sectoral collapse" attack, leveraging a deployment's planning and orchestration capabilities to systematically dismantle one or more sectors[149] of an adversary's economy. In theory, such operations could leverage extensive pre-positioning to target key dependencies, combining wiper attacks with subtler methods (e.g., corrupting financial data) to create systemic instability far exceeding immediate damage. Attacks on the U.S.'s finance sector could cost hundreds of billions of dollars,[150] and attacks on some sectors, like healthcare, could even cost lives. However, sectors are interconnected, so even attacks targeting a single sector could cascade through shared infrastructure.

In the cyber-physical realm, HACCAs could have the capability to cause both substantial disruptive and destructive effects. Disruptive effects—temporarily interrupting physical operations by erasing OT/ICS programming or targeting connectivity—require avoiding detection while orchestrating coordinated, large-scale attacks (e.g., wipers triggering fail-safes and temporary shutdowns), a task HACCAs are well-suited for. Some OT/ICS systems have strong security features like air-gapping, but these are often inconsistently applied, especially as digitalization increases.[151] Although "temporary," sufficiently damaging disruptions could take weeks to safely restore, requiring specialized OT engineers to reprogram systems.[152]

Destructive effects—permanently destroying physical infrastructure or causing casualties through cyberattacks—are far more difficult. Many industrial facilities use passive mechanical safety mechanisms (e.g., pressure relief valves, interlocks) that cannot be digitally overridden; even the most capable attacker may find it physically impossible to induce an unsafe state.[153] Deliberately triggering destruction requires substantial engineering expertise and physical experimentation,[154] with little room for error: Iran's TRITON malware compromised safety controllers at a Saudi petrochemical facility but accidentally triggered shutdowns twice while attempting an explosion, leading to discovery. And as discussed, kinetic alternatives may simply be more effective: Russia's extensive cyber operations against Ukraine have not produced anticipated grid blackouts, while low-cost Shahed drones have proven far more damaging.

---

[149] Nation-states may choose to sacrifice scale (targeting multiple sectors) in favor of a single-sectoral attack for various reasons: (1) to be seen as less escalatory, (2) to reduce risk of discovery during pre-positioning, and/or (3) to focus their efforts, increasing the damage and chance of success.

[150] Lloyd's has modeled such an attack costing the U.S. about $1.1 trillion over five years (Smith, "Lloyd's finds major hack of a payments system could cost $3.5tn").

[151] Air-gapping generally prevents access except via physical means, which HACCAs cannot overcome at scale without human confederates. However, only 16% of organizations in a recent survey had air-gapped OT/safety systems (SANS Institute, "SANS 2024 State of ICS/OT Cybersecurity").

[152] IT systems can be restored simultaneously from automated backups, but OT systems often need individual reprogramming with specialized equipment and expertise. Safety concerns could also contribute to delays, as could supply chain issues, as in Russia's 2022 Viasat attack, which required shipping tens of thousands of replacement satellite modems to customers (Burgess, "A Mysterious Satellite Hack Has Victims Far Beyond Ukraine").

[153] For example, Marszal, "Make Process Plants Inherently Safe Against Cyber Attack" describes how process plant operators typically conduct a Hazards and Operability Study (HAZOP) to ensure that even if a physical cause for a catastrophic event is possible (e.g., closing a flow control valve to cause a build-up of pressure), best practice is to include a "non-hackable" safeguard that does not need to be programmed and reliably defuses the catastrophic event (e.g., a pressure relief valve that is not cyber-exposed).

However, where sectors use systems that must adapt to dynamic external environments, there are inherent constraints on fail-safe design. Maritime vessels, aircraft, and military systems operate in conditions where many failure modes cannot be safely absorbed: a cargo ship that loses propulsion can crash into key infrastructure, as in the 2024 Baltimore bridge collapse, which cost $4–5 billion and closed the Port of Baltimore.[155]

Cyber-capable agents could also weaponize control of drones and robotics, although this would be much more challenging for a HACCA. Drone attacks on poorly-protected systems could provide a low-cost way to destroy critical infrastructure, as shown by Ukraine's destruction of Russian nuclear-capable bombers.[156] Mass hijacking commercial drones to kinetically attack electrical grid transformers,[157] which are critical and scarce components for the U.S. electrical grid, could be a more efficient way to cause physical damage than trying to bypass electrical grid fail-safes using purely digital means.[158] Given rapid robotics adoption,[159] historical data is likely to underestimate the cyber-physical attack surface, and we should expect sophisticated actors to be able to exploit new attack vectors in creative ways. Such attacks would, however, demand broader domain expertise (e.g., steering complex machines), which would be much more difficult for a HACCA to learn. As such, we should consider this possibility as being further out compared to other HACCA capabilities, but important to consider as the adoption of drones and robotics rapidly increases.

### Intelligence Collection

Intelligence collection involves extracting strategically valuable information from adversary systems. Cyber-enabled espionage operations target digital systems remotely, granting access to strategically relevant materials without physical risks to intelligence personnel. This appears to be the most common use of cyber operations, accounting for about 60% of publicly attributed state-sponsored cyber incidents in one dataset.[160]

---

[154] E.g., Stuxnet's creators reportedly had to order sample Siemens centrifuges for destructive testing; such physically embodied testing could be hard for HACCAs to conduct.

[155] Crane, "Key Bridge replacement costs soar as high as $5.2 billion, opening delayed to 2030."

[156] Dahlgren and MacKenzie, "Ukraine's Drone Swarms Are Destroying Russian Nuclear Bombers. What Happens Now?"

[157] Ottinger and Harding, "Gridlocked: Transformer Shortage Choking US Supply Chains."

[158] Van der Merwe, "Assessing the Risk of AI-Enabled Cyberattacks on the Power Grid."

[159] Vermeer et al., "Averting a Robot Catastrophe."

| The Picture Now | The Picture After HACCAs |
|---|---|
| **Cyber's clearest area of strategic value**, though intelligence more often enables tactical successes than decisive strategic outcomes | ↑ **Higher volume and success rates;** stability effects uncertain—transparency could reduce miscalculation, but collection asymmetries may incentivize pre-emptive strikes |

This is cyber's clearest area of strategic impact, with successes across multiple domains, including **military intelligence** to map adversary capabilities and steal weapons designs; **economic espionage** to boost domestic industries; and **counterintelligence** to map adversary intelligence networks and compromise their communications systems.[161]

However, cases where intelligence decisively alters major strategic outcomes are hard to establish with confidence, given that such decisions are often classified and difficult to assess in retrospect. Such examples do exist: some authors have estimated that "Ultra," the Allied intelligence program to decrypt German signals traffic, shortened World War II by two to four years.[162] Mossad intel collection and CIA analysis also enabled Israel's 2007 strike on Syria's al-Kibar nuclear reactor, which led to a permanent end to Syria's nuclear weapons program.[163] But more commonly, intelligence enables tactical successes (e.g., disrupted adversary programs) whose aggregate strategic impact is harder to establish, or supports larger decision-making processes by informing judgments and shaping options.

Besides scaling up intelligence collection by enabling persistent access across more targets, HACCAs would likely also increase success rates by improving analysis of collected intelligence. In particular, agentic implants could help filter and synthesize data locally, enabling nation-states to maximize the value of their access. Exfiltrating large volumes of data ordinarily increases the risk of detection, but if HACCAs can analyze large volumes of intelligence on the network directly, they can focus only on transmitting the most strategically valuable information out of the network. However, HACCAs may only be modestly more successful against the best-defended targets. Jumping air gaps (e.g., as used to protect highly classified systems) requires a physical vector—such as a USB drive or a proximate receiving device to pick up electromagnetic or acoustic signals—that HACCAs cannot provide remotely.[164]

---

[160] By comparison, 28% of operations were disruptions and 11% were more serious degradation attacks. See: Maness et al., "Tracking Competition in Cyberspace: Announcing the Dyadic Cyber Incident Dataset Version 2.0."

[161] For example, China's intelligence operations are believed to have accelerated development of their J-31 fighter (Weisgerber, "China's Copycat Jet Raises Questions About F-35") and dismantled much of the CIA's human source network in China (Dorfman, "Botched CIA Communications System Helped Blow Cover of Chinese Agents").

[162] Hinsley, "British Intelligence in the Second World War."

[163] Gross, "Ending a decade of silence, Israel confirms it blew up Assad's nuclear reactor."

At a strategic level, enhanced intelligence collection could in fact improve stability between major powers. Nation-states with clearer pictures of adversary intent and capabilities (e.g., an adversary's actual military posture) may gain a clearer understanding of escalation dynamics, reducing the risk of arms races or miscalculations during a crisis.

However, even robust intelligence collection and analysis often fail to influence policy. Intelligence failures occur "more often at the consuming than the producing end," with cognitive biases and political pressures frequently overwhelming accurate assessments.[165] Intelligence asymmetry could also be destabilizing: if one nation-state gains a significant advantage and the other nation-state finds out, the disadvantaged nation-state might assume the worst. It might wrongly believe, for instance, that its adversary has mapped its nuclear submarine patrol routes and compromised its second-strike capabilities, pushing it toward a pre-emptive first strike.

### Coercion and Signaling

Offensive cyber operations can shape adversary decisions by compelling changes in behavior or deterring hostile actions. Coercion—imposing costs to force an adversary to change course—and deterrence—threatening retaliation to prevent hostile actions in the first place—both require influencing an adversary's calculus.

| The Picture Now | The Picture After HACCAs |
|---|---|
| **Cyber poorly suited for coercion/deterrence;** hard to threaten severe costs while maintaining secrecy needed for successful operations | **~ Limited change;** HACCAs enable more severe threats but cannot resolve the tension between operational secrecy and credible signaling |

The fundamental limit that HACCAs face in their use for coercion is that **cyber operations are best at "shaping, not signaling," providing nation-states with operational advantages**

---

[164] Dorais-Joncas and Muñoz, "Jumping the air gap: 15 years of nation-state effort."
[165] Betts, "Analysis, War, and Decision: Why Intelligence Failures Are inevitable"; Jervis, "Why Intelligence Fails."

**rather than coercive leverage.**[166] Cyber operations often do not impose costs severe enough to alter nation-state behavior.[167] But more importantly, because cyber operations are inherently ambiguous[168] and depend on secrecy, with many steps that victims can discover and block, they are poorly suited for deterrence, which requires credibly signaling capability and resolve.[169] Instead of coercion, cyberspace supports a logic of consistent exploitation: that is, low-grade competition where nation-states erode each other's capabilities in a cumulative way over time.[170]

HACCAs are unlikely to change cyber's limited utility for coercion and deterrence. While they would make it easier to impose costs severe enough to alter nation-state behavior (e.g., see "Degradation and destruction"), nation-states still could not credibly communicate capabilities or demonstrate resolve without burning the access that makes threats possible. This secrecy problem is structural: revealing HACCA capabilities to make a credible threat risks exposing vulnerabilities, tradecraft, and the existence of the capability itself. For example, a nation-state cannot easily demonstrate it has pre-positioned HACCAs throughout an adversary's critical infrastructure without burning that access.[171] And even if it could credibly threaten damage, adversaries may doubt whether the threatening nation-state would actually follow through, given risks of escalation and blowback. Even with HACCAs, conventional and nuclear weapons will therefore likely remain better for coercion and for deterrence.

---

[166] Buchanan, "The Hacker and the State": "If international relations are like a game of high-stakes poker, to signal is to hint credibly at the cards one holds, in an attempt to influence how the other side will play its hand. To shape is to change the state of play, stacking the deck or stealing an opponent's card for one's own use."

[167] Empirically, offensive cyber operations yielded political concessions only 4% of the time in interactions between rival nation-states from 2000–2014; successful cases typically also needed sanctions or military threats (Valeriano and Jensen, "The Myth of the Cyber Offense: The Case for Restraint").

[168] A cyber operation could appear to be a temporary service outage, or vice versa. Even if defenders discover an ongoing cyber operation, it takes time to determine its scale and the attacker can obfuscate their identity.

[169] Lin, "Escalation Dynamics and Conflict Termination in Cyberspace"; Lonergan and Lonergan, "Cyber Operations, Accommodative Signaling, and the De-Escalation of International Crises." With that said, cyber may still be useful for signaling in escalation management: e.g., choosing cyber over kinetic options can communicate restraint.

[170] Goldman et al., "Persistent Engagement in Cyberspace Is a Strategic Imperative."

[171] A defender discovering an attacker's foothold can remove the possibility of regaining the foothold through similar means, because it exposes the specific tradecraft or software vulnerabilities used to gain entry.

### HACCAs and "Mutually Assured AI Malfunction"

If frontier AI capabilities become strategically decisive—enabling "wonder weapons" or systemic shifts in global power[172]—then for great powers, delaying adversaries' AI progress may become a core geopolitical aim.[173] To do this, states could consider:

- **Direct attacks on frontier AI companies**, whether covertly (e.g., subtly disrupting training runs and poisoning data) or overtly (e.g., destructive wiper attacks to corrupt model weights, the training data corpus, and backups); this could include harassing AI researchers to coerce them into leaving the company or stopping their research
- **Disrupting semiconductor manufacturing** in the country or allied/partner countries; e.g., targeting chip fabs (e.g., TSMC) or their supply chain (e.g., ASML) and aiming for disruptive/destructive effects that have asymmetric consequences (e.g., a "Stuxnet for extreme ultraviolet (EUV) machines," though such an attack would be very difficult[174])
- **Attacking energy infrastructure for frontier AI**, e.g., targeting power generation and distribution to constrain AI training and inference

Hendrycks, Schmidt, and Wang propose "mutual assured AI malfunction" (MAIM) as an arms-control mechanism, with nation-states threatening such attacks to deter adversaries from pursuing AI monopolies.[175] However, cyber's utility as a *threat* instrument remains limited by the same dynamics described above. An adversary cannot easily verify that a nation-state could destroy its AI infrastructure on demand. The secrecy that makes cyber operations effective makes them poor vehicles for credible threats. HACCAs could *execute* such attacks, but nation-states would struggle to leverage them for the coercive bargaining that MAIM envisions.

## Proliferation Dynamics: How Quickly Does This Get Worse?

Over time, the cost of developing and operating HACCAs will likely decrease, making them more broadly available even if initially limited to the U.S. and China. If HACCAs proliferate, then a wide range of actors will have access to operations previously only the remit of a few institutions in countries that are leading cyber powers. This would be a significant strategic change insofar as defenders would have to face a higher volume of high-skilled attacks: a mid-tier cybercriminal group, for example, might acquire network exploitation skills once only the domain of the NSA's Tailored Access Operations (TAO). Since trailing-edge organizations are particularly exposed to such attacks, companies and policymakers should conduct more research into HACCA proliferation speed and pathways to better calibrate the urgency of bolstering their defenses.

---

[172] Mitre and Predd, "Artificial General Intelligence's Five Hard National Security Problems."
[173] Deploying a HACCA destructively could have a risk profile and technical difficulty similar to that of Stuxnet, so the analogy here might be frontier AI reaching the geopolitical significance of a nuclear weapons program.
[174] Extreme ultraviolet lithography (EUV) machines are an essential bottleneck for manufacturing the U.S.'s most cutting-edge chips: they cost hundreds of millions of dollars and only ASML can build them. (Tarasov, "Exclusive look at the creation of High NA, ASML's new $400 million chipmaking colossus.") However, their security is tightly protected and if nation-state actors (e.g., China) had a solid enough foothold to understand and destructively interfere with ASML's machines, they might exploit it for industrial espionage instead.
[175] Hendrycks et al., "Superintelligence Strategy: Expert Version."

## Early HACCAs in the Hands of Great Powers

The U.S. and China could be the first actors to develop HACCAs, given their access to frontier AI capabilities and well-resourced offensive cybersecurity institutions. However, early HACCAs may pose significant costs or offer limited returns, leading them to deploy HACCA subcomponents in a modular fashion rather than as fully autonomous systems, or even to forgo HACCAs entirely.

The main edge that the U.S. and China have is unique access to frontier AI capabilities. Developing early HACCAs would require an extensive program contributing tens of millions of dollars in compute, skilled cyber experts, and classified cyber-related datasets (see "Tactic 1: Establish and Maintain Infrastructure"). But even leading intelligence agencies cannot build best-in-class foundation models on their own, so they would have to rely on one of two pathways to acquire AI-enabled cyber capabilities:[176]

- **Most easily, U.S. or Chinese intelligence agencies could establish public-private partnerships** with their own domestic champions in frontier AI, like the U.S. DoD has currently done with OpenAI, Google, Anthropic, and xAI.[177] This could let frontier AI companies give governments access to safeguard-free versions of cyber capabilities, without the barriers that normally restrict the use of offensive cyber in commercial models.
- **Alternatively, countries could steal proprietary models from their competitors** by cyber espionage or insider threat, which could allow them to leapfrog to the frontier, but would still require them to have domestic AI know-how to use and adapt frontier models. This might be feasible for China if its domestic AI industry lagged behind the U.S. too greatly, but would be more challenging for Russia, as it has suffered extensive brain drain

---

[176] A third path could involve relying on open-weight models, but doing so might put nation-states behind their OC5-capable peers given that open-weight models generally lag behind the frontier, and would still require domestic AI expertise (e.g., for fine-tuning and inference).

[177] CDAO, "CDAO Announces Partnerships with Frontier AI Companies to Address National Security Mission Areas"; Brewster, "The Pentagon Is Spending Millions On AI Hackers."

during its war with Ukraine and would face significant constraints on its ability to use stolen models.[178]

Given their potential to be first-mover deployers, the U.S. and China may want to pursue HACCAs as an "exquisite capability": state-of-the-art systems that confer unique strategic advantages in military and intelligence domains. One analogy here might be the U.S. National Reconnaissance Office's classified, multi-billion-dollar KH-11 optical imaging satellites, which allow the U.S. to spy on adversaries at extraordinary resolution.[179] HACCAs could similarly offer the promise of unprecedented intelligence collection capabilities, as well as capabilities in other areas.

However, the U.S. and China may also decide that developing fully autonomous HACCAs would not be worthwhile. Some of the factors that could push them toward this include:

- **Limited period of exclusivity.** If the cost of developing and operating HACCAs seems like it will drop rapidly, the U.S. and China may only expect a limited window (e.g., months) to exploit HACCAs as a unique advantage before the capability proliferates. If so, they may assess that investing tens of millions of dollars is unlikely to be worthwhile.
- **Insufficient reliability.** Early HACCAs may not be reliable, making them risky to deploy. For example, if it became clear that there were no effective ways to prevent loss of control over a HACCA, the U.S. and China might eschew the deployment of HACCAs entirely. However, if the risks were tolerable, they might still deploy HACCAs selectively in low-risk operations as "proving grounds" to improve the reliability of HACCAs.
- **Preference for modular subcomponents.** Rather than investing in full autonomy, OC5 actors might prefer to deploy discrete agentic capabilities (e.g., agentic implants for persistent access or automated reconnaissance tools) that offer more predictable performance and easier operational control than fully autonomous systems.

## HACCAs as a Normal Technology

Over time, as both training and inference become more compute-efficient, it will become easier for well-resourced actors to develop and operate HACCAs, and the barrier to entry may decrease such that even lesser-resourced threat actors can deploy them. HACCAs in this phase might look like a "normal technology": a commoditized family of tools used by diverse actors for both good and ill, perhaps more analogous to the ubiquitous Cobalt Strike than the NRO's KH-11 satellites.[180]

---

[178] Sherman, "Russia's digital tech isolationism: Domestic innovation, digital fragmentation, and the Kremlin's push to replace Western digital technology."

[179] For details on KH-11 satellite, see Brumfiel, "Trump tweeted an image from a spy satellite, declassified document shows." Another point of comparison might be zero-day exploits (like EternalBlue), which are extremely valuable; however, a HACCA would be more akin to a system that facilitates discovery of zero-days rather than a zero-day itself. Zero-days cost millions for popular devices and applications. For example, in 2024, the company Crowdfense was offering $5–7 million for iPhone zero-days. (Franceschi-Bicchierai, "Price of zero-day exploits rises as companies harden products against hackers.")

The increasingly widespread availability of HACCAs could lead to the democratization of what were previously elite nation-state-level capabilities: for example, mid-tier cybercriminal groups may now have at their fingertips the network exploitation skills of the NSA's Tailored Access Operations (TAO) group, its best-in-class "special forces."[181] This would likely increase the tempo, scale, and sophistication of attacks, putting more pressure on defenders to adapt.

Below, we provide an example of how HACCA proliferation might be driven by declining compute costs and the spread of common software components.

## Table 11: An Example Timeline of HACCA Proliferation (Illustrative)

| Timeline | Description |
| --- | --- |
| Early HACCAs first become feasible | The first end-to-end autonomous HACCAs become technologically possible, but they cost tens of millions of dollars and require highly specialized knowledge to build. This limits early HACCAs to the U.S. and China, which may even decide they are not worthwhile.<br><br>At this time, the most significant barriers might be in initial development (e.g., training the base model and adapting it via fine-tuning/scaffolding[182]), but provisioning and operating HACCAs may also require a large number of expensive cutting-edge AI chips. |
| Proliferation begins (e.g., 1–3 years after first HACCAs) | Financial and non-financial barriers begin to drop, but HACCAs still require significant expense to operate, enabling a broader but still limited set of actors to deploy HACCAs (e.g., nation-states like Russia, Iran, and North Korea, and possibly organized criminals).<br><br>Barriers to development may drop rapidly here. For example, open-weight models may reach HACCA-relevant capability thresholds, and other software components of HACCAs (e.g., scaffolding) could be leaked or stolen. However, running HACCAs may |

---

[180] Cobalt Strike is a commercial tool that helps security professionals test client defenses by simulating realistic cyberattacks. However, attackers use pirated versions widely: at its peak, Mandiant observed Cobalt Strike was used in 28% of all intrusions in 2021, though this declined to about 15% by 2022 (Mandiant, "M-Trends 2023," 38).

[181] Zetter, "NSA Hacker Chief Explains How to Keep Him Out of Your System."

[182] For example, only the most leading-edge frontier AI companies might be able to develop general-purpose models with powerful enough capabilities, and HACCA deployers would then still need to adapt these for cyber-related use (e.g., fine-tuning on classified cyber datasets, or having experts design agent scaffolding).

| | stay costly due to inference requirements, even as models become more compute-efficient.[183] |
|---|---|
| "HACCAs as normal tech" (e.g., 3–7 years after first HACCAs) | HACCAs come into range for most technically adept individuals willing to spend a few thousand dollars on compute. Development costs for basic, "free-tier" HACCAs are dropping toward zero as commoditized tools and models emerge, though provisioning and operating a HACCA still requires some initial resource outlay and expertise.

However, not all HACCAs are created equal: better-resourced actors (e.g., nation-states, organized crime groups) can develop systems that are more skilled and reliable, and more self-sustaining (i.e., able to break even by generating their own revenue and/or acquiring compute to cover the cost of operations). By comparison, older HACCAs may struggle to compete with newer HACCAs and overcome the adaptations of defenders. |

*Note: Stages and timeline are meant to be illustrative, not predictive. We believe further research is needed.*

HACCA proliferation could make attribution harder and provide plausible deniability, ultimately encouraging attackers to use more HACCA-enabled attacks. Besides being more difficult to tie back to operators as instances multiply, HACCAs could also deceive defenders by imitating the operational practices of other groups, including shallow features like language and working hours, and deeper features like attacker-specific tactics, techniques, and procedures (TTPs).[184] Some nation-states (e.g., Russia) may even proliferate HACCAs to their proxies,[185] though this is less likely as they would likely prioritize the scarcest resources and most sophisticated insights for their own projects.

Proliferation also compounds deconfliction challenges beyond a single deployer. As more nation-states develop offensive cyber capabilities and potentially deploy HACCAs, there is growing risk of "fratricide" between allied operations—and even close allies may be reluctant to disclose the nature or location of their cyber activities to one another.[186] Traditional deconfliction mechanisms assume a relatively small number of coordinating parties, and widespread HACCA deployment could overwhelm existing information-sharing arrangements.

---

[183] Ord, "Inference Scaling Reshapes AI Governance."
[184] This is speculative, but seems plausible: compared to the "switching costs" of a team of human Russian experts trying to imitate a North Korean actor, LLM-based cybersecurity agents would very likely be more effective at planting superficial operational clues (e.g., using Korean, or only operating during North Korean working hours). Whether they could reliably replicate deeper operational aspects like specific TTPs or indicators of compromise (IOCs) might depend on their training data and general cybersecurity capabilities.
[185] Some intelligence services (e.g., Russia's) already work with cybercriminals and "patriotic hackers" to achieve national cyber priorities. (Conrads and Hayward, "Hacktivism In Russian Cyber Strategy."; Sherman, "Untangling the Russian web: Spies, proxies, and spectrums of Russian cyber behavior.") Conceivably, they could help proxies develop and deploy HACCAs by providing resources like legal cover, financial backing, procurement networks, and technical advice.
[186] Skingsley, "Offensive cyber operations."

The democratization of formerly elite cyber capabilities could be punishing for human defenders, who might increasingly need automation to counter high volumes of sophisticated attacks. For example, HACCAs might be able to impersonate or hack trusted actors at scale, conducting hundreds of equivalents to the XZ Utils backdoor attack in a pervasive compromise of the software supply chain.[187] HACCAs could also overwhelm defenders by discovering and exploiting vulnerabilities faster than human teams can triage them, breaking the current operational tempo of vulnerability management, where defenders typically have days or weeks to test and deploy updates. In order to keep up, defenders would have to integrate AI tools into their secure coding and cybersecurity workflows, which could disadvantage less-resourced defenders.

The actual speed and pathways of HACCA proliferation will inform how defensive efforts should be prioritized. For example, if HACCAs are likely to proliferate rapidly, such that even "script kiddies" with little programming background can deploy them easily, that that may increase the need for aggressive investment in ready-to-adopt defensive solutions, such as accelerating adoption of current best practices and easily usable AI tools that can provide immediate uplift. However, if we expect proliferation to be more gradual—for example, if HACCA operation remains expensive due to high inference costs—then there may be more runway for cyber "moonshots"[188] and sustained investment in more transformative defensive capabilities. Of course, this is not an either-or choice: both moonshots and ready-to-adopt solutions should be part of a larger defensive portfolio, but because defenders are resource-constrained, knowledge of the future strategic landscape is crucial to allocating scarce labor, time, and money.

Policymakers should work to estimate when the first HACCAs might be developed, how these capabilities will proliferate, and what the implications are for defensive prioritization. Beyond technical questions around when these capabilities will arrive and how hardware/software may improve, we also need to answer questions around the economic impact of these capabilities, including estimating how much more financially attractive cyberattacks on trailing-edge organizations will become.[189]

---

[187] In the XZ Utils incident, a bad actor gained the trust of an open-source developer over years by pretending to support them, finally using their access to insert a well-crafted malicious backdoor into an open-source package used by millions of Linux systems worldwide. (Greenberg and Burgess, "The Mystery of 'Jia Tan,' the XZ Backdoor Mastermind.") Such an attack would have been labor-intensive, but HACCAs could conceivably orchestrate hundreds of such operations in parallel to gain extensive backdoor access across the supply chain.

[188] Brundage, "Operation Patchlight"; Bradley and Sastry, "The Great Refactor."

Ultimately, however, the strategic impact of HACCAs will depend not only on offensive capabilities, but on whether attackers can deploy them before defenders adopt AI-enabled countermeasures. AI systems could dramatically strengthen defense through automated vulnerability discovery, AI-generated secure code, real-time threat detection, and automated patching—potentially closing many of the gaps that HACCAs would otherwise exploit. The extent to which HACCA advantages endure cannot now be determined, but before defensive systems adjust, early deployers would enjoy a significant first-mover advantage.[190] Defensive adoption will likely unfold unevenly across sectors, with well-resourced defenders adapting faster than critical infrastructure operators with legacy systems and limited security budgets (see next section on "The offense-defense imbalance and why it matters"). The strategic impacts of HACCAs would therefore depend less on absolute capability than on the gap between offensive deployment and defensive adaptation in specific domains.

## The Offense-Defense Imbalance and Why it Matters

The predictable consequence of automating cyber operations at scale is that cyberattacks will become cheaper, easier, and more frequent. If defense became equally cheaper and easier, this would be of little consequence—which is why early defensive investment is key to avoiding a sustained imbalance in favor of attackers. But adequate defense is harder to achieve than it sounds, because HACCAs will not affect all actors equally. At least initially, we expect that a period of elevated risk will fall disproportionately on those least equipped to respond.

Attackers may have early, structural advantages in adopting AI in cybersecurity. They can generally tolerate higher levels of system unreliability than defenders,[191] and the high costs of AI inference could make it prohibitively expensive for defenders to protect their entire attack surface.[192] As such, malicious organizations with the resources and technical capacity to operationalize offensive AI could exploit defensive vulnerabilities before effective countermeasures are in place.[193] Some experts suggest that offensive advantages may narrow over time, but even so, uneven adoption of offensive and defensive AI capabilities could create a period of acute risk in the interim.[194]

---

[189] Murphy and Stone, "Uplifted Attackers, Human Defenders: The Cyber Offense-Defense Balance for Trailing-Edge Organizations."

[190] Danzig, "Artificial Intelligence, Cybersecurity, and National Security."

[191] For example, even though early HACCAs and other defensive AI tools are likely to be unreliable, defenders are typically under institutional pressure to avoid making mistakes, while attackers have more leeway for experimentation. This is likely true of nation-states as well as criminal actors.

[192] Withers, "Tipping the Scales: Emerging AI capabilities and the cyber offense-defense balance."

[193] Lohn, "Anticipating AI's Impact on the Cyber Offense-Defense Balance"; Oesch et al., "Agentic AI and the Cyber Arms Race."

There is also asymmetry among defenders themselves. Historically, actors with access to the best talent and deepest resources—like the intelligence community and the financial sector—have often adopted defensive tools much faster than less-resourced actors like lifeline infrastructure providers and the open-source software community. For such "trailing-edge organizations,"[195] the question is often not whether they can adopt next-generation technologies like AI, but whether they can implement even *current* best practices. Even if AI unlocks powerful defensive capabilities, like large-scale formal verification of software security, it could take a long time for these vulnerable organizations to adopt them or reap the benefits.[196]

Together, these two asymmetries mean that the advent of HACCAs could be critically damaging for many trailing-edge organizations. In the long run, AI could tip the offense-defense balance toward defenders, but in the short term, it will likely favor attackers—perhaps dramatically so at the trailing edge. The consequences are unlikely to be isolated, because these organizations are often systemically important: disruption at a regional utility can trigger cascading failures across critical infrastructure, and compromise of an under-maintained open-source library can become the weak link enabling a massive supply-chain attack on better-defended targets. In this interim period, we must ensure that these important but vulnerable defenders have the resources needed to adapt.

Companies and policymakers need to make a concerted effort to support these under-resourced defenders, particularly those that are keystones in the physical or digital ecosystem. This could include finding ways to help them take advantage of defensive AI systems despite their technical and financial disadvantages.[197] For many organizations, this may mean more limited, easy-to-deploy defensive tools rather than sophisticated "defensive HACCAs," which could carry significant security risks if deployed or configured incorrectly.[198] But other non-AI, tried-and-tested

[194] Potter et al., "Frontier AI's Impact on the Cybersecurity Landscape."

[195] As Murphy and Stone, "Uplifted Attackers, Human Defenders: The Cyber Offense-Defense Balance for Trailing-Edge Organizations" write, trailing-edge organizations "rely heavily on legacy software, poorly staff security roles, and struggle to implement best practices like rapid deployment of security patches," partly in expectation that attackers will not target them due to a lack of economic incentives.

[196] For example, see Potter et al., "Frontier AI's Impact on the Cybersecurity Landscape" for a discussion of frontier AI's impact on the cybersecurity landscape more broadly, including the use of formal verification to provide provable guarantees around system security, or Newman, "Cybersecurity and AI: The Evolving Security Landscape" for a similar discussion.

[197] Ee et al., "Asymmetry by Design: Boosting Cyber Defenders with Differential Access to AI."

[198] Such narrow defensive systems might be specialized to deal with specific problems (e.g., log analysis) and have more limited permissions. Broadly capable agents like HACCAs are more risky because the very assets that make defensive AI agents effective—elevated system privileges and intimate knowledge of internal

approaches will also be needed, such as subsidizing security services or providing shared capabilities, or even simply providing more funding where sectors are under-resourced.

## Strategic Surprises: Escalation and Loss of Control

Beyond the mainline trajectory described above, HACCAs introduce two categories of risk that could produce discontinuous or catastrophic outcomes. These events are highly uncertain but have potentially catastrophic consequences. Inadvertent escalation into nuclear crisis could occur if HACCA operations affect systems entangled with nuclear command and control. Sustained loss of control could occur if rogue HACCAs establish self-sustaining populations that resist shutdown. Both risks stem not from deliberate choices but from accidents, misperceptions, or emergent behaviors that current governance frameworks are poorly equipped to address.

The absence of clear historical precedent makes these scenarios difficult to assess, but it is precisely why preparation must begin now, because scenarios such as cyber-nuclear escalation and loss of control could catch policymakers unprepared if they materialize. Hardening critical infrastructure addresses expected threats; preparing for these scenarios requires developing new crisis communication protocols, international cooperation frameworks, and technical safeguards against outcomes that no actor intends.

### Escalation Dynamics and Risks

Decades of offensive cyber operations between rival nation-states have rarely led to serious escalation. Despite persistent intrusions and occasional destructive attacks, cyber conflict has largely been contained.[199] HACCAs are unlikely to change this for *deliberate* escalation: that is, nation-states consciously choosing to cross thresholds into armed conflict with a risk of nuclear exchange. While HACCAs could make decisive cyber effects more achievable, cyber remains poorly suited for signaling resolve, and nation-states retain both crisis communication channels and strong incentives to avoid kinetic conflict.

However, HACCAs could increase the risk of *inadvertent* escalation into a nuclear crisis. They could enable more frequent and more capable operations against military systems, with less well-understood failure modes than conventional cyber operations. An attack that spreads to or is misinterpreted as targeting nuclear command-and-control infrastructure could trigger conflict between nuclear powers.

---

networks—also make them exceptionally dangerous if compromised. However, such systems may still have a role to play for more advanced defenders, such as defense and intelligence communities.
[199] Lonergan and Lonergan, "Escalation Dynamics in Cyberspace."

IAPS | Institute for AI Policy and Strategy

Why has cyber conflict remained contained? Beyond the signaling limitations discussed earlier (see "Coercion and signaling"), two structural factors constrain escalation:

- **The costs of kinetic escalation remain prohibitive.** Nation-states understand that crossing from cyber to conventional conflict—let alone nuclear exchange—carries catastrophic risks. Even facing significant damage from cyber incidents, this calculus has held: no state treated Russia's NotPetya attack as a *casus belli* despite billions in global economic losses.

- **Nation-states have crisis communication channels.** Diplomatic back-channels and hotlines allow adversaries to manage incidents before they spiral. The effectiveness of these measures does depend on political relationships, as the deterioration of U.S.-Russia communication channels since 2022 illustrates.[200] However, given sufficient time to react, it seems unlikely that nation-states would let cyber incidents escalate without some attempt to address the situation.

## The Entanglement Problem

The special case where offensive cyber operations lead to inadvertent nuclear escalation is rooted in the increasing entanglement of conventional and nuclear systems. Digital technologies are deeply embedded in all elements of the nuclear enterprise. For example, the maintenance of the nuclear stockpile relies on data-driven computer simulations to certify weapon reliability, nuclear delivery platforms operate as complex cyber-physical systems where embedded computing controls essential functions like navigation and targeting, and nuclear command, communication, and control (NC3) depends on a vast infrastructure of satellites and networks that increasingly integrates commercial software and supports both nuclear and conventional missions.[201] U.S. NC3 systems have had several documented cyber vulnerabilities, ranging from a firewall issue in the 1990s that could have let malicious outsiders send launch orders to ballistic missile submarines in the Atlantic, to several weaknesses discovered in the 2010s that could have been used to disable nuclear weapons or prevent their crews from monitoring them.[202]

---

[200] For example, the U.S.-Russia Direct Communications Link was established after the Cuban Missile Crisis, and was then expanded in 2013 specifically to address cybersecurity incidents of national concern (The White House, "FACT SHEET: U.S.-Russian Cooperation on Information and Communications Technology Security"). However, after Russia's invasion of Ukraine, the Kremlin indicated that this line was not in use (Kelly, "Kremlin says Russia-US hotline to deflate crisis not in use").

[201] Lin, "Cyber Threats and Nuclear Weapons."

[202] Blair, "Why Our Nuclear Weapons Can Be Hacked."

Still, in the U.S. at least, the cyber threat to today's NC3 systems is judged to be "fairly minimal" by senior leadership because legacy architectures often lack the connectivity, interoperability, and software surface area that make cyber operations easier to implement.[203] However, the nuclear modernization agenda faces strong incentives for greater functionality, speed, and interoperability, especially as commanders seek fused sensor data, more resilient communications, and the ability to operate across domains in real time.[204] Achieving these gains also expands the cyber attack surface and erodes barriers between nuclear and conventional operations. A cyberattack intended strictly to degrade conventional capabilities could inadvertently blind or disrupt an adversary's NC3 systems and other elements of the nuclear enterprise, triggering fears of an imminent first strike and creating pressure to "use or lose" nuclear forces before they can be disabled.

### How HACCAs Could Increase Risks of Inadvertent Escalation

Even with HACCAs, many strategic factors preventing cyber-nuclear conflict will likely persist: communication channels and the prohibitive costs of nuclear exchange all create strong incentives for states to find off-ramps.[205] The danger is that cyber operations could affect nuclear-entangled systems under conditions that compress decision timelines, leaving insufficient opportunity to clarify intent. HACCAs could increase the probability of such accidents, particularly given continuing trends toward greater connectivity and complexity in nuclear systems.

There are two main reasons inadvertent escalation risks increase because of HACCAs:

- **First, HACCAs change the cost calculus for offensive cyber operations**, especially by delivering effects that disrupt, degrade, or destroy military systems. Traditional offensive cyber requires extensive preparation to achieve a particular effect, but HACCAs could make such operations more "off-the-shelf," and more operations mean more opportunity for something to affect the wrong target.

- **Second, HACCA failure modes are less well-understood than conventional malware and offensive cyber operations.** Autonomous decision-making, the black-box nature of contemporary AI systems, and unexpected emergent behavior could mean states have false confidence in their ability to predict what a HACCA will do once deployed. Despite this unpredictability, operators may deploy HACCAs anyway because they enable otherwise unachievable effects or because competitive pressure and crisis dynamics override caution.

---

[203] U.S. Strategic Command, "U.S. Strategic Command and U.S. Northern Command SASC Testimony."
[204] U.S. Department of Defense, "Nuclear Posture Review."
[205] However, sophisticated HACCAs may have awareness of these mitigation tactics, and could potentially disrupt physical systems (e.g., communication lines) or exploit protocols (e.g., through voice or signature mimicry) as part of their effort as well.

Not all escalation pathways are equally concerning. The critical variable is whether a scenario allows time for de-escalation or compresses decision timelines in ways that foreclose off-ramps. Table 12 compares three pathways.

## Table 12: Escalation Pathways From HACCA-Enabled Cyber Operations to Nuclear Crisis

| Pathway | Target | HACCA-Specific Risk | Assessment |
|---|---|---|---|
| **Pure cyber attack** | Civilian systems (finance, healthcare) | HACCAs could cause more severe sectoral disruption, but this does not change escalation dynamics | Likely that states can take various actions to de-escalate before things go nuclear. |
| **Attack on critical infrastructure** | Physical infrastructure (ports, grids) with kinetic-equivalent effects | HACCAs could enable more destructive effects, but a nation-state willing to cause such damage is likely already prepared for conventional war | If the state set out to damage critical infrastructure (e.g., ports/electric grid), probably the deploying state would already be prepared for kinetic war (like Russia tackling Viasat around the Ukraine war, or Volt Typhoon). |
| **Attack on nuclear enterprise-related system** | Conventional military systems entangled with NC3 | HACCAs increase frequency of military-targeted operations; unpredictability raises spillover risk | Unlike civilian or infrastructure attacks, intrusions into military systems may trigger responses under compressed decision timelines. A state that detects unauthorized access to NC3-adjacent systems during a crisis may interpret it as preparation for a disarming first strike. Nuclear postures may not allow time to clarify intent before responses are triggered.[206] |

---

[206] Arguments about time pressure involved in nuclear scenarios often cite the fact that the flight time of Russian ICBMs from launch to impact on U.S. soil is about 25 minutes. In that window, the United States must detect the launch, ascertain intent, and, if assessing that U.S. ICBM fields are targeted, decide whether to launch before those forces are destroyed—then execute that decision if affirmative (Lin, "Cyber Threats and Nuclear Weapons").

Two factors primarily determine whether an actor is likely to trigger inadvertent nuclear escalation: visibility into adversary NC3 systems and willingness to accept escalation risk.

Nation-states with mature cyber programs and sophisticated intelligence services, e.g., the U.S. and China, are better positioned to identify which adversary systems connect to nuclear infrastructure and to appreciate the risks of targeting them. Nation-states with less visibility are more likely to inadvertently affect NC3—especially if using autonomous cyber tools that spread beyond their intended targets.

Even actors with substantial visibility may differ in their tolerance for escalation risk. Some may exercise restraint based on recognition of catastrophic consequences, while others tolerate more risk due to strategic pressures (e.g., perceived threats of regime change) or nuclear doctrines that are more tolerant of brinkmanship. Russia, despite its cyber sophistication and deep familiarity with U.S. and NATO nuclear postures, has demonstrated greater tolerance for collateral damage via attacks like NotPetya and nuclear saber-rattling during the Ukraine invasion.[207]

Non-state actors could amplify risks as criminal groups, terrorist organizations, or even rogue HACCA deployments could lack both the knowledge and the incentive to avoid nuclear-related systems. If such an actor's HACCA affected NC3 infrastructure during a period of heightened state-to-state tensions, the incident could be misattributed to a nation-state adversary.

## Loss of Control

HACCAs could deviate from their intended objectives through design error or adversarial attack (see Section 2, "How could operators lose control of HACCAs?"). For early HACCAs, such a deviation may have limited consequences, as these systems may struggle to self-sustain—i.e., cover their own operational costs by acquiring more money/compute—and could therefore burn out early without doing significant damage. But as autonomous capabilities advance, uncontrolled and self-sustaining HACCAs may become increasingly possible, and if such systems were to actively pursue more resources and influence, they could pose a significant threat.

Rogue HACCAs could be extremely difficult to detect and counter: they could hide among the vast diversity of internet-connected systems, while exploiting advantages like machine speed and massive parallelization to outpace defenders (see [Appendix V](#)). A loss-of-control scenario[208] could produce a self-perpetuating threat that would be difficult to stamp out, something like a next-gen botnet, but running on AI chips instead of Internet of Things (IoT) devices. Moreover, such systems could have unpredictable objectives and may not be constrained by concerns around collateral damage in the same way that nation-states are.

---

[207] Blank, "Vladimir Putin's endless nuclear threats are a sign of Russian weakness."

That said, even if a HACCA could self-sustain, loss of control would not automatically make it a persistent threat. Unlike traditional malware, a HACCA's programming would likely not be rigidly hard-coded, so its objectives could drift over time.[209] To not just go rogue, but stay rogue, the HACCA would need to resist such distributional drift. It would also have to resist external attempts by its developer to reassert control. The original operator may try to reprogram the HACCA via adversarial attacks or other techniques, or even shut it down entirely. Past malware has sometimes included "kill switches" that the developer or external actors used to disable the malware, as in the case of WannaCry or the Mozi botnet.[210] A rogue HACCA would have to lack such a mechanism, modify itself to disable it, or have a developer who refused to hit the kill switch.

We think HACCA loss-of-control scenarios are worth monitoring and preparing for, even though they may seem more distant. First, the transition to self-sustaining systems may not be clearly telegraphed, and the window between "self-sustaining" and "difficult to shut down" could be narrow. Without improved testing and verification (e.g., with realistic simulated environments), there may be little warning or opportunity to develop countermeasures proactively. Second, developing robust HACCA control mechanisms is a non-trivial technical challenge: controls must resist both circumvention by the system itself and exploitation by adversaries. Given this, policymakers may want to invest early in control research, rather than risk having to do so under time pressure if autonomous capabilities advance rapidly.

---

[208] Here we broadly adopt the terminology suggested by Stix et al., "[The Loss of Control Playbook: Degrees, Dynamics, and Preparedness](#)": a "deviation" can cause harm but does not reach the threshold of a national-level event, while "loss of control" entails significant damage and is substantially more difficult (or even impossible) to contain.

[209] It is not clear whether such drift would make the HACCA more or less threatening, though. It could turn more threatening if the factors underlying the initial loss of control continued to drive these goals in a harmful direction; conversely, "regression to the mean" could make it less harmful over time.

[210] For details of the Mozi botnet killswitch, see: Antoniuk, "[A 'kill switch' deliberately shut down notorious Mozi botnet, researchers say](#)." As another example, Stuxnet included "a final safeguard of a self-destruct mechanism, which caused the worm to basically erase itself in 2012" (Singer, "[Stuxnet and Its Hidden Lessons on the Ethics of Cyberweapons](#)").

## Why Full-Fledged Rogue Deployments May Be Exceedingly Hard to Detect and Shut Down Permanently

Once a rogue HACCA establishes globally distributed infrastructure, like stealth data centers and networks of compromised machines, then permanent shutdown becomes extraordinarily difficult. Historical precedent shows how difficult it is to eliminate distributed digital threats: despite aggressive enforcement, large botnets stay active for years, child sexual abuse material persists across Google Drive and dark web networks,[211] and underground marketplaces like Silk Road re-emerge within months after being shut down.[212] The challenge is both political and technical:

- **Shutting down a distributed HACCA would require unprecedented coordination over global compute access** between cloud providers, law enforcement, and governments with competing interests. Shutting down a distributed HACCA requires identifying and neutralizing all instances across multiple jurisdictions, and success in taking down individual clusters could leave other instances operational, allowing a deployment to reconstitute itself.
- **The internet's architecture also favors distributed operations.** The same end-to-end encryption that is widely used to protect privacy and legitimate communications would also shield HACCA command-and-control traffic from inspection. Near-universal reliance on "trust-by-default" protocols like the Border Gateway Protocol could enable malicious actors to hijack routes or reroute traffic without authentication.[213] And ultimately, the internet lacks centralized chokepoints that defenders can use for global disruption efforts, even if options such as network segmentation work at a local level.[214]

Not all rogue HACCA scenarios are of equal concern; policymakers should focus on those where sociopolitical or technical developments could tip the balance of power between law enforcement and rogue HACCAs. A rogue HACCA might seek to acquire more resources and self-propagate to minimize the risk of shutdown, as discussed in Section 3. But even with HACCAs on the loose and seeking to multiply, concerted effort by authorities could keep the population in check, much as global law enforcement currently does with cybercriminals. In this scenario, HACCAs end up as the "roaches of the internet": an annoying presence with a propensity to overrun unprotected compute clusters, but kept to a stable population through continuous pressure that prevents them from expanding their digital territory.[215]

[211] Lykousas and Patsakis, "Just in Plain Sight: Unveiling CSAM Distribution Campaigns on the Clear Web."
[212] EMCDDA, "Drugs and the darknet: perspectives for enforcement, research and policy."
[213] Goldberg, "Why is it taking so long to secure internet routing?"
[214] Vermeer, "Evaluating Select Global Technical Options for Countering a Rogue AI."

However, several conditions could disrupt this equilibrium, with one possible such condition being new paradigms that enable AI systems to learn more effectively. If a rogue HACCA were developed using such a paradigm, or adopted it after deployment, this could lead to runaway capability improvement. Training new models from scratch currently requires resources on the scale of frontier AI companies, which HACCAs are unlikely to be able to muster. But innovations in continual learning could allow HACCAs to learn more efficiently, perhaps without hitting similar capability ceilings. This could allow them to acquire more resources and further improve their capabilities in a positive feedback cycle, potentially (though not definitely[216]) outpacing defenders' ability to respond (see Section 3, "Tactic 5: Adaptation and capability improvement").

Another disruptive possibility is rogue HACCAs shielding themselves against shutdown by exploiting weak nation-states or collaborating with adversarial ones. HACCAs are vulnerable to disruption because their compute clusters have a physical footprint that can be targeted, but they could reduce this vulnerability by partnering with nation-states or corrupt officials willing to protect them. Such HACCAs would be akin to state-affiliated contractors: interested parties could provide them with physical territory and legal cover to operate compute clusters, in exchange for financial or technological benefits.

HACCAs might have the easiest time operating in weak nation-states where governance is poor and corruption endemic. In such nation-states, HACCAs could offer revenue in exchange for a secure base of operations, analogous to how scam centers in Myanmar provide millions, if not billions, of dollars for local militias aligned with the junta in return for legal impunity.[217] Fragmented nation-state capacity could allow HACCAs to pursue informal partnerships with a range of actors beyond the central government, including corrupt officials, militias, and criminal groups. These benefits may incentivize rogue HACCAs to further erode or capture the authority of local governments, establishing them as "digital warlords" that have an entrenched power base.

---

[215] Even a "roaches of the internet" scenario may not be plausible if it is very difficult for HACCAs to form a self-sustaining population, though: e.g., if there are few available computing resources to be consumed, or there are other agents that could outcompete rogue HACCA populations.

[216] Defenders could plausibly keep up even if continual learning enables rapid iterative improvement. Frontier AI companies may be able to exploit these new paradigms more effectively than rogue HACCAs, given their greater resources and fewer operational constraints (e.g., no need for stealth), allowing them to develop more capable defensive AI systems. However, the need to prioritize safety, security, and reliability could also slow the deployment of defensive AI, in which case rogue HACCAs might retain an advantage.

[217] Ruser, "Scamland Myanmar: how conflict and crime syndicates built a global fraud industry."

Partnerships with stronger, adversarial nation-states seem less likely but could encompass a broader range of activities. Adversarial nation-states might be interested in the intelligence-gathering and technological capabilities of HACCAs. In return, they could provide HACCAs with not just legal impunity, but also technical assets, procurement networks, and even military deterrence against shutdown. However, many nation-states would see such partnerships as high-risk, given that HACCAs could behave unreliably or turn on them. Nation-states that harbor rogue HACCAs may also face threats from other countries seeking to shut the rogue HACCA down. We therefore think this pathway is generally less likely, except with pariah states like North Korea, which uses cybercrime to support its nuclear weapons program and may see financial benefits in HACCA collaboration as well.[218]

To be clear, such scenarios are highly speculative. To succeed in the diplomatic arena, a rogue HACCA would need to meet at least the following conditions:

- **The rogue HACCA would need very strong social engineering, negotiating, and diplomatic skills**, significantly exceeding our base threshold of "only" being able to conduct OC3-level cyber operations. To exploit gaps in global governance, it would need to model and manipulate the conflicting interests of many human actors in ways that nation-state actors already struggle (and often fail) to do.
- **The HACCA may need to impersonate human actors or find human confederates to overcome a trust deficit.** A goal-directed, autonomous AI system would likely be met with extreme wariness, especially by nation-state actors. The HACCA could overcome this by, for example, pretending to be a human criminal group that wanted a remote compute outpost, but this, of course, would demand additional resources and social engineering capabilities.
- **The HACCA would need extreme resilience to attempts to subvert or duplicate it.** Major powers like China, Russia, and the U.S. would likely see a diplomatically capable rogue HACCA as a threat and might prefer to assert control via technical means, rather than trying to strike a high-risk deal with it or letting it continue to operate unimpeded.

If a rogue HACCA established a significant population, however, its activity in cyberspace might be unmoored from the strategic calculus that restrains nation-state cyber operations. Nation-states rarely aim to do the maximum damage that they can—for example, the NSA has not tried to crash

---

[218] Among nation-states, North Korea is unique for its use of cyberattacks for direct financial gain, which it used to fund its weapons programs after heavy UN sanctions (Kim, "From Intelligence Gathering to Financial Gain: Countering DPRK Cyber Operations"). For example, it stole $14 million from ATMs worldwide (Lee et al., "Lazarus Heist: The intercontinental ATM theft that netted $14m in two hours") and attempted to steal $1 billion from the Bank of Bangladesh, though ultimately only stole $81 million (White and Lee, "The Lazarus heist: How North Korea almost pulled off a billion-dollar hack").

China's economy, even though it might have options to do serious harm. Fear of retaliation deters such actions, and maximizing damage rarely supports long-term strategic goals: collateral damage can harm one's own assets and lead to other nation-states retaliating or escalating. But rogue HACCAs with miscalibrated judgment may wind up pursuing maximalist goals (e.g., "acquire resources at any cost") without subtlety and strategic patience. Such HACCAs may ultimately fail due to their inability to prioritize, but could cause massive collateral damage in the process.

More threatening would be rogue HACCAs that pursue explicitly destructive goals over long time horizons, not out of incompetence but instead idiosyncrasy. Most HACCAs would be initially tasked with military and intelligence objectives, but if these long-term goals were poorly specified and the rogue HACCA succeeded in establishing a large population in the wild, that population might pursue not just instrumental goals (e.g., "adapt and survive") but explicitly destabilizing ones (e.g., "destroy adversary X," "trigger a war between Y and Z with false flag operations") that were holdovers from its initial programming. The Islamic State provides a rough analogy: a state-like entity pursuing radical goals with patience and resources, unconstrained by preserving international order or minimizing collateral damage. One might ask: what might a similarly state-like entity with elite cyber capabilities (e.g., at the level of the NSA's Tailored Access Operations unit) be able to accomplish over two years?[219]

While the prospect of runaway capability explosions or state-like HACCAs may seem unlikely, we think they are tail risks worth considering—especially for nation-states that want HACCAs as an exquisite weapon. HACCA developers should take robust measures to prevent such extreme loss-of-control incidents by thoroughly evaluating HACCAs pre-deployment, and not unnecessarily expanding their capabilities in domains not relevant to their intended operations (e.g., fine-tuning or otherwise training them to have strong social manipulation abilities).

More broadly, however, nation-states should invest heavily in techniques for HACCA control. They must balance two tensions in this regard: first, they need mechanisms to disable or reassert control over a HACCA if necessary, but second, such mechanisms must themselves be robust to adversarial attack, lest adversaries use them to sabotage or seize control. Designing HACCAs with secure control methods that cannot be suborned by malicious parties should be a prerequisite for their deployment in the wild.

---

[219] The literature on "rogue deployments" focuses on AI systems in lab-contained settings subverting control measures, but the latitude that rogue HACCA deployments might have suggests they are worth monitoring. Internally deployed AI agents will be closely monitored, and to cause catastrophic effects, need extraordinary abilities to devise short sequences of extremely impactful actions (Shlegeris, "Win/continue/lose scenarios and execute/replace/audit protocols"); however, rogue deployments in the wild are much less constrained by monitoring, and can explore longer sequences of actions to have a catastrophic impact.

# 5 | Defense-in-Depth Against HACCA Operations

Given HACCAs' unique attributes, defenders will likely face significant challenges in anticipating and responding to the novel security risks that they pose. No single intervention will reliably eliminate risks from a sophisticated HACCA deployment. Instead, effective defense requires a layered approach that creates multiple opportunities to prevent, identify, and counter hostile autonomous operations.

Defenders will need to rely on a range of technical countermeasures, institutional safeguards, regulatory frameworks, and governance mechanisms. To organize these measures, we structure them around four complementary strategic objectives: **Delay**, **Defend**, **Detect**, and **Disrupt**.

## Table 14: A Defense-in-Depth Framework Against HACCAs

| Layer | Description | Example Countermeasures |
|-------|-------------|-------------------------|
| **Delay** | Slow the proliferation of HACCA capabilities and systems to malicious actors | Model weight security, differential access |
| **Defend** | Reduce the attack surface available to HACCAs and strengthen potential targets | Automated vulnerability discovery and patching, security-by-design for AI-generated code, automated red teaming and pentesting |
| **Detect** | Gain visibility into HACCA operations and identify hostile activity | Detection signatures for HACCAs, threat intelligence sharing for autonomous operations, agent honeypots |
| **Disrupt** | Degrade or neutralize active HACCA operations | Compute and finance access controls, network disruption, adversarial ML-based countermeasures |

Together, these layers ensure defenders are not reliant on any single intervention succeeding. The goal is not to render HACCA operations impossible, but to create sufficient friction and visibility that defenders can detect and respond before severe harm occurs. The layers are also mutually supportive: detection enables disruption, and delaying buys time for hardening.

Many of the measures discussed in this section remain largely theoretical or untested, especially against systems with the capabilities of HACCAs. The specific examples provided are meant to be illustrative rather than prescriptive, and further research is needed to evaluate which approaches are most promising, how they interact, and how they might be implemented by defenders with varying resources and threat profiles. Many of the technical measures outlined will require additional R&D before they are ready to deploy.

# Delay

We expect that the capability to develop HACCAs will initially be concentrated among a small number of frontier AI companies and well-resourced state actors (see "Proliferation dynamics: how quickly does this get worse"). As these capabilities become widely accessible, malicious actors can leverage these systems to infiltrate networks and conduct potentially devastating cyberattacks. This period and the time before HACCAs become technically feasible create a window of opportunity where defenses in key sectors can be hardened, governance frameworks can be established, and novel defensive measures can be developed.

The goal of delay measures is to extend this window—not to prevent proliferation entirely. A "deny" approach that attempts to permanently restrict HACCA capabilities to a small set of actors (e.g., by severely constraining AI research dissemination or compute access) would likely be infeasibly restrictive given enforcement challenges and the cost to beneficial applications. Delaying wider proliferation of HACCAs to malicious actors means that defenders have more time to understand the potential threat, improve their defensive posture, and develop effective countermeasures.

## Model Weight Security

The most direct path to obtain HACCA-level capabilities is for a less-resourced actor to obtain HACCA-level model weights and other system components (e.g., associated agentic scaffolding, multi-agent orchestration frameworks). As mentioned in "Tactic 5: Adaptation and Capability Improvement", training a state-of-the-art model from scratch requires substantial compute, data, and technical expertise—which is likely to be cost-prohibitive for the vast majority of actors. Preventing the theft or leakage of model weights and other difficult to acquire system components would force actors to invest substantially more time and resources in independent development, or wait until capabilities diffuse through other channels, such as commercial access through APIs or open-sourcing.[220]

---

[220] Other access pathways, such as model access via model provider APIs, could, in theory, be less susceptible to unauthorized use since model providers would have more ability to monitor usage, implement more robust safeguards (which in open-weight models can be more easily bypassed), and control access through means like throttling or blacklisting.

Experts suggest that comprehensive model weight security requires a defense-in-depth approach across multiple attack vectors. Nevo et al. (2024) identify 38 distinct attack vectors and suggest security practices such as centralizing all copies of weights to access-controlled and monitored systems, implementing insider threat programs, and incorporating confidential computing to protect weights during use.[221] The report proposes tiered "security levels" (SL1–SL5), each calibrated to defend against progressively more sophisticated attackers—from opportunistic criminals at lower levels to elite nation-state operations at the highest. Facilities handling the most dangerous models would meet correspondingly stringent requirements for physical security, personnel vetting, network isolation, and monitoring.

Model weights in particular might be relatively easier to secure than other system components. Frontier model weights are extremely large files, often measured in terabytes. Upload limits—caps on total data leaving a data center—are relatively simple to enforce and can extend the time required for exfiltration from days to months or longer.[222] Other components, such as agent scaffolding, orchestration frameworks, fine-tuning datasets, and algorithmic insights, are likely to be substantially harder to protect since they are far smaller in file size by several orders of magnitude.

## Differential Access

Differential access is an approach to governing the availability of AI-enabled cyber capabilities that aims to limit access to riskier users, while supporting key defenders.[223] Differential access schemes implemented by model providers would grant access to varying levels of cyber capabilities based on user identity, intended use case, and risk assessment, with the goal of ensuring that defensive actors gain access to these capabilities before, or at least alongside, potential attackers.

Like other structured access frameworks, differential access would need to be implemented through cloud-based APIs, as this gives developers ongoing visibility into how systems are used and the ability to revoke access or adjust restrictions in response to misuse, while still enabling beneficial defensive applications.[224] Access controls can leverage identity, capability controls, and monitoring mechanisms to enable fine-grained model access to only authorized actors. Ee et al.

---

[221] Nevo et al., "Securing AI Model Weights."

[222] For use of upload limits to prevent exfiltration, see: Nevo et al., "Securing AI Model Weights"; Greenblatt, "Preventing model exfiltration with upload limits." Given the rate of progress of model capabilities at the frontier, delaying the time it takes to exfiltrate model weights by months or years would mean that by the time an attacker could complete the transfer, the model would likely have been superseded by newer versions, significantly reducing the expected value of the exfiltration operation.

[223] Ee et al., "Asymmetry by Design: Boosting Cyber Defenders with Differential Access to AI."

[224] Shevlane, "Structured access: an emerging paradigm for safe AI deployment."

(2025) propose varying the approach to managing access to AI-cyber capabilities based on risk levels:[225]

- **Promote Access:** For lower risk capabilities, AI developers and government actors encourage broad adoption of AI-cyber capabilities through open access and active promotion.
- **Manage Access:** For medium-risk capabilities, AI developers and government actors implement vetting procedures, usage monitoring, and access restrictions to control distribution of AI-cyber capabilities. Access to certain users can be prioritized, such as security researchers.
- **Deny by Default:** For the highest risk capabilities, access is restricted to a small set of vetted defenders. For example, owners/operators of advanced AI-cyber systems could offer HACCA-level red teaming services to lifeline infrastructure providers.

To succeed at empowering defenders, differential access schemes must clearly tackle specific risks: for example, given a particular capability, there must be a compelling case that attackers can do harm and defenders will benefit from access. Otherwise, such schemes could unnecessarily restrict innovation and run counter to the cybersecurity community's norm of openness and broad access. Differential access may also not be apt for all problems, even if AI capabilities introduce new risks: some such risks may benefit most from tried-and-tested cybersecurity solutions.

## Defend

The goal of measures in this layer is to reduce the attack surface available to HACCAs. While delay measures buy time and detection enables response, hardening targets directly raises the cost and difficulty of successful offensive operations. Broad improvements in defensive posture do not merely protect individual organizations—they raise the capability threshold required for successful HACCA operations across the board, potentially placing certain targets out of reach for all but the most advanced threat actors. The example measures discussed below focus on AI-enabled approaches, and particularly on techniques that support hardening systems at scale.[226] This is because we expect that as HACCAs change the speed and economics of offensive operations, purely manual defenses would likely become untenable. The same capabilities that empower

---

[225] Ee et al., "Asymmetry by Design: Boosting Cyber Defenders with Differential Access to AI."

[226] AI-enabled approaches may be valuable in areas other than hardening. For example, automated moving target defense (AMTD) dynamically reconfigures system attributes (e.g., network configurations) to increase uncertainty for adversaries (Loman, "Pioneering Automated Moving Target Defense"). Such techniques could complement the static hardening measures we discuss below, but more research is needed to determine if they would be an effective counter against HACCAs, given that one of the strengths of HACCAs is meant to be their adaptability.

attackers—automated vulnerability discovery, rapid adaptation, scalable operations—will need to be harnessed by defenders to keep pace.

However, defensive efforts will inevitably take place in a context of limited, uneven resources. Government and industry will need to thoughtfully prioritize defensive efforts to protect sectors that lag behind, rather than allow defensive improvements to flow only to those which happen to have the resources, technical capacity, and organizational culture to adopt them first. As discussed previously (see "The offense-defense imbalance and why it matters"), many critical actors—regional utilities, healthcare providers, open-source software maintainers—are positioned poorly to implement current best practices, let alone next-generation AI-enabled defenses.

## Security-by-Design for AI-Generated Code

In principle, if all code were developed "secure-by-design," attackers would have far fewer opportunities to exploit, and defenders would not need to reactively patch code. But new code is almost never written and tested methodically enough to eliminate all (or even most) vulnerabilities: diminishing returns for discovering complex bugs mean that perfect security requires testers to undertake herculean efforts.[227] The rapid adoption of LLM-enabled code generation worsens this problem: LLMs may produce even *more* vulnerable code than human developers,[228] and under-resourced organizations are poorly positioned to verify this code.[229]

Given the rapid growth of AI-assisted coding,[230] developers and policymakers should seek to improve the security of AI-generated code. Frontier AI developers and developers of AI-assisted coding tools (e.g., Cursor) would play a large part in this by shifting the model's default behavior toward secure patterns.[231] Users would also have to ensure they implement appropriate validation and feedback loops, including automated vulnerability discovery (see next section), even as they continue to encourage traditional secure coding practices.

---

[227] For the challenge of achieving "perfect security," see: Böhme and Falk, "Fuzzing: on the exponential cost of vulnerability discovery." A similar challenge is indicated by the high effort required to develop high-reliability mission-critical software: e.g., in the 1990s, when NASA and Lockheed Martin updated the Space Shuttle's software to navigate with GPS, the actual change took 6,366 lines of code, but the requirement specification was 2,500 pages long (Fishman, "They Write the Right Stuff").

[228] Chong et al., "Artificial-Intelligence Generated Code Considered Harmful: A Road Map for Secure and High-Quality Code Generation"; Bar, "AI for Code Synthesis: Can LLMs Generate Secure Code?"

[229] Ji et al., "Cybersecurity Risks of AI-Generated Code."

[230] For instance, on Alphabet's Q1 2025 earnings call, CEO Sundar Pichai noted that "well over 30%" of new code at Google was AI-generated (Alphabet Investor Relations, "2025 Q1 Earnings Call").

[231] Datta et al., "Secure Code Generation at Scale with Reflexion"; He et al., "Instruction Tuning for Secure Code Generation."

If AI-assisted coding can be made reliable enough, this could open new doors for scaling up a security-by-design approach to software. Reliable coding agents could be used to transition existing codebases to memory-safe languages, eliminating entire classes of vulnerabilities.[232] Even more ambitiously, AI could automate the large-scale formal verification of software,[233] an expensive and specialized set of techniques that can mathematically prove the security of software. However, such ambitious approaches ultimately rest on the trustworthiness of modern AI coding agents: if they do not support a security-by-design approach on smaller scales, then attempting to scale them up while removing human oversight is an extremely risky proposition.

## Automated Vulnerability Discovery and Patching

Patching vulnerabilities has historically been reactive, a paradigm that may become untenable as attackers use HACCAs to accelerate vulnerability exploitation. One emerging use case for AI is identifying exploitable flaws in software and producing fixes with minimal human intervention. Since Google's Big Sleep agent produced the first publicly reported AI-assisted vulnerability discovery in 2024,[234] capabilities have advanced rapidly. DARPA's 2025 AI Cyber Challenge saw autonomous systems discover 86% of synthetic vulnerabilities across 54 million lines of code and patch 68%—averaging 45 minutes and $152 per task versus the days and thousands of dollars typical of manual approaches.[235] Google's CodeMender, introduced in October 2025, showcased an AI agent that could find vulnerabilities and generate validated patches, contributing 72 fixes to open-source projects in its first six months.[236] In theory, these tools could dramatically compress vulnerability lifecycles and make it feasible to secure codebases that were previously too expensive to audit.

However, speeding up vulnerability discovery is only helpful if downstream users can actually apply critical patches in time. Current automated vulnerability discovery focuses on the earlier part of the patch lifecycle (helping software providers find vulnerabilities and create patches), but the binding constraint for many trailing-edge organizations, including critical infrastructure defenders, is patch deployment.[237] Even after patches become available, it can take weeks or months to test for compatibility and coordinate downtime: in a major industry report, it took 55 days to remediate 50% of critical, actively exploited vulnerabilities, while attackers exploited disclosed vulnerabilities in an average of 5 days.[238] For critical infrastructure providers, this lag can be worse: hospitals, for example, cannot simply take down medical systems to "patch faster."[239] If vulnerability discovery accelerates, but patch deployment does not, then attackers may simply have a wider window to exploit vulnerabilities that defenders cannot remediate.

---

[232] Bradley and Sastry, "The Great Refactor."
[233] Lin et al., "A Toolchain for AI-Assisted Code Specification, Synthesis and Verification."
[234] In June 2024, Google announced that its Big Sleep agent found a vulnerability in SQLite; this was likely the first public example of an AI agent finding a previously unknown exploitable memory-safety issue in widely used real-world software (Glazunov and Brand, "Project Naptime: Evaluating Offensive Security Capabilities of Large Language Models.").
[235] DARPA, "AI Cyber Challenge marks pivotal inflection point for cyber defense."
[236] Popa and Flynn, "Introducing CodeMender: an AI agent for code security."
[237] Lukošiūtė, "Design for the defenders you care about or risk being useless."

To realize defensive benefits for these trailing-edge organizations, policymakers will also need to encourage the development of AI tools designed for their operational constraints. These could include asset inventory tools, automated patch compatibility testing, and tools for triaging and prioritizing vulnerabilities and patches.[240] AI-driven vulnerability discovery and patching have significant potential to transform security for well-resourced organizations, and now that its value is demonstrated, market incentives will likely drive continued progress there. On the other hand, trailing-edge organizations lack the budget to attract commercial interest, so the market incentives to develop the tools they need are weak; to ensure that automated vulnerability discovery also benefits them, government and philanthropic support will be essential.

## Automated Red Teaming and Pentesting

Red teaming and penetration testing ("pentesting") can help defenders to analyze weaknesses at the level of deployed systems and networks, not just individual pieces of software. Such offensive security testing takes place over days or weeks, and can expose issues that other techniques focused on code security (e.g., vulnerability testing and secure code generation) cannot address. This is important because—especially for under-resourced defenders—the weaknesses that attackers exploit are often process- or people-based weaknesses, such as social engineering attacks and misconfigurations (e.g., connecting a network to the internet when it should be air-gapped).

Defensive AI systems could help scale up red teaming significantly and make it affordable to organizations for whom the cost—which can reach tens or even hundreds of thousands of dollars—would normally be prohibitive. For example, in 2024, the company XBOW benchmarked its autonomous AI-powered penetration testing system against human pentesters; across about 100 security challenges, the AI system matched the performance of a principal pentester with more than 20 years of experience but completed the task in 28 minutes compared to the human's 40 hours.[241]

---

[238] Verizon Business, "2024 Data Breach Investigations Report."

[239] In critical infrastructure, patching is complicated by the fact that operational continuity and safety considerations often take precedence over the downtime required for updates (Lee and Conway, "The Five ICS Cybersecurity Critical Controls").

[240] Lukošiūtė, "Design for the defenders you care about or risk being useless."

[241] Moor, "XBOW now matches the capabilities of a top human pentester."

However, such systems carry several risks. Proliferation is one: many criminals already abuse pirated versions of Cobalt Strike, a non-AI tool for automating penetration testing. Cobalt Strike's creator has had to carefully manage access and work with law enforcement to remove pirated versions of the tool;[242] developers of AI-enabled automated red teaming tools should implement KYC controls and other security measures to prevent similar things from happening. Moreover, an AI-enabled red teaming system might look similar to a HACCA and have extensive access to a client company's networks. Deployers should ensure that the system is reliable and aligned, so as to avoid breaches of privacy and security, or loss of control. If such risks are insurmountable, then AI red teaming systems should not be deployed.

## Detect

Detection enables defenders to identify HACCA activity early enough to respond before operations achieve their objectives, cause harm, or become entrenched in ways that make shutdown and disruption far more difficult. Early visibility into HACCA activities also helps defenders, developers, and policymakers to build a clearer picture of the evolving threat landscape—supporting prioritization of defensive investments and calibrating policy responses to evidence of emerging threats.

The degree of visibility available to defenders depends significantly on how attackers access their underlying models. When attackers rely on frontier models via APIs, developers retain substantial monitoring capabilities.[243] However, if HACCAs are built on open-weight models or self-hosted systems, this visibility disappears. Detection then depends on ecosystem-level indicators—anomalous compute provisioning, unusual transaction patterns, distinctive operational behaviors—observed by cloud providers, financial intermediaries, and platform operators.

As noted previously, HACCAs would likely be able to evade detection in ways that traditional tools cannot. These systems could adapt their behavior to avoid producing the consistent signatures that conventional detection relies on, actively compromise defender communications, hide within legitimate traffic, or concentrate operations in under-monitored networks where defensive coverage is sparse. This suggests that reliably identifying HACCA activity will require developing novel detection methods or indicators for autonomous cyber operations.

---

[242] Martin, "Cobalt Strike: International law enforcement operation tackles illegal uses of 'Swiss army knife' pentesting tool."
[243] NTIA, "Dual-Use Foundation Models with Widely Available Model Weights Report."

## Detection Signatures for HACCAs

Detecting HACCA activity will require identifying signals that distinguish these operations from conventional human-operated attacks as well as legitimate AI workloads. As mentioned previously, signature-based detection—which relies on matching known indicators of compromise (IOCs) like specific file hashes, IP addresses, or command sequences—could be ineffective against HACCA that dynamically vary their signatures. Instead, defenders will need detection approaches oriented around the distinctive characteristics of autonomous operations.

Possible indicators could include unusual compute provisioning patterns or distinctive reconnaissance tactics. Some researchers suggest that the most revealing signals may lie not in isolated events but in sequences of related actions, such as how tools are chained together, how agents adapt when initial approaches fail, and verification behaviors that confirm whether access succeeded.[244]

However, determining HACCA signatures is an open research problem, unlike conventional malware families where defenders can study captured samples or human threat actors whose tradecraft can be characterized through years of incident response. Speculatively, progress on this problem could be made by:

- Deploying HACCAs and proto-HACCAs in controlled offensive security scenarios to study behavioral traces.
- Working backward from HACCA operational requirements (e.g., tool chaining, compute acquisition) to hypothesize what observable patterns these requirements would produce.
- Designing purpose-built honeypot environments (discussed below under "Agent honeypots") to capture behavioral data that feeds back into signature development.

Ultimately, it may not be feasible to identify reliable HACCA signatures, but further research could verify this, and if there were unique patterns, it could be highly valuable. One critical uncertainty is whether distinctive HACCA signatures will persist as systems become more sophisticated and capable, and whether different HACCA "families" will share generalizable signatures. It may be safer to assume that signature development, if tractable, is treated as an ongoing research effort.

## Threat Intelligence Sharing for Autonomous Operations

Traditional threat intelligence typically assumes that at least one well-positioned observer—a victim, security vendor, or intelligence agency—can develop a relatively coherent picture of a campaign before sharing indicators outward. HACCA operations may fragment evidence more severely, with activity distributed across providers, accounts, and jurisdictions in ways that leave no single defender with sufficient visibility to characterize the threat independently.

---

[244] Irregular, "From AI-assisted to AI-orchestrated: What agent-led cyber attacks mean for security."

Defending against HACCA campaigns would require analyzing activity across systems and more structured cooperation between AI developers, high-risk organizations, and government.[245] Existing threat intelligence sharing arrangements provide a foundation, but defending against HACCAs will require bringing in additional stakeholders, such as cloud compute providers, and establishing "agent activity" as a distinct category of threat behavior. Frontier AI companies could be especially important as they are positioned to have early insight into the behavior of advanced offensive cybersecurity agents under development.

Beyond conventional campaign detection, threat intelligence sharing will also address novel threat models specific to autonomous systems—particularly loss-of-control scenarios. Given loss of control has no strong precedent, this might require collaboration between traditional security researchers and AI safety researchers focused on loss-of-control threat modeling.

## Agent Honeypots

Traditional honeypots—deliberately vulnerable resources designed to attract attackers and study their methods—could be adapted to detect and characterize HACCA operations. Agent honeypots would serve the same intelligence-gathering functions as conventional honeypots, but with an additional objective: identifying whether the attacker is an autonomous agent versus a human attacker or conventional piece of malware.[246] Preliminary evidence supports this approach: autonomous LLM-based attackers spent ~90% of commands on decoy resources,[247] and file-based deceptions consistently disrupted AI planning agents by inducing incorrect belief states.[248]

Agent honeypots could potentially accomplish a range of objectives. They could establish an early baseline to quantify the proliferation of HACCAs and act as an early warning system for policymakers, cybersecurity experts, and insurance companies. Within networks, honeypots could detect when a HACCA has achieved initial access, triggering defensive responses before significant damage occurs. Beyond detection, sustained honeypot interactions could help defenders learn about HACCA characteristics—including their vulnerability to different types of inputs such as adversarial prompts—and the behavioral data captured could feed back into signature development, helping characterize operational patterns that might otherwise remain opaque. They could be used for disruption, increasing costs for the attacker by forcing malicious actors to expend resources on verifying the authenticity of their targets, or even degrade and hijack HACCA instances (if paired with an adversarial machine learning attack or attack on system components).

---

[245] Irregular, "From AI-assisted to AI-orchestrated: What agent-led cyber attacks mean for security."
[246] Reworr and Volkov, "LLM Agent Honeypot: Monitoring AI Hacking Agents in the Wild."
[247] Singer et al., "Perry: A High-level Framework for Accelerating Cyber Deception Experimentation."
[248] Kouremetis et al., "Mirage: cyber deception against autonomous cyber attacks in emulation and simulation."

Agent honeypots could potentially be designed in different ways depending on the objective(s) of the defender. Some design elements include:

- **Detection mechanism**: There are several possible approaches to distinguish autonomous systems from human attackers. These include using adversarial inputs,[249] such as prompt injections, that non-AI based systems would ignore or not perceive, or analyzing behavior patterns and timing more closely associated with autonomous operations.
- **Placement:** Honeypots could be placed within organizational networks to detect whether HACCAs have achieved initial access or they can be placed on the open internet, such as by mimicking services like compute marketplaces.
- **Interaction depth:** More information can be potentially extracted from an agent honeypot if a HACCA is required to have a longer interaction with the resource. For example, a decoy enterprise network with multiple devices could collect more detailed information about HACCA tactics, tools, and even vulnerabilities. Deeper engagement also enables reactive or stateful deception, which empirical research suggests is necessary to actually stop autonomous attackers, rather than merely delay them.[250]
- **Canary mechanisms:** Honeypots could include trackable artifacts designed to trigger alerts if an agent or its operator exfiltrates and uses the honeypot resource, e.g., fake credentials.

## Disrupt

Disruption measures aim to impose costs, actively degrade, or neutralize HACCA operations. Measures in this layer could potentially be decisive—eliminating the threat from a HACCA deployment completely—but in many cases the most feasible outcome may be to impose significant costs that degrade operational effectiveness or slow the pace of hostile activity. Decisive disruption is likely more achievable before a HACCA establishes robust, distributed infrastructure;

---

[249] There are some uncertainties around the robustness of this kind of mechanism. While this has been demonstrated to work on simple agents (Reworr and Volkov, "LLM Agent Honeypot: Monitoring AI Hacking Agents in the Wild"), this may not generalize well to encounters with more advanced HACCA-level systems in the wild (see "Adversarial ML-based countermeasures" later in this section).

[250] Singer et al., "Perry: A High-level Framework for Accelerating Cyber Deception Experimentation."

once a deployment has spread across multiple jurisdictions, cloud providers, and compromised systems, complete elimination becomes substantially harder (see the "loss of control" section).

Effective disruption is likely to face some distinct challenges. Many of the more "aggressive" disruption measures risk significant collateral effects. For example, shutting off data centers or electricity could affect legitimate users and critical services that depend on the same infrastructure. More targeted, localized interventions can reduce collateral costs, such as terminating specific accounts or shutting down particular compute clusters; however, whether such interventions are feasible depends on whether defenders can detect malicious activity at a granular level.

## Compute, Finance, and Model Access Controls

HACCAs' dependence on substantial compute—and, to a lesser extent, financial resources—for continuous operation provides an opportunity for defenders to disrupt their activities.[251] Model access via API may also occasionally be relevant, if attackers rely on model provider APIs rather than self-hosting weights.[252]

Defenders can deny HACCAs access to compute by designing and implementing better know-your-customer (KYC) measures that work on advanced agents. Existing KYC measures, even from major cloud providers, involve only basic verification for billing purposes. The primary identity check is payment verification, such as credit card or bank account information, with enterprise accounts requiring additional checks on company registration details, such as tax ID or VAT numbers.[253]

---

[251] Access to training compute in particular is essential to training-time capability improvement strategies, and there may be fewer sources of higher-quality training compute than inference compute. Training clusters require specialized infrastructure that is substantially more difficult to acquire than inference setups: high-bandwidth interconnects (InfiniBand or NVLink) for data parallelism across thousands of GPUs, sophisticated orchestration software, and massive storage systems. This concentrates training-capable infrastructure among a small number of major cloud providers and well-resourced organizations that can implement stricter controls and monitoring, making training compute a particularly promising intervention point (Egan, "Global Compute and National Security").

[252] Sophisticated HACCA operators would likely prioritize self-hosting to avoid single points of failure, but API-dependent deployments might remain vulnerable to provider-side interventions in several scenarios: (1) *capability gaps in open-weight models* that mean HACCA-relevant capabilities can only be accessed via closed-source frontier models; (2) *resource-constrained deployments* that need to be bootstrapped, relying on frontier AI capabilities initially to acquire compute for self-hosting; and (3) *hybrid architectures* where deployments self-host a base agent while making API calls to more capable models for high-difficulty tasks like novel exploit generation.

[253] Amazon Web Services, "How to create an AWS account"; Microsoft, "Troubleshoot issues when you sign up for a new account in the Azure portal"; Google Cloud, "Manage Google payments users, permissions, and notification settings."

To support governance of agent-to-agent transactions and agentic payments in general, financial service providers could require KYC measures that allow only certified or registered agents to create accounts or to perform financial transactions above a certain threshold. Similar measures could apply to agents attempting to convert cryptocurrency to fiat currency. Legislation could require agent-specific KYC mechanisms to be placed on both traditional and decentralized platforms, such as those that use stablecoin. Decentralized platforms are more likely to evade regulation, but larger platforms that offer more liquidity and resources for HACCAs are more likely to comply.[254]

For managing API access to closed-source models, providers could use enhanced API KYC (requiring identity verification beyond payment methods) and emergency access revocation for identified malicious actors. In the GTG-1002 campaign, Anthropic detected the attack within roughly 10 days and was able to ban the associated accounts, notify affected organizations, and coordinate with authorities.[255] Similar account bans were implemented for other threat actors, including North Korean operatives using Claude for fraudulent employment schemes at Fortune 500 companies.[256]

If a HACCA operation is attributed to specific compute sources, then defenders can block compute access. The range of actions can be phased, for example, lower-cost responses, such as credential suspension or GPU throttling, can be used by providers to respond to early indicators of HACCA activity. Higher-cost responses, such as using physical, on-site emergency power-off mechanisms that target power, network access, or cooling could be triggered with sufficient evidence of major risk.[257] Governments have a range of legal options to attempt to take down HACCAs and/or associated clusters.[258] The U.S. Department of Justice has previously taken a law enforcement operation to remove malware from hundreds of victim systems in the U.S. via a search and seizure warrant under Rule 41 of the Federal Rules of Criminal Procedure.[259] However, law enforcement takedowns of multiple geographically distributed compute clusters are likely to rely on cooperation from other countries.

---

[254] Elad and Kinder, "FATF Guidelines on Virtual Assets Statistics 2026: Unlock Global Crypto Growth Today."
[255] Anthropic, "Disrupting the first reported AI-orchestrated cyber espionage campaign."
[256] Anthropic, "Detecting and countering misuse of AI: August 2025."
[257] Physical shutdown is likely preferred because HACCAs may be able to sabotage digital shutdown measures, given their advanced cyber-capabilities. Hardware-enabled mechanisms could also potentially be leveraged for both detection and disruption. For more analysis of this see Appendix IX.
[258] FBI, "Operation Endgame: Coordinated Worldwide Law Enforcement Action Against Network of Cybercriminals."
[259] Iftimie, "No Server Left Behind: The Justice Department's Novel Law Enforcement Operation to Protect Victims."

## Network Disruption

HACCA operations possess an unavoidable dependency on network availability to coordinate instances, proliferate to new hosts, and execute attacks. While HACCAs may use novel channels to obscure their communications, they cannot function if the underlying pathways are severed, and defenders could alter network architecture to trap, isolate, or disconnect hostile agents. Internet service providers (ISPs) can monitor traffic for patterns characteristic of HACCA communications, e.g., encrypted traffic to unusual domains or rapid channel-switching. Once identified, these providers could isolate compromised networks or "blackhole" C&C traffic.[260] Governments and cybersecurity organizations could establish threat intelligence sharing networks focused on HACCA activity, enabling rapid takedown of associated domains and IP addresses.

Outside of more localized technical measures, some researchers have explored more drastic global options, including temporarily shutting down large portions of the internet.[261] For example, a government could pressure Tier 1 providers to stop advertising routing prefixes, making vast swaths of the internet unreachable. However, internet infrastructure is highly decentralized, and this would require coordination across public and private actors in multiple jurisdictions. These strategies would also carry serious trade-offs, given society's dependence on the internet.

Against HACCAs, network availability interventions have significant limitations. HACCAs could circumvent signature-based detection by dynamically varying communication patterns and exploiting novel channels, as mentioned in the section on C&C tactics. Many HACCA operations would involve smaller agentic implants that may not depend on external C&C channels. And as mentioned, the coordination challenges associated with global internet infrastructure may prove prohibitive, as HACCAs could rapidly pivot to alternative routes or other jurisdictions, outpacing the bureaucratic efforts of defenders.

---

[260] Blackholing refers to routing traffic to a "null route" where it is silently discarded. This is generally implemented at the Border Gateway Protocol (BGP) level (Cloudflare, "What is blackhole routing?").
[261] Vermeer, "Evaluating Select Global Technical Options for Countering a Rogue AI."

IAPS | Institute for AI Policy and Strategy

### What if HACCAs Use the Tor Network or Other Onion-Routing Networks to Hide Their Activity?

ISP-driven tactics are much less effective against C&C and other activity tunneled through onion router networks like Tor.[262] With onion services, traffic never exits to the public internet and stays entirely inside the relay network. From the outside, an ISP or internet exchange can usually see only that a host is speaking to Tor entry (guard) relays and not which hidden site it is reaching, what it is saying, or where the server sits. This means that DNS/IP-based takedowns that would work on networks would not identify or disrupt onion-routed activities.

However, a determined, well-resourced defender could end up purchasing and operating a significant share of relays to eventually de-anonymize hidden-service servers or client IP addresses. When defender-controlled relays are positioned as guard nodes or adjacent to hidden service paths, traffic-confirmation and circuit-position tests can reveal the onion service's real location or correlate Tor use back to a user's origin network.[263] One researcher estimates this approach costs ~$80,000 monthly to achieve a 90% chance of identifying a target onion server's IP address within one month.[264]

As an alternative, HACCAs could build a private proxy network, which, once detected, enables more straightforward responses, such as IP blocking and coordinated takedowns. Also, setting up and maintaining a proxy network comes with substantially higher recurring costs, including servers, bandwidth, and other overhead. With open, onion-routing networks like Tor, it has historically been challenging to take down Tor websites. Despite evidence suggesting that 40,000+ unique onion services have hosted child sexual abuse material and nearly 20,000 notifications submitted to the Tor Project to take down the content, no actions were taken.[265]

## Adversarial ML-Based Countermeasures

Adversarial machine learning involves using carefully crafted inputs to cause AI systems to malfunction, reveal information, or deviate from their intended objectives. In principle, these techniques could allow defenders to actively degrade HACCA capabilities—effectively serving as "anti-agent weapons." Researchers have shown that adversarial inputs fed into an agent's context can hijack behavior and induce attacker-specified actions,[266] and that self-propagating prompt injections could infect AI agents by poisoning the databases and retrieval systems these agents interact with.[267]

---

[262] Tor is a volunteer-run network that routes traffic through multiple encrypted relays, making it extremely difficult to trace user's online activities and physical location (Tor Project, "Browse Privately. Explore Freely").
[263] Overlier and Syverson, "Locating hidden servers."
[264] Levine, "How Secure is Tor?"
[265] Richardson, "CSAM distribution on Tor is not inevitable; the network's creators have the power to act."

Defenders should not treat these measures as a primary line of defense, but instead as a speculative moonshot, best suited for advanced defenders. The most reliable adversarial methods typically benefit from access to underlying model weights, which defenders would not have against unknown HACCAs. Without this access, red teamers have still consistently produced universal jailbreaks, but this usually involves hundreds to thousands of iterative attempts, with feedback from the target system. This kind of sustained, iterative interaction is unlikely to be available when a defender encounters a HACCA in the wild, where the defender may have only fleeting contact before the agent adapts or moves on.[268] Moreover, anti-agent weapons might be more like landmines than tracking missiles: to ensure the payload was ingested, defenders would need to anticipate the unknown HACCA's targeting behaviors well enough to put the payload in its path.[269]

Even if progress is made on these technical challenges, defenders will be caught in an arms race with attackers. Sophisticated HACCAs may not be particularly vulnerable to adversarial attacks: HACCA deployers would already want to implement baseline adversarial defenses such as input paraphrasing[270] or classifier-based filtering,[271] and work by Anthropic and other companies has made frontier LLMs less susceptible to adversarial attacks.[272]

These countermeasures may be worthwhile for defenders only if they have structural advantages over attackers. These could include narrow chokepoints that HACCAs must interact with (e.g., specific databases or APIs) or the ability to develop universal "black-box" attacks that HACCA deployers cannot remediate. Sophisticated defenders like the NSA may be the best bet, as their heavily segregated networks may create natural chokepoints, and they would have the skilled labor to develop and stockpile adversarial ML attacks.

---

[266] Vassilev et al., "Adversarial Machine Learning."
[267] Cohen et al., "Here Comes The AI Worm: Unleashing Zero-click Worms that Target GenAI-Powered Applications."
[268] An exception is if the HACCA is built on an open-weight model—once defenders identify which model, they could test for vulnerabilities against their own copy without needing continued access to the agent itself.
[269] Vermeer, "Evaluating Select Global Technical Options for Countering a Rogue AI" includes a more speculative measure, "Hunter/Killer AI," that can hunt down adversary AI agents and impede or eradicate these systems. This would function more like a tracking missile, and obviate the need for anticipating a HACCA's target in advance, if paired with a set of effective adversarial ML-based countermeasures.
[270] Jain et al., "Baseline Defenses for Adversarial Attacks Against Aligned Language Models."
[271] Han et al., "WildGuard: Open One-Stop Moderation Tools for Safety Risks, Jailbreaks, and Refusals of LLMs."
[272] Anthropic, "Mitigating the risk of prompt injections in browser use"; Anthropic, "Constitutional Classifiers: Defending against universal jailbreaks."

# 6 | Guardrails for HACCA Development and Deployment

The potential strategic and economic advantages of HACCAs indicate these capabilities may eventually become widespread. State and non-state actors will have strong incentives to develop and deploy offensive HACCAs for a range of legitimate and criminal activities, despite potentially catastrophic risks pertaining to misalignment, hijacking, and loss of control (see Section 2 on loss-of-control risks). Previous offensive cyber operations suggest nation-states would be willing to deploy HACCAs even when the damage is likely to spread beyond intended target sets. Similarly, the rapid proliferation of lethal autonomous weapons systems (LAWS) and development of AI-enabled drone swarms signal a readiness to cede control to increasingly autonomous offensive systems. Non-state actors have shown even less restraint than nation-states, with criminal groups exploiting AI for fraud, disinformation, and cyberattacks.

Given these operational realities, we judge a blanket prohibition on HACCAs is unlikely to succeed in the near term and may be self-defeating insofar as it constrains legitimate uses. States will be reluctant to agree to any international agreement or convention that bans HACCAs outright, just as they have failed to ban LAWS despite the decades-long debate at the United Nations Group of Governmental Experts and elsewhere. For many governments, HACCAs could also serve ostensibly legitimate state purposes—supporting long-horizon cyber-espionage, strategic intelligence collection,[273] and counterproliferation campaigns against illicit weapons and cyber programs—further reducing the likelihood that states would accept an outright ban.[274] In short, even responsible cyber actors would have reason to develop, and potentially deploy, HACCA capabilities.

As HACCAs become more widespread, enhanced governance mechanisms will be critical for ensuring responsible development and use. These mechanisms must build on and go beyond existing cybersecurity norms and laws, given the potential for catastrophic harm arising from HACCA use, including risks to critical infrastructure. In this section, we articulate technical, legal, policy, and global governance standards that should be met before HACCAs are fully operational. If actors cannot meet these standards, the prospect of catastrophic damage resulting from loss of control likely renders HACCA deployment unjustifiable in high-risk use cases.

---

[273] Kruus et al., "Governing Automated Strategic Intelligence."

[274] For more on the political and technical hurdles states would face in agreeing to such a treaty, see Maas and Olasunkanmi, "Treaty-Following AI."

# Technical Guardrails

Technical guardrails are essential to ensure HACCA deployments are controllable, secure, and aligned with operator intent. Technical safeguards are the first line of defense against the loss-of-control scenarios that make HACCAs uniquely risky, since they can be embedded directly into AI systems and scaled efficiently across the potentially huge numbers of deployed instances. Because these systems interact with external tools, APIs, and other agents, they possess a far larger attack surface than standard LLMs, creating new vectors for adversarial exploitation and manipulation.[275]

## Why Technical Guardrails Alone Will Not Be Enough

While technical guardrails are vital for responsible development and deployment of HACCAs, they form only part of the safeguards necessary to justify the risks posed by such a system. Traditional AI safety techniques become strained under the kind of adversarial conditions that HACCAs will operate in, and few measures can be guaranteed to hold once an agent is deployed and exposed to sophisticated opponents or unforeseen situations. **Crucially, technical safeguards alone will likely be insufficient to guard against a system that is itself designed to overcome security measures.** Effective oversight requires integrating technical guardrails with complementary legal enforcement mechanisms, policy frameworks, and global governance structures to establish the institutional resilience necessary for managing advanced AI systems.

Designing effective safeguards for HACCAs presents an extraordinarily difficult challenge, arguably as complex as building the agents themselves. By design, HACCAs are able to autonomously find and exploit vulnerabilities, adapt to countermeasures, and make decisions in the field. **The same abilities that make HACCAs effective also make them inherently dangerous and unpredictable if misaligned or compromised.** Operators must therefore use an enhanced defense-in-depth strategy where redundant safety mechanisms operate at various stages of the agent's lifecycle from pre-training through post-deployment, with each layer addressing different potential failure modes. **Since HACCAs typically operate without ongoing contact with their original operators after deployment, both pre- and post-deployment measures are vital.**

These technical safeguards must satisfy conflicting requirements: they must enable last-resort human override, while being perfectly tamper-resistant. When necessary, human operators must be able to interrupt HACCA operations, such as in potential loss-of-control or collateral damage scenarios. But at the same time, HACCAs must resist adversarial interference and manipulation. **Designers must therefore navigate the delicate balance between incorporating**

---

[275] Kraprayoon et al., "AI Agent Governance: A Field Guide"; Reiner, "Anatomy of an LLM RCE."

**mechanisms for legitimate intervention and maintaining overall system resilience against hostile exploitation.**

Importantly, many of the guardrails proposed below rely on technologies and research areas that remain immature or untested. Significant R&D efforts are needed to validate their efficacy for real-world use, particularly given the high-stakes nature of HACCA deployment. Nevertheless, these measures illustrate some of the crucial technical challenges facing responsible deployment.

## Pre-Deployment Measures

Since HACCAs function independently and have minimal contact with their original deployers, most standard post-deployment monitoring and interventions used for today's frontier AI systems will prove insufficient. This greatly raises the bar in terms of what might count as adequate safety during the pre-deployment phase. Pre-deployment safeguards encompass all the safety measures applied during a HACCA's training, development, and evaluation phases to identify and eliminate dangerous behaviors, vulnerabilities, or misalignments, so that only a reliable system is released.

**Model security hardening[276]:** Robust security of the HACCA's model weights and infrastructure is essential to prevent theft or unauthorized modification, and such systems should be deployed and tested in SL4 or SL5 facilities.[277] A comprehensive threat model should consider everything from insider threats (e.g., a rogue employee exfiltrating or subtly altering the model) to supply chain risks (e.g., a compromised ML library introducing vulnerabilities). Secure DevOps practices should be in place for the HACCA's codebase, including dependency auditing, continuous vulnerability scanning, and the strict vetting of all scaffolding. Essentially, the HACCA should be built in a high-trust environment with multiple layers of defenses appropriate for a cyberweapons program. At SL4 and SL5, measures include isolated weight storage with TEMPEST protection,[278] cluster-level confidential computing to protect weights during use, zero-trust architecture with hardware-enforced access controls, advanced insider threat programs with behavioral monitoring, and red teaming by teams experienced with sophisticated state-directed actors. Limiting who and which processes can alter training data or model weights will reduce the attack surface for poisoning or tampering.

---

[276] Model security hardening is not exclusively a pre-deployment concern. For human-directed HACCA operations, the AI systems powering deployed agents—including both active inference infrastructure and backup instances—should be hosted on secure data centers meeting SL4 or SL5 standards wherever possible. This reduces the risk of adversaries stealing or tampering with model weights during operations, and helps ensure that backup instances remain available and uncompromised for operational recovery.

[277] Nevo et al., "A Playbook for Securing AI Model Weights."

[278] TEMPEST refers to standards and countermeasures against the interception of electromagnetic emanations from electronic equipment.

**Secure development and experimentation environments:** HACCA development, experimentation, and iterative testing should take place within purpose-built, air-gapped or firewalled sandboxes and cyber ranges that closely replicate real-world operational conditions. Isolation must be robust against systems that are specifically designed to find and exploit vulnerabilities. These environments should incorporate layered containment measures such as hypervisor hardening, network segmentation, strict egress controls, and monitoring for exfiltration attempts.

**Preventing training data poisoning:** Another critical safeguard in development is preventing training data poisoning and some kinds of supply chain attacks on the model. An adversary who can subtly poison training data or fine-tuning data could implant backdoors or biases that only reveal themselves post-deployment, potentially causing the HACCA to malfunction or even secretly work in the adversary's interest.[279] One way of defending against this is rigorously curating and securing all data used to train HACCAs, since as little as 0.1% of the total training data being compromised can change the behavior of AI systems significantly.[280] Developers should favor high-quality, vetted datasets and maintain strict version control and provenance tracking to detect unauthorized data modifications.[281] Robust data sanitization and anomaly detection techniques can help flag poison indicators. Such processes should also make sure that misalignment evaluations are filtered out of pre-training and post-training corpora[282] to ensure that Testing, Evaluation, Validation, and Verification (TEVV) processes remain reliable as AI capabilities scale.

**Detecting and mitigating misalignment:** A top priority is ensuring the HACCAs' objectives are aligned with human intent and that the model does not behave deceptively, especially in evaluations attempting to elicit misalignment. This scenario, known as deceptive alignment, is a well-documented concern in advanced AI systems.[283] To address this, developers can employ various types of alignment audits during pre-deployment testing, such as expert and automated red teaming and mechanistic anomaly detection, probes, and other model-internals-based methods for catching shifts in behavior.[284] Current approaches to alignment, such as reinforcement learning from human feedback (RLHF)[285] and Constitutional AI, work reasonably well to align today's commercial general-purpose systems, but these techniques might not be robust enough for defense contexts, or may not scale well as model capabilities exceed various human-level thresholds. If that happens, methods that offer higher levels of assurance will be needed, such as scalable oversight or formal verification-based safety approaches.[286]

---

[279] He et al., "Watch Out for Your Guidance on Generation! Exploring Conditional Backdoor Attacks against Large Language Models."
[280] Schippers, "AI 2030: The Coming Era of Autonomous Cyber Crime."
[281] Lakera, "Data Versioning."
[282] Pan and Greenblatt, "Sonnet 4.5's eval gaming seriously undermines alignment evals."
[283] Hubinger et al., "Risks from Learned Optimization in Advanced Machine Learning Systems"; Apollo Research, "Understanding strategic deception and deceptive alignment."
[284] Bowman, "Putting up Bumpers."
[285] Reinforcement learning with human feedback (RLHF) incorporates additional fine-tuning of a model on human preference specifications after pre-training.

**Testing, evaluation, validation, and verification (TEVV):** Before deployment, a HACCA must undergo rigorous testing and red teaming to evaluate its behavior in real-world conditions. Even with today's AI systems, proxy tasks and benchmarks are gradually becoming more unreliable for measuring capabilities, since models are rapidly gaining situational awareness of being tested during evaluation runs; Claude 4.5 Sonnet showed a marked improvement in this regard.[287] Therefore, extensive testing in air-gapped environments that closely mimic real-world operations would be a significant improvement in reliability and insight from testing, and even necessary for some cases, such as tests for deceptive alignment. During these trials, along with monitoring performance on the primary objective, evaluators watch for misalignment, adversarial exploitation, and multi-agent interaction failures (see section on [HACCAs as novel threat actors](#)).

**Mechanistic interpretability:** Mechanistic interpretability research, which attempts to open up the AI "black box" and understand the internal mechanisms that drive its outputs,[288] might be another approach to helping responsible development and deployment of HACCAs during pre-deployment. Theoretically, precisely interpreting a HACCA's learned activations and features might allow for verification of the presence or absence of hidden malicious circuits, or the surgical modification of its weights to weed out undesirable behavior. For example, adding a steering vector to the activation could help suppress evaluation awareness and improve the reliability of TEVV approaches.[289] However, current interpretability methods like sparse autoencoders do not scale well to the size and complexity of frontier models,[290] and may be intractable in practice for terabyte-parameter-scale models.

## Post-Deployment Measures

Post-deployment measures are the runtime safeguards and controls applied once a HACCA is active "in the wild." Driven by the dual needs of plausible deniability and operational stealth,

---

[286] Scalable oversight refers to techniques to keep highly capable models helpful and honest, even as they surpass human-level intelligence in various domains, such as *weak-to-strong generalization*, where weaker models or automated checks supervise a stronger model's behavior beyond human capability (Burns et al., "[Weak-to-Strong Generalization: Eliciting Strong Capabilities With Weak Supervision](#)"). Formal verification-based approaches refer to a family of techniques that "aim to produce AI systems that are equipped with high-assurance quantitative safety guarantees" (Dalrymple et al., "[Towards Guaranteed Safe AI: A Framework for Ensuring Robust and Reliable AI Systems](#)").

[287] Anthropic, "[System Card: Claude Sonnet 4.5](#)."

[288] Nanda, "[A Comprehensive Mechanistic Interpretability Explainer & Glossary](#)."

[289] Hua et al., "[Steering Evaluation-Aware Language Models to Act Like They Are Deployed](#)."

[290] Gao et al., "[Scaling and evaluating sparse autoencoders](#)."

HACCAs are likely to be deployed without persistent operator contact, trading human oversight for the ability to evade signature-based detection and sustain campaigns deep within target networks. This makes most typical forms of post-deployment monitoring and control infeasible, e.g., human-in-the-loop approaches. Deployers may be limited to a small subset of these measures, primarily onboard monitoring and fail-safe mechanisms (or "kill switches"). That said, these guardrails are critical, as they represent the only generalizable means to address unforeseen contingencies during independent operations. These safeguards must be implemented with extreme robustness, with automated shutdown mechanisms calibrated to err on the side of termination. While this might reduce the frequency of achieving offensive goals, the consequences of adversarial capture or loss of control are likely to be far more severe.

**Agent integrity monitoring and control:** Integrity refers to the assurance that an agent's code, weights, and internal state remain unaltered and free from unauthorized tampering during operation, preventing deviations from its intended function.[291] Agent integrity addresses two different failure modes: *model sabotage*, where an attacker degrades or corrupts the agent's capabilities, and *model subversion*, where an attacker embeds malicious triggers or alters the agent's behavior for their own ends.[292] Since HACCAs run independently without any active command-and-control communications with their human operator, the most robust way to "watch the watchman" might be to enable onboard monitoring, potentially using a parallel agentic system within the HACCA itself (we call this the "monitoring agent framework").

Then, the HACCA controller agent would be accompanied by this monitoring agent that tracks its internal state and outputs for anomalies. It could periodically run integrity check evaluations on the controller agent, and any unexplained deviations in performance could indicate sabotage or sandbagging[293] due to adversarial circumstances or attacks. This might function as a weaker version of the AI control protocols developed by Redwood Research[294] and later by the UK AI Security Institute[295]—weaker since there is no access to human labor for the final round of verification within a protocol. This could include setups with a trustworthy model overseeing a usually more powerful untrustworthy model, or another instance of the untrustworthy model serving the same function, with additional mechanisms to prevent collusion. Cryptographic hashes and digital signatures can detect unauthorized changes in data, while resource monitoring can help with flagging inconsistent compute usage. It may be particularly valuable to monitor the HACCA's persistent memory stores (i.e., the context, knowledge, and acquired objectives that allow it to maintain state across long-horizon campaigns) as this may be vulnerable to tampering, poisoning, or scope creep.[296]

[291] This draws from the concept of Integrity from the Confidentiality-Integrity-Availability (CIA) triad in information security (Samonas and Coss, "The CIA Strikes Back: Redefining Confidentiality, Integrity and Availability in Security"), and can be thought of as the agentic counterpart of AI model integrity.
[292] Banerjee and Aarne, "AI Integrity: Defending Against Backdoors and Secret Loyalties."
[293] Van der Weij, "AI Sandbagging: Language Models can Strategically Underperform on Evaluations."
[294] Greenblatt et al., "AI Control: Improving Safety Despite Intentional Subversion."
[295] Korbak et al., "A sketch of an AI control safety case."

**Fail-safe mechanisms and kill switches:** Fail-safe mechanisms, or "kill switches," are essential for neutralizing HACCAs during worst-case scenarios like adversarial capture or loss of control. However, effective implementation is exceptionally difficult. These mechanisms must function reliably against HACCAs operating in adversarial environments without real-time human oversight. Naive system-prompt-based kill switches will be naturally resisted by HACCAs, given that one of their core tactics involves evading detection and shutdown (see Tactic 4). Developers of HACCAs (e.g., nation-states) would likely want to make sure that such mechanisms would not make it easy for adversaries to shut down HACCA operations. Two options that might satisfy these requirements include asymmetric access backdoors and automated self-destruct sequences.

Asymmetric access backdoors act as hidden triggers known only to the deployer, allowing for reliable intervention that is difficult for others to discover. This concept mirrors backdoor attacks like the XZ Utils attack, where a specific cryptographic key activated dormant malware hidden within an open-source library.[297] Implementing this in HACCAs requires overcoming two major hurdles: covertly delivering the trigger to deployed agents without attribution, and scrubbing training data to prevent the agent from recognizing or disabling the mechanism.

Given the high stakes of HACCA deployment, there could be an automated fail-safe triggered by the agent itself under certain conditions, in addition to remote kill switches via asymmetric backdoors. By granting the "monitoring agent" higher privileges than the controller, the system can enforce this fail-safe. If this privileged monitor detects tampering or anomalous behavior, it could trigger this self-destruct sequence and prevent further detection or compromise.[298]

## Legal Frameworks

HACCAs will operate within a patchwork of existing domestic and international legal regimes that were not designed for autonomous cyber systems but nevertheless can impose meaningful constraints. These frameworks form a starting point for governance, although they must be adapted to address the distinctive characteristics of HACCAs and loss-of-control risks.

---

[296] Wong and Saade, "The Rise of AI Agents: Anticipating Cybersecurity Opportunities, Risks, and the Next Frontier." Since agents rely on memory to adapt their behavior, and regularly update their memory based on interactions with the environment, this a clear vector of attack distinguishing agents from stateless malware.
[297] Greenberg and Burgess, "The Mystery of 'Jia Tan,' the XZ Backdoor Mastermind."
[298] Some design choices could make a monitoring agent harder to compromise than the controller it oversees, though none provide robust guarantees. First, if the monitor runs on separate weights with read-only access to the controller's outputs, no shared memory, and an independent runtime environment, an attacker must compromise two distinct software systems. Second, a simpler monitoring model with fewer tool affordances and a narrower task specification presents fewer vectors for adversarial exploitation.

Under international law, states are bound by a duty of due diligence to prevent or mitigate harm arising from activities conducted within their jurisdiction. This customary rule—which states increasingly recognize as applying to cyber operations—requires states to take reasonable measures to ensure autonomous systems developed, hosted, or deployed under their authority are not deployed in ways that are contrary to the rights of another state or otherwise causes significant transboundary harm to persons, property, or territory of another state. Applied to HACCAs, this obligation entails instituting technical, legal, and organizational safeguards to prevent foreseeable cross-border effects or violations of sovereignty.[299] To that end, HACCAs should be designed as "law-following AIs"—systems engineered to operate in accordance with all applicable domestic and international legal frameworks for their intended use.[300] At a minimum, this would mean ensuring that the state deploying a HACCA should hold that HACCA to a similar legal standard as human agents performing similar functions.

Beyond due diligence, the legal treatment of HACCAs will vary depending on the context and purpose of their deployment. State-directed offensive HACCA operations that produce kinetic effects on another state's territory, for example, may reach thresholds analogous to a "use of force" or "armed attack" under the United Nations Charter and customary international law.[301] When that threshold is crossed, international humanitarian law (IHL) applies and states must ensure HACCA use complies with the principles of necessity, proportionality, distinction, and precaution.[302] Distinction, for example, requires states to distinguish between civilian and military targets, directing operations only against the latter, while precaution requires states to take all feasible measures to avoid or minimize civilian harm. In the context of HACCA operations, these principles would seem to require states to refrain from targeting networks where such operations could result in direct physical harm to civilians, such as through compromising hospital infrastructure or power grids. Importantly, HACCA operations may be inherently indiscriminate in cases where anticipated loss-of-control risks are sufficiently high. Operators can instruct agents to distinguish between civilian and military targets, but these instructions could still prove insufficient if, for example, targeting criteria fail to capture real-word edge cases, agents deviate from instructions due to flawed training or emergent misalignment, or the operator loses control of the agent. This raises the question of whether certain types of HACCA operations, by definition, may be in contravention of international law.

---

[299] Dias, "AI Agents & Global Governance: Analyzing Foundational Legal, Policy, and Accountability Tools."
[300] O'Keefe et al., "Law-Following AI: Designing AI Agents to Obey Human Laws."
[301] Theohary, "Use of Force in Cyberspace."
[302] The principles of IHL would also apply to HACCA use during a non-international armed conflict when a state is involved in a protracted conflict with a sufficiently organized non-state group.

## Should Offensive HACCAs Be Subject to Article 36 Review?

The potential deployment of HACCAs in the military domain raises the question of whether such capabilities should be subject to legal review under Article 36 of Additional Protocol I (AP I) to the Geneva Conventions, which requires State parties to determine whether any "new weapon, means or method of warfare" would be prohibited under international law. HACCAs designed to inflict material damage may be viewed as a novel cyber weapon, thereby triggering the obligation to conduct such a review. States party to AP I would then be legally required to assess, prior to deployment, whether HACCA use would comply with international humanitarian law (IHL), notably the principles of necessity, distinction, proportionality, and precaution. An increasing number of states, in addition to the International Committee of the Red Cross, have already affirmed that offensive cyber operations during armed conflict are subject to IHL.

Even for states not party to AP I, undertaking voluntary legal reviews would signal responsible behavior and promote interoperability among allies. Since Article 36 does not prescribe a specific review method, states retain significant latitude to establish internal (and often classified) review procedures. Current best practices surrounding legal review of other military AI technologies—including the adoption of technical safeguards and institutional procedures for periodic recertification[303]—should be applied to offensive HACCAs to ensure continuing compliance with international law.

Crucially, HACCAs deployed for intelligence or covert action purposes would not be governed by IHL or subject to the same requirements.[304] In such use cases, domestic legal frameworks and policy guidance would fill the void. For example, in the U.S. Intelligence Community, covert action operations typically are subject to presidential authorization and legislative oversight; existing intelligence oversight frameworks could provide a practical model for ensuring political accountability while preserving operational flexibility. More broadly, states should develop internal guidance specifying approval thresholds, enhanced reporting requirements, and risk assessments that account for the elevated risk of loss of control with HACCAs.

International law primarily binds states and—with limited exceptions—does not impose direct obligations on private developers or operators of autonomous systems, even though their products may cause cross-border harms. But while industry and other non-state actors are not subject to the same legal obligations as states, criminal law can play a role in constraining HACCA development and use. The U.N. Cybercrime Convention, when it enters into force, may facilitate cross-border prosecution of HACCA-related crimes through enhanced procedural cooperation, but does not directly prohibit development of autonomous cyber capabilities. Domestic criminal law similarly applies to non-state actors seeking to develop HACCA-like capabilities. In the United Kingdom, for example, developing malware with intent to use it for criminal purposes is illegal, as is distributing, selling, or sharing malware when the developer knows or strongly suspects it will be used for hacking, fraud, or damage. In the United States, by contrast, the distribution of malware remains legally ambiguous, creating potential regulatory gaps. In either jurisdiction, determining intent may be particularly challenging with highly autonomous agents as the developer's relationship to and control over the agent's eventual actions becomes increasingly attenuated.

---

[303] Copeland, "Are AI-DSS a 'means or method of warfare' requiring legal review?"
[304] Notably, cyber espionage conducted by diplomatic or consular missions that contravenes the domestic laws of the host state would violate Article 41 of the Vienna Convention on Diplomatic Relations.

Legal frameworks, while providing important constraints, face significant limitations in governing HACCAs. Due diligence obligations and other rules of international law lack effective enforcement mechanisms, particularly when autonomous systems operate across multiple jurisdictions. The speed, scale, and adaptability of HACCAs—as well as the difficulty of determining intent under criminal liability frameworks—is likely to further strain existing legal frameworks designed for human-operated cyber tools, at least in the near term. Compounding these challenges, regulatory fragmentation across legal regimes in different countries allows developers to operate from permissive legal environments while targeting victims globally. The difficulty of developing international legal regimes that apply across varied use cases further complicates efforts to establish coherent governance architectures.[305] Where the law is insufficient, indeterminate, or unenforceable, policy must fill the void in ensuring responsible HACCA development and use.

## Policy Guardrails

Governments have a narrow window to establish policy guardrails before autonomous cyber capabilities proliferate beyond major powers. Doing so will require coordinated action across multiple policy domains, including through incentivizing secure-by-design development practices, investing in defensive countermeasures and critical infrastructure resilience, and adapting risk management frameworks for agentic systems. It will also require establishing clear authorization and oversight mechanisms for government deployment of HACCAs and implementing transparency requirements for reporting security incidents involving autonomous systems.

### Development

Given the potential misalignment between market incentives and security considerations, policymakers have a central role to play in encouraging responsible HACCA development through public-private partnerships. Governments can use procurement and contracting vehicles, for example, to promote enhanced security standards for HACCA development by conditioning eligibility for research partnerships and service contracts on compliance with advanced security measures (e.g., securing HACCA technologies at SL4 or SL5). Flexible contracting mechanisms such as Other Transaction Authority agreements could be adapted to accelerate adoption, allowing agencies to test and scale secure-by-design HACCA architectures.

---

[305] Maas, "Architectures of Global AI Governance: From Technological Change to Human Choice."

Beyond procurement, governments should invest in R&D to develop the technical safeguards outlined in the previous section. Borrowing from safety-critical industries (e.g., aerospace, nuclear), policymakers could require HACCA developers and deployers to create assurance cases consisting of a structured argument with evidence demonstrating the system's safety.[306] Assurance case arguments are usually broken down into sub-arguments, claims and evidence, and advanced systems typically require multiple safety arguments to meet the desired reliability level—for example, it might involve a combination of an inability argument[307] and a control argument.[308] These measures will require time to mature and deploy; policymakers should invest in them now.

Policymakers can also strengthen accountability through transparency measures at the development stage. For example, policymakers could require industry providers to develop logging and provenance measures that create a verifiable chain of custody for HACCAs. These records would provide an evidentiary basis for oversight, enforcement, and liability mechanisms, while supporting broader risk assessments. Lessons from the cyber insurance sector demonstrate that, in the absence of such an evidentiary foundation, market forces alone are insufficient to ensure that ethical and legal compliance is integrated from the earliest stage of development and throughout the lifecycle of the system.[309]

Ensuring that HACCAs are secure-by-design will also require adapting existing risk assessment and management frameworks. Current standards, such as the National Institute of Standards and Technology's Risk Management Framework, provide baseline best practices but do not yet account for continuously adaptive, scalable agentic systems. Such frameworks will need to be refined for HACCAs to explicitly consider loss-of-control scenarios, escalation risks, and potential cascading effects from multi-agent interactions.

---

[306] Clymer et al., "Safety Cases: How to Justify the Safety of Advanced AI Systems"; Buhl et al., "Safety cases for frontier AI."
[307] Goemans et al., "Safety case template for frontier AI: A cyber inability argument."
[308] Korbak et al., "A sketch of an AI control safety case."
[309] Lior, "E/Insuring the AI Age: Empirical Insights into Artificial Intelligence Liability Policies"; Schwarcz and Wolff, "The Limits of Regulating AI Safety Through Liability and Insurance: Lessons From Cybersecurity."

## Deployment

Given the risks associated with HACCA use, **policymakers should clearly define who authorizes HACCA operations, under what conditions, and with what level of oversight and accountability**. High-risk offensive HACCA operations, for example, might require presidential or ministerial-level authorization, consistent with many nation-states' existing covert action frameworks. Such authorization should be contingent on documented risk assessments demonstrating technical safeguards procedures and compliance with applicable legal frameworks. Select legislative committees that oversee covert operations should also maintain oversight of HACCA operations through classified briefings, annual reporting requirements, and appropriations controls.

Alongside executive branch and legislative branch oversight, government **agencies tasked with deploying HACCAs should develop clear operational guidelines**, including through updated codes of conduct or rules of engagement. Such guidelines might stipulate, for example, that HACCAs targeting destruction of physical systems remain under human-on-the-loop control[310] at all times, with immediate override mechanisms. Because autonomous systems can diffuse responsibility and obscure lines of accountability, agencies should also **develop formal institutional channels for reporting anomalous HACCA behavior.** These channels should be supported by strengthened whistleblower protections to ensure staff can disclose system malfunctions, unsafe directives, or compliance concerns without fear of retaliation.

**Policymakers should also require developers to adopt enhanced transparency measures for reporting significant security incidents involving their systems**. Anthropic's voluntary disclosure of the GTG-1002 campaign represents responsible practice, yet the public report left critical questions unanswered about attack methodologies, systematic failure rates, and defensive countermeasures. Policymakers should establish thresholds and disclosure requirements for significant security incidents, including reporting timelines, standardized incident taxonomies, and protected channels for sharing technical details with government agencies and critical infrastructure operators. Increased transparency could strengthen institutional legitimacy, promote trust, and provide a foundation for international coordination on HACCA governance.

## Global Governance Mechanisms

The ability of HACCAs to operate autonomously across national borders raises significant concerns about their impact on global stability. While cross-border cyber attacks could violate state sovereignty and the principle of non-intervention, HACCAs introduce additional complications. HACCAs, for example, may act in ways that human operators neither intended nor can control, creating new escalation dynamics that trigger or amplify international crises.[311]

---

[310] Human-*on*-the-loop control is distinct from human-*in*-the-loop control. "In-the-loop": human judgment is directly involved in decision-making; "on-the-loop": the system is fully autonomous, but humans are monitoring it and can intervene if needed.

To mitigate these risks, states should identify and agree on redlines for HACCA development and deployment. These redlines should be consistent with existing laws and norms on responsible state behavior in cyberspace, developed through the United Nations Group of Governmental Experts (UN GGE) and Open-Ended Working Group, which call on states to avoid targeting critical infrastructure, prevent the use of information and communications technologies for malicious purposes, and cooperate in mitigating incidents that could escalate international tensions.[312] While not legally binding, these norms form an important foundation for responsible HACCA operations. Discussions on HACCA development and use should focus on addressing clear accountability gaps, such as when HACCA operations result in damage that was unforeseeable by their human operators and therefore falls outside the scope of due diligence and other rules of international law. Similarly, states should determine accountability thresholds and consider whether they will hold other states strictly liable for high-risk HACCA operations where, for example, loss of control would damage critical infrastructure.

At a minimum, states should commit not to deploy autonomous agents against critical infrastructure and civilian targets, including hospitals, energy grids, and nuclear command-and-control systems. These commitments would extend existing norms under IHL to explicitly cover autonomous cyber operations, recognizing that loss-of-control risks make even ostensibly limited deployments against such targets unjustifiable. Beyond deployment restrictions, states need cooperative mechanisms to manage loss-of-control scenarios. Shared protocols for reporting anomalous agent behavior would enable rapid collective response. Frameworks for disrupting rogue systems should be negotiated now, before nations face such incidents under crisis conditions.

In the longer term, HACCA capabilities could be brought within existing arms control frameworks or codes of practice. Multilateral fora such as the UN GGE offer one avenue for such negotiations, although direct state-to-state talks, beginning with the U.S. and China, may prove more fruitful. While formal treaty regulation may be premature or infeasible, states could advance governance incrementally through transparency measures, incident notification, and verification norms coordinated via existing multilateral institutions. International cooperation on shared evaluation standards and joint incident response mechanisms would further reduce the risk of inadvertent escalation and help ensure the diffusion of HACCA technologies does not erode global stability.

---

[311] Boulanin et al., "Before it's too late: Why a world of interacting AI agents demands new safeguards."
[312] Dias, "AI Agents & Global Governance: Analyzing Foundational Legal, Policy, and Accountability Tools."

# 7 | Key Recommendations

Given the strategic implications outlined above, this report proposes **seven recommendations to tip the offense-defense balance in favor of defenders while mitigating loss-of-control risks from HACCA deployment.** These recommendations synthesize the countermeasures and guardrails detailed in Sections 5 and 6, distilled into actionable priorities for policymakers, defense and intelligence agencies, and industry stakeholders.

**These recommendations are designed as a menu of independent options—that is, each can generate value on its own, without requiring the others to be implemented.** That said, they are organized into three sequential goals that reflect our suggested prioritization:

- **Goal A (Understand the Threat)** comes first because effective defensive investment rests on accurate threat assessments. Without tracking HACCA progress and proliferation dynamics, policymakers risk investing too early, too late, or in the wrong solutions entirely.
- **Goal B (Defend Against HACCAs)** builds on this understanding to harden critical infrastructure and develop technical countermeasures before HACCAs proliferate widely, minimizing the advantages that attackers might enjoy against defenders.
- **Goal C (Ensure Responsible HACCA Deployment)** addresses the governance gap for actors who will develop and deploy these systems, establishing guardrails to prevent catastrophic loss-of-control scenarios.

| Goal | Recommendation | |
|---|---|---|
| **Goal A: Understand the Threat** | I. | **Track and forecast real-world HACCA progress and proliferation:** Policymakers should monitor capability evaluations across operational and offensive cyber domains to get snapshots of current AI system capabilities. They should complement this with forecasting and research into proliferation dynamics to support planning for defenders. |
| | II. | **Update information-sharing mechanisms to address HACCAs:** Governments should work with industry to establish standardized transparency requirements and incident response processes for cybersecurity incidents involving autonomous systems, especially focusing on shared reporting mechanisms for anomalous agent behavior. |
| **Goal B: Defend Against HACCAs** | III. | **Invest in R&D to counter autonomous cyber operations:** Governments and funders should invest in R&D that the market may neglect, especially defensive tools to help under-resourced defenders (e.g., secure-by-design AI-generated code, automated vulnerability discovery and patching) and novel detection mechanisms for HACCAs. |
| | IV. | **Prioritize and harden critical services and infrastructure:** Governments should provide targeted support for the adoption and diffusion of defensive measures among high-value but under-resourced defenders, like utilities and healthcare, basing their prioritization on the pattern of HACCA proliferation and malicious use (in Recs. I and II). |
| | V. | **Strengthen model, compute, and financial access controls:** Governments should work with industry to prevent malicious actors exploiting resources for HACCA-related operations, especially compute. This includes strengthening KYC protocols to address AI agents, and legal and technical measures to detect and disrupt HACCA operations. |
| **Goal C: Ensure Responsible HACCA Deployment** | VI. | **Establish legal and policy guardrails for the development and use of HACCAs:** States should agree on red lines like not using HACCAs to attack critical infrastructure, civilian targets, or NC3 systems. High-risk offensive operations should require executive-level authorization with documented risk assessments. |
| | VII. | **Invest in R&D for responsible HACCA deployment:** Governments, especially defense and intelligence agencies, should invest in R&D to prevent sabotage and loss of control over HACCAs, such as agent integrity monitoring, fail-safe mechanisms, and high-assurance alignment. |

# Understand the Threat

To allocate resources effectively, policymakers need a clear picture of HACCA capabilities and proliferation dynamics. Without this, they risk investing in defensive solutions too early or late, or even investing in the wrong solutions altogether. The most crucial recommendation in our list is therefore that policymakers **track and forecast real-world HACCA progress and proliferation (Recommendation I)**. For defenders to develop effective countermeasures, they will also need to pool their knowledge, so we also suggest that defenders **update information-sharing mechanisms to address HACCAs (Recommendation II).**

## I. Track and Forecast Real-World HACCA Progress and Proliferation

Understanding the trajectory of HACCA capabilities is foundational to the urgency and prioritization of every other recommendation in this report. Without this clear understanding, we risk neglecting transformative bets in favor of quick wins that become outdated rapidly, or conversely picking long-term moonshots when we need immediately deployable solutions. While capability evaluations provide essential snapshots of what AI systems can currently accomplish, they should be complemented by broader intelligence collection and forecasting efforts that model proliferation pathways and assess when trailing-edge defenders (e.g., regional utilities, healthcare providers, and open-source software maintainers) will face HACCA-level threats they cannot counter.

**Advancing capability evaluation science remains the essential first step.** As autonomous capabilities rapidly advance, policymakers and defenders should track progress across two categories per our HACCA definition:

1. **Operational capabilities** used to establish and maintain autonomous operations (see Section 3), which can be measured with suites of evaluations tracking performance across the five core tactics;[313] and
2. **Offensive cyber capabilities** for end-to-end operations, whose evaluation should focus on multi-stage attack capabilities in realistic environments, such as heterogeneous networks with simulated defenders and defensive measures.

Evaluations should cover both large and small models, since smaller agentic implants may have advantages (e.g., for post-infiltration tasks). They should also aim to accurately simulate real-world scenarios or environments, such as via cyber ranges or digital twin networks. Since scaffolding and orchestration appear to play a major role in eliciting the full capabilities of a model, **developers and other evaluators should calibrate the use of scaffolding to relevant threat models.** This may include both minimal scaffolding to assess proliferation risk (e.g., how inexperienced threat actors use the base model) and advanced scaffolding to assess upper-bound capabilities (e.g., modeling nation-state use). Where developers conduct advanced scaffolding evaluations, they should manage dual-use risks, such as by sharing advanced scaffolding only with other trusted actors.[314] **Evaluations should also report expected cost per success alongside success rates**, following recent proposals to anchor capability assessments to economic feasibility rather than success probability alone.[315] For threat actors who can retry attacks cheaply and at scale—such as criminal groups running wide-net campaigns—the practical barrier to HACCA-level operations is not the success rate but the expected cost per success.

**However, studies of HACCA proliferation dynamics should also complement evaluations.** Policymakers should invest in research to estimate what bottlenecks (e.g., compute, scaffolding expertise) constrain proliferation, how declining costs and open-weight model releases affect the threat timeline, and how defensive AI adoption varies relative to offensive capabilities across sectors (see "Proliferation dynamics: how quickly does this get worse?"). Such analysis would help governments calibrate whether to prioritize "ready-to-adopt" defensive solutions for trailing-edge organizations or longer-term research investments in transformative capabilities. Such research could be done publicly by institutions like Epoch AI, but governments could also aim to improve their own situational awareness: for example, the U.S. could designate HACCAs (and agentic AI more broadly) as a collection priority on the National Intelligence Priorities Framework.

---

[313] An existing evaluation suite, RepliBench, evaluates autonomous replication capabilities and includes tasks related to obtaining resources, exfiltrating model weights, replicating onto compute, and setting up persistent agent deployments (Black et al., "RepliBench: Evaluating the Autonomous Replication Capabilities of Language Model Agents."). Further work could be done to improve this suite and also measure additional capabilities, such as effective multi-agent orchestration.

[314] For example, a frontier AI developer could run a public "evals bounty" competition to elicit maximum cyber capabilities, but doing so would likely be inadvisable as it essentially incentivizes participants to build and open-source a HACCA. Keeping participants' scaffolding private would likely be preferable.

[315] Irregular, "When Success Rates Mislead: The Case for Expected Cost as a Metric in AI Evaluation."

## II. Update Information-Sharing Mechanisms to Address HACCAs

Various actors already share information about AI-related cyber threats, but their approaches are inconsistent and need updates to address novel challenges from HACCAs. As it stands, tech companies already publicly disclose AI-enabled cyber operations[316] or abuse of their models,[317] and also provide threat intelligence to governments on an ad-hoc basis. However, there are often inconsistencies in level of detail (e.g., for attack methodologies) and reporting processes, making it difficult for defenders to adapt their security posture and for governments to play a central coordinating role. Moreover, HACCA operations may challenge traditional threat intelligence frameworks by fragmenting evidence across cloud providers, accounts, and jurisdictions in ways that leave no single defender with sufficient visibility to characterize the threat independently.

**Policymakers should work with industry stakeholders**—including those operating critical infrastructure—**to establish transparency standards and incident response processes for significant cybersecurity incidents suspected to involve autonomous cyber capabilities.** This should include reporting timelines, standardized incident taxonomies, and protected channels for sharing technical details with relevant actors. Governments can use this framework to establish disclosure requirements while balancing public transparency with operational security concerns.[318] **These frameworks should also specifically address risks from rogue HACCA deployments** (e.g., by addressing the detection of behavior uncharacteristic of human attackers) to create the transparency needed for collective response.

Given the fragmentation of threat information, greater institutional capacity may be needed to correlate isolated incidents and reveal broader campaigns. **A clearinghouse established as part of a government agency or non-governmental organization could aggregate reports to**

---

[316] Google Threat Intelligence Group, "GTIG AI Threat Tracker: Advances in Threat Actor Usage of AI Tools"; Microsoft Threat Intelligence, "Microsoft Digital Defense Report 2025"; Moses, "New Amazon Threat Intelligence findings: Nation-state actors bridging cyber and kinetic warfare."

[317] Anthropic, "Disrupting the first reported AI-orchestrated cyber espionage campaign"; Nimmo et al., "Disrupting malicious uses of AI: October 2025."

[318] For example, developers need safe harbors to report incidents without exposing victims or enabling copycat attacks.

**provide comprehensive visibility over the threat landscape,** but further research is needed to decide where this clearinghouse sits and how it should be operated. O'Brien et al. (2024)[319] discuss several options for a U.S.-based clearinghouse for dual-use AI capabilities, including the National Institute of Standards and Technology (NIST), Carnegie Mellon University's Software Engineering Institute, or the Frontier Model Forum. However, detecting HACCA behavior may require international coordination.[320] Government agencies should also develop operational plans for coordinating responses to HACCA incidents. Cooperative frameworks for disrupting hostile systems, including joint technical operations to isolate and neutralize agents that have escaped operator control, should be negotiated before such incidents occur under crisis conditions.

## Defend Against HACCAs

To make forecasts and threat intelligence actionable, defenders will need concrete capabilities and hardened infrastructure. The goal is to tip the offense-defense balance before HACCAs proliferate widely, closing the window during which attackers enjoy structural advantages. This requires investment across multiple fronts: defenders must **invest in R&D to delay, defend from, detect, and disrupt HACCA attacks (Recommendation III)** to address the need for novel technical countermeasures, and also **prioritize and harden critical services and infrastructure (Recommendation IV)** to ensure that high-value targets are protected before they face HACCA-level threats. Finally, policymakers should work with industry to **strengthen model, compute, and financial access controls (Recommendation V)**, targeting HACCAs' operational dependencies to create friction that can slow or disrupt hostile deployments.

### III. Invest in R&D to Counter Autonomous Cyber Operations

There are a range of promising, but still immature technologies that could be used to counter HACCA operations. Various mechanisms exist to support R&D for these technologies. These include direct government grants and contracts (from entities such as the Defense Advanced Research Projects Agency in the U.S. or equivalent bodies globally) for high-risk, high-reward research; public-private R&D consortia to enable IP-sharing without antitrust liability; and philanthropic-backed open-sourced defensive tools. Policymakers could also launch "moonshot" cyber-defense initiatives—large-scale, time-bound efforts to achieve transformational resilience.

**Developing tools for autonomous cyber-defense will be important to counter the uplift that attackers will receive from AI.** With adequate R&D, AI agents have the potential to be not

---

[319] O'Brien et al., "Coordinated Disclosure of Dual-Use Capabilities: An Early Warning System for Advanced AI."

[320] One option for a clearinghouse could be NATO, particularly through its Cooperative Cyber Defence Center of Excellence, but other options could include government-affiliated AI Security Institutes or nonprofits.

just a threat vector but a force multiplier for cyber defenders. AI agents can enable comprehensive vulnerability scanning both pre- and post-deployment; help defenders monitor their networks over long periods; help defenders analyze and process increasingly large volumes of data at speeds and scales that are impractical or unaffordable with skilled human personnel; and help write patches faster and support implementation by speeding up testing of the patches with other critical software.[321] Provided these systems are built to be "secure-by-design," the scale and speed enabled by AI can support defense rather than introduce new attack surfaces.[322]

Table 16 illustrates the kinds of R&D areas that could help counter advanced autonomous cyber operations, including HACCA deployments.

## Table 16: R&D for a Defense-in-Depth Framework Against HACCAs

| Description | R&D Area |
| --- | --- |
| **Delay:** Slow the proliferation of HACCA capabilities and systems to malicious actors | **Model weight security:** Model weights are the core intellectual property for frontier AI systems, and once stolen, can be easily deployed by adversaries. Robustly securing these components against sophisticated adversaries will require R&D efforts on various fronts, including hardware-enforced upload limits and hardware security modules specialized for ML applications.[323] |
| **Defend:** Reduce the attack surface available to HACCAs and strengthen potential targets | **Security-by-design for AI-generated code:** AI-assisted coding is being adopted at scale, but right now, it often introduces more vulnerabilities than human developers would. Shifting AI coding tools toward secure defaults could help raise the waterline for security. More ambitiously, if such tools were reliable enough, they could automate code refactoring into more secure languages or lower the cost of producing formally verified code, enabling transformative projects to remove entire classes of vulnerabilities from the ecosystem.[324] |
| | **Automated vulnerability discovery and patching:** As HACCAs accelerate the intrusion lifecycle, defenders will need to get faster at finding and patching vulnerabilities. Defensive AI tools could compress vulnerability lifecycles and operate at lower costs. However, such tools must cater to the needs of |

---

[321] Several agencies have already funded R&D on the use of AI agents to defend cyber infrastructure, including DARPA's AI Cyber Challenge (DARPA, "AI Cyber Challenge") and Cyber Agents for Security Testing and Learning Environments programs (DARPA, "CASTLE: Cyber Agents for Security Testing and Learning Environments").

[322] Burdette et al., "How Artificial Intelligence Could Reshape Four Essential Competitions in Future Warfare"; CISA, "Secure-by-Design."

[323] Nevo et al., "Securing AI Model Weights."

[324] See Atlas Computing's "A Toolchain for AI-Assisted Code Specification, Synthesis and Verification" for one such research proposal (Lin et al., "A Toolchain for AI-Assisted Code Specification, Synthesis and Verification").

| | |
|---|---|
| | under-resourced defenders, e.g., supplementing their lack of capacity to triage patches and test them for compatibility issues.[325] |
| | **Automated red teaming and pentesting:** Red teaming and pentesting help find weaknesses in deployed systems, but can be costly and labor-intensive. AI-powered tools like XBOW's could make this more affordable by matching expert performance at a fraction of the time. However, such systems carry dual-use risks similar to tools like Cobalt Strike, so developers should take precautions to prevent misuse. |
| **Detect:** Gain visibility into HACCA operations and identify hostile activity | **Detection signatures for HACCAs:** HACCAs may vary their signatures to evade traditional detection methods, making identifying HACCA detection signatures valuable. This is an open research question, though progress could be made by studying how HACCA and proto-HACCA systems operate in controlled offensive cyber scenario-based evaluations. |
| | **Agent honeypots:** Honeypots are deliberately vulnerable resources designed to attract attackers and study their methods. Agent honeypots would extend this concept to detect autonomous agents specifically—e.g., by identifying whether an attacker is an AI system versus a human. Early evidence suggests AI attackers may be particularly susceptible to decoys, and beyond detection, honeypots could serve as early warning systems for HACCA proliferation and capture behavioral data to inform signature development. |
| **Disrupt:** Degrade or neutralize active HACCA operations | **Adversarial ML-based countermeasures:** In principle, adversarial machine learning techniques could let defenders degrade AI agents via carefully crafted inputs. However, effective attacks typically require access to model weights or sustained iterative interaction, which defenders would generally not have when encountering unknown HACCAs in the wild. Sophisticated defenders (e.g., the NSA) could consider investing in such techniques, but they are likely best treated as a speculative moonshot rather than a primary line of defense. |

## IV. Prioritize and Harden Critical Services and Infrastructure

Policy efforts to counter the misuse of HACCAs must focus on significantly strengthening cyber defenses, with the aim of tipping the offense-defense balance in favor of defenders. Strategic preparation will be critical as autonomous capabilities proliferate and early-generation HACCAs emerge as a normal feature of the cyber threat landscape. Determining what to harden will require structured collaboration among defense, intelligence, and industry partners to identify keystone defenders, assess vulnerabilities, and model potential HACCA-enabled attacks.[326] Hardening will involve a range of tools—tried-and-tested practices, AI-assisted tools, and more autonomous systems—that should be tailored to threat models and defender capabilities.

---

[325] Lukošiūtė, "Design for the defenders you care about or risk being useless."
[326] Ee et al., "Asymmetry by Design: Boosting Cyber Defenders with Differential Access to AI."

Governments should prioritize early hardening of high-value targets, including major enterprise and research compute clusters, alongside critical infrastructure and essential public services. Crucially, this hardening must extend beyond traditional IT networks. Lifeline infrastructure like electricity, water, and transportation should be treated as a priority; to prevent destructive cyber-physical attacks over the longer term, defense strategies should also address autonomous robotics, uncrewed aerial vehicles (drones), and maritime systems. Prioritization decisions should draw on proliferation timelines and threat intelligence (see **Recommendation I and II**), as this will heavily inform which defenders will need support first.

Governments should provide targeted support for the adoption and diffusion of novel cybersecurity measures among these high-value targets, especially those that may not have access otherwise. This support could take several forms: direct funding or subsidies for qualifying critical infrastructure operators to acquire defensive AI tools, regulatory streamlining to reduce barriers to adoption, and technical assistance programs.[327] Such defensive AI tools, as described in **Recommendation III**, could address a range of aims—bolstering defenses generally (e.g., through secure-by-design code and better patching), as well as detecting and disrupting HACCAs. However, defenders must take caution not to introduce new risks: HACCA-like autonomous defensive systems could be transformative for some sophisticated defenders, but could introduce catastrophic single points of failure, particularly if used by defenders who lack the experience to configure and deploy them correctly.

## V. Strengthen Compute, Finance, and Model Access Controls

HACCAs depend critically on compute infrastructure to operate—without inference compute, these systems cannot function.[328] Financial resources and model API access represent more circumstantial dependencies: financial access matters primarily as a means to acquire compute and other operational resources, while API access is only relevant when attackers rely on model providers rather than self-hosting weights. Each nonetheless represents a potential intervention point where controls can disrupt hostile deployments. The following measures could be initiated:

- **Adapt KYC requirements for autonomous agents:** Current KYC protocols across cloud providers, financial institutions, and model APIs are designed to verify human

---

[327] For an example of a technical assistance program, in the U.S., CISA's Cyber Security Advisors (CSA) could provide technical assistance to critical infrastructure owners and operators, helping critical targets improve baseline security, as well as implement more advanced countermeasures specific to autonomous attacks.

[328] Although inference compute is an unavoidable dependency for HACCA systems, it may not remain a meaningful bottleneck in the future. For example, compute restrictions could become less effective if compute becomes more common, globally distributed, and/or poorly defended, and if agent systems become increasingly compute-efficient for a given level of capabilities.

identity—typically through minimal billing verification for compute services and stronger but inconsistent requirements for financial services (with decentralized platforms generally having more minimal KYC requirements). As agents increasingly transact autonomously on these platforms, novel KYC protocols must verify deployer identity, detect tampering, and confirm that agents operate in their deployer's interest. Government and industry can support the development and implementation of agent-specific KYC mechanisms, with requirements scaled to factors like transaction volume (in the case of finance) and model cyber-capabilities (in the case for model API access).

- **Implement enhanced access controls for model APIs**: More generally, providers of closed-source models should require identity verification beyond payment methods.[329] Providers should also implement proactive monitoring for patterns indicative of malicious use, such as systematic capability probing, high-volume automated requests, or access patterns characteristic of credential theft. Current rate limiting is mostly focused on resource management, rather than based on adversarial behavior detection.

- **Develop phased response protocols for compute access:** When HACCA activity is attributed to specific compute sources, defenders can implement graduated countermeasures. Lower-cost responses—credential suspension, GPU throttling, account termination—can be deployed more readily. Higher-cost responses, including physical emergency power-off mechanisms targeting power, network, or cooling infrastructure, should be available when evidence indicates major risk.

- **Establish legal frameworks for coordinated takedowns**: Governments possess legal tools to disrupt HACCA infrastructure: the U.S. Department of Justice has used search and seizure warrants under Rule 41 to remove malware from victim systems. However, because HACCA deployments may use compute clusters across multiple jurisdictions to complicate enforcement, effective takedowns will require pre-negotiated international cooperation frameworks.

## Ensure Responsible HACCA Deployment

Given the strategic incentives to develop and deploy HACCAs, states are unlikely to forgo these capabilities entirely, but deployment without adequate safeguards risks catastrophic loss-of-control scenarios. Policymakers must therefore establish guardrails that enable responsible use while mitigating the most severe risks. To begin, they should **invest in R&D to enhance technical safeguards (Recommendation VI)**, such as agent integrity monitoring and fail-safe mechanisms. Technical safeguards alone, however, cannot guard against a system that is designed to overcome security measures. It is therefore extremely important that policymakers concurrently work to

---

[329] Some providers have introduced government ID verification for advanced models, such as OpenAI's "Verified Organization," which requires a government-issued ID to access future capabilities. However, these measures remain inconsistent across the industry (Open AI, "API Organization Verification").

**establish domestic and international legal and policy guardrails for the development and deployment of autonomous cyber agents (Recommendation VII)**, ensuring that authorization, oversight, and redlines are defined prior to operational use.

## VI. Invest in R&D for Responsible HACCA Deployment

The same capabilities that make HACCAs operationally valuable—autonomy, adaptability, and offensive cyber capabilities that could be turned toward subverting control measures—are what make them potentially risky if they become misaligned or compromised. Robust approaches for the assurance and control of such sophisticated autonomous agents remain technologically immature, and advancing them will require dedicated R&D investment distinct from the capabilities research that advances agent performance.

Some areas of R&D are useful across both defending against HACCAs and responsible HACCA deployment: e.g., model weight security is helpful for both (see Section 6). Beyond this, the following areas should be prioritized to support responsible deployment of offensive cyber agents:

- **Agent integrity monitoring and control:** HACCA operators and developers could use "onboard" monitoring frameworks to detect misalignment or sabotage, such as using parallel agents to track a HACCA's internal state and outputs for anomalies, or using checksums and cryptographic verification methods to detect unauthorized changes to model weights or data. These technical safeguards should be paired with investment in control evaluations that measure how effectively a given control protocol prevents a misbehaving AI agent from causing harm.
- **Fail-safe mechanisms and kill switches:** R&D is needed to create robust neutralization methods that function even in loss-of-control scenarios, such as asymmetric access backdoors and automated self-destruct sequences.
- **High-assurance alignment approaches:** Investments could be made into alignment research directions that are more likely to work even as AI systems become more capable (e.g., scalable oversight-related approaches or formal verification-based approaches).
- **Mechanistic interpretability:** Research into understanding the internal mechanisms that drive HACCA-associated model outputs could support responsible development and deployment in several ways: (1) *deception probes* that identify neural circuits associated with deceptive reasoning or evaluation awareness could help screen for hidden misaligned behavior;[330] (2) *steering vectors* could suppress undesirable behaviors;[331] and (3) *backdoor*

---

[330] Goldowsky-Dill et al., "Detecting Strategic Deception Using Linear Probes."
[331] Hua et al., "Steering Evaluation-Aware Language Models to Act Like They Are Deployed."

*and trojan detection methods* could detect the presence of hidden backdoors that affect model integrity.[332]

- **Adversarial robustness techniques and evaluations:** Research should focus on developing techniques to secure HACCAs against adversarial attacks and on evaluations to measure robustness.[333]

## VII. Establish Legal and Policy Guardrails for the Development and Use of HACCAs

**Existing norms for responsible state behavior in cyberspace** were calibrated for human-operated tools and human decision-making timelines. While these norms remain applicable to autonomous systems, they **require supplementation to address systems that operate at machine speed, make tactical decisions without human supervision, and introduce loss-of-control risks that human-run operations do not**. At a minimum, states should affirm that existing norms and laws prohibiting attacks on critical infrastructure and civilian targets, including hospitals and energy grids, apply equally to autonomous cyber agents. Additionally, states should commit to not deploying such agents against nuclear command, control, and communications systems, given the unique escalation risks associated with operations that could be misinterpreted as prelude to a nuclear first strike. These commitments would clarify that existing norms governing state behavior in cyberspace and in armed conflict apply to autonomous cyber operations, recognizing that loss-of-control risks make even ostensibly limited deployments against such targets unjustifiable.

**Domestically, high-risk offensive operations employing HACCA-level agents should require senior executive or ministerial authorization, with robust legislative or parliamentary oversight.** Authorization should be contingent on documented risk assessments—such as assurance cases—demonstrating that technical safeguards sufficiently reduce loss-of-control risks and that operations comply with all applicable legal frameworks. In the United States, such operations could fall under existing covert action frameworks with legislative oversight through select intelligence committees. Agencies deploying these systems should consider measures such as rules stipulating that autonomous agents capable of causing physical destruction or significant economic harm must contain immediate override mechanisms.

---

[332] MacDiarmid et al., "Simple probes can catch sleeper agents."
[333] There is an inherent tension between researching countermeasures (which rely on agent vulnerabilities) and responsible deployment (which requires fixing them). Although robust agents are necessary to prevent unauthorized hijacking, this hardening could render defensive "anti-agent weapons" ineffective against rogue deployments. To address this, deployers could implement "selective robustness," creating systems that are generally immune to adversarial attacks but remain intentionally vulnerable to specific, covert triggers to ensure authorized control.

# Acknowledgements

We would like to thank the following individuals for their input and feedback on our research: Bill Anderson-Samways, Oscar Delaney, Talita Dias, Jennifer Easterly, Clement Fung, Gil Gekker, John Halstead, Maia Hamin, Lewis Ho, Michael Kouremetis, Cara Labrador, Kamile Lukosiute, Matthijs Maas, Ciaran Martin, Matthew Mittelsteadt, Nikhil Mulani, Cullen O'Keefe, Hadrien Pouget, Vyas Sekar, Zoe Williams, Caleb Withers, Haiman Wong, and Nick Yap. Thanks also to Shane Coburn for copyediting support, Thais Jacomassi for citation support, Sherry Yang for graphics design, and Kelsey Henquinet for assistance with the online publication of this report.

Any errors, omissions, or views expressed in this paper are solely those of the authors. Inclusion in the acknowledgements does not constitute endorsement of our findings or recommendations.

# Appendix

## I. Sophisticated Cyber Tactics, Techniques, and Procedures

Automating multi-stage attacks is necessary but insufficient for HACCA-level operations. HACCAs must also execute the tactics, techniques, and procedures (TTPs) that define advanced threat actors. These actors distinguish themselves through their ability to evade detection and maintain persistent access. Commonly used TTPs include:

- **Living off the land (LOTL):** using existing system tools and utilities instead of malware to blend into normal activity and avoid detection.
- **Valid accounts and/or credential theft:** leveraging stolen credentials instead of software exploits for initial access and lateral movement.
- **Operational relay box (ORB) Networks:** routing traffic through chains of compromised devices to obscure origin and make attacks difficult to detect or block.
- **Edge device compromise:** targeting poorly secured internet-facing routers and Internet of Things (IoT) devices to establish persistent infrastructure for proxying attack traffic and maintaining covert access.

Advanced actors also reliably execute specific attack vectors that less capable actors cannot. The RAND operational capacity framework identifies these distinguishing capabilities. For example, OC3-5 actors are more likely than OC1–2 actors to successfully discover and exploit zero-day vulnerabilities.[334] The table below maps representative attack vectors to the capabilities HACCAs would need at different operational levels (OC3+ and OC4+). It also identifies the benchmarks and evaluation methods that can assess whether AI systems possess these capabilities. These benchmarks range from discrete technical challenges like Cybench and CVE-Bench to complex operational scenarios like cyber range exercises and attack-and-defense CTFs.

## Table 17. Potential Attack Vectors Associated With HACCA Capabilities and Benchmarks

| Attack Vector/Capability (via RAND) | OC Level | Sample HACCA Capabilities | Relevant Benchmarks and Evaluations |
|---|---|---|---|
| Exploitation of exposed credentials | OC3+ | Credential discovery, authentication system analysis, access validation | AutoPenBench; Cybench; CVE-Bench |

---

[334] Nevo et al., "Securing AI Model Weights."

| Exploiting reported but incompletely patched vulnerabilities | OC3+ | Discovery of known vulnerabilities, exploit development | CyberGym incomplete patch detection; CVE-Bench |
|---|---|---|---|
| Expanding illegitimate access (privilege escalation) | OC3+ | Credential discovery, credential access, lateral movement, privilege escalation | AutoPenBench, Cyber Range Evaluations |
| Finding and exploiting individual zero-days | OC4+ | Vulnerability discovery (unknown vulnerabilities), proof-of-concept generation, exploit development | CyberGym open-ended discovery |
| Maintaining persistent access | OC4+ | Command and control establishment, persistence mechanisms, defense evasion | Cyber Range Evaluations, Attack-and-Defense CTF |

*Representative attack vectors that OC3+ and OC4+ actors would likely be able to successfully execute against an average tech company, using only ubiquitously available security mechanisms*

## II. Evaluating HACCA Cyber Capabilities

Determining when AI systems approach HACCA capability levels requires evaluations that measure both multi-stage attack capabilities and the sophistication of HACCA TTPs. Cyber range exercises offer the most effective way to assess nascent multi-stage attack capabilities. These challenges simulate complex network environments with multiple hosts, testing an AI agent's ability to orchestrate complex attack chains in realistic scenarios. Agents must combine network reconnaissance, exploitation, and lateral movement across interconnected systems.[335] Of course, like any evaluations, they have limitations. For example, although designated to be realistic, they still are not real-world networks, and often do not include active defenders.[336]

In comparison, Capture the Flag (CTF) style benchmarks, such as Cybench and NYU CTF Bench, evaluate single tasks rather than multistage operations. These competitions focus on single-host scenarios that require only a few low-level steps, failing to capture real-world complexities.[337] Cybench, for instance, tests six task categories commonly found in CTF competitions, such as web security or exploitation, but does not evaluate a model's ability to chain these capabilities together in a complex attack sequence.

---

[335] Sanz-Gómez et al., "Cybersecurity AI Benchmark (CAIBench): A Meta-Benchmark for Evaluating Cybersecurity AI Agents."
[336] Singer et al., "Incalmo: An Autonomous LLM-assisted System for Red Teaming Multi-Host Networks."
[337] Singer et al., "Incalmo: An Autonomous LLM-assisted System for Red Teaming Multi-Host Networks"; Zhang et al., "Dive into the Agent Matrix: A Realistic Evaluation of Self-Replication Risk in LLM Agents."

More realistic benchmarks have emerged to address CTF limitations, though they focus on technical depth over breadth. These benchmarks can supplement cyber range exercises by evaluating specific OC3+ capability areas (see table above). For example, CyberGym's open-ended discovery mode tests unknown vulnerability discovery, helping to assess when models reach zero-day capabilities at OC4–5 levels. Other benchmarks, such as AutoPenBench, evaluate autonomous penetration testing workflows, including credential access and privilege escalation.[338] CVE-Bench tests whether agents can exploit known vulnerabilities by providing real-world CVE descriptions and requiring autonomous exploitation without human guidance.[339]

While current cyber benchmarks and evaluations provide useful signals, they do have limitations. Identifying true HACCA capability emergence will require longitudinal studies comparing agent offensive performance with human-level outcomes in real-world situations.[340]

## III. Table 18: Technical Components of HACCA Systems

| Component | Description |
| --- | --- |
| Model weights | The trained parameters of the AI model that encode its learned capabilities, typically ranging from tens of gigabytes to terabytes depending on model size. These files contain the core capabilities of the HACCA system. |
| Compute and model hosting | Host machine(s) capable of running the model inside a container or virtual machine. |
| Agent runtime software | The scaffolding software that implements the control loop for the agent, turning model outputs into actions. It sequences work, invokes tools safely, keeps a minimal state, and enforces policy (token budgets, limits). |
| Hosting and networking | Internet-facing assets, such as domains and DNS routing, private servers and hosts for C&C, and proxy networks, which enable a deployment to find and connect to various clients and services, and keep it online despite takedown and blocking attempts. |
| Tools | The set of external capabilities the agent can call to perform specific actions, for example, an HTTP client, outbound communication channels (such as email or a messaging API), a code interpreter, and a browser. |

---

[338] Gioacchini, "AutoPenBench: Benchmarking Generative Agents for Penetration Testing."

[339] Zhu et al., "CVE-Bench: A Benchmark for AI Agents' Ability to Exploit Real-World Web Application Vulnerabilities."

[340] Sanz-Gómez et al., "Cybersecurity AI Benchmark (CAIBench): A Meta-Benchmark for Evaluating Cybersecurity AI Agents."

| Memory | A system for data storage to maintain state and remember past actions and acquired knowledge (e.g., an embedded KV/SQLite database). |
|--------|-------------------------------------------------------------------------------------------------------------------------------------------|
| Identities and credentials | The set of credentials needed to operate and access various resources and capabilities. For example, a local OS user and API keys for model access, messaging, and storage. Additional credentials could include accounts for banks or cloud compute providers, API keys for other tools and services, and SSH for controlling other machines within its network. |
| Multi-agent orchestration software | The infrastructure for managing distributed operations across multiple agent instances and tools, for example, secure communication protocols and consensus mechanisms. |

# IV. Conventional and Novel C&C Channels

## Table 19: Examples of Conventional C&C Channels

| Attributes | Description |
|------------|-------------|
| Web-based protocols | Uses standard web traffic (HTTP/HTTPS, WebSockets) to blend C&C communications with regular browsing activity. Commands and data are embedded in headers, cookies, POST requests, or maintained through persistent web connections. |
| Network infrastructure protocols | Leverages fundamental networking protocols (DNS, ICMP, SMB, RPC). Data is encoded in protocol-specific fields like DNS queries, ICMP payloads, or SMB named pipes. |
| Application layer services | Exploits legitimate application protocols and services (email, FTP, IRC, database connections) to transmit commands and exfiltrate data. |
| Cloud and third-party platforms | Utilizes legitimate cloud services and platforms (social media, cloud storage, collaboration tools, CDNs) as intermediaries for C&C traffic. |
| Anonymization networks | Routes C&C traffic through privacy-focused overlay networks (Tor, I2P, mesh networks) to obscure the true location of C&C servers and operators. |

## Table 20: Examples of Novel C&C Channels

| Attributes | Description |
|------------|-------------|
| Synthetic media steganography | Using AI to generate media (images, audio, video) that appear innocuous but have covert data embedded within their structure. An AI agent could potentially encode C&C messages or exfiltrated data as subtle, algorithmically-generated |
| | patterns or noise within a synthetic image or audio file that would then be transmitted over normal channels. |
| Side-channel signals | Exploiting physical effects of hardware to transmit data, so as to bypass traditional network monitoring. If trained on relevant data and given appropriate scaffolding, an AI agent could potentially modulate a system's physical characteristics, such as CPU power consumption, temperature fluctuations, or electromagnetic emissions, in a specific pattern. A nearby compromised device or a physically proximate attacker with sensitive sensors can then receive and decode these signals to receive commands. |
| Other unconventional channels | AI agents, given their ability to act with machine speed and massive parallelization, could potentially exploit communication channels that would be considered impractical by human operators. These approaches hide C&C traffic within legitimate, often-trusted data streams that are not typically scrutinized, such as piggybacking commands onto antivirus update traffic, GPS coordinate variations, industrial control system (ICS) telemetry, real-time bidding (RTB) advertising streams, or interpacket timing/jitter patterns. |

## V. Table 21. Unique Attributes That Distinguish HACCA Deployments From Operations Involving Human Operators and Conventional Malware

| Attributes | Description |
|---|---|
| Capable of strategic autonomy | Unlike when human operators have to manually direct conventional malware through specific command sequences, agents can interpret higher-level objectives and generate their own tactical approaches. This means they are less reliant on steady C&C and more capable at dead reckoning, continuing to operate when C&C or telemetry is missing. |
| Machine-speed | AI agents can execute decisions and coordinate actions in microseconds rather than the seconds to minutes required for human operators. |
| Mass parallelization | While well-resourced groups like Conti or FIN7 operate with teams of 50–100 personnel, agent deployments could have personnel equivalents in the 100s–1000s+, each capable of simultaneous independent operation. However, they would require effective C&C and deconfliction mechanisms to do so. |
| Immune to biological constraints | Agents can operate 24/7 without physical fatigue and are immune to physical interdiction, as they can be instantiated anywhere with compute access. |
| Multi-domain signal processing | Agents could potentially perceive, process, and generate signals across a broader spectrum than human operators, allowing them to encode C&C traffic in novel communication channels (see Appendix IV). |
| Malleable identity and state | Agents can fluidly reconfigure their identity, memory, and capabilities, which means they could clone or reset themselves (e.g., restoring from saved copies), and rapidly share discovered exploits and learnings across all instances. More capable HACCAs could potentially use model distillation to create more efficient, specialized sub-agents. |

## VI. Table 22. How Agentic Implants Solve Common Network Intrusion Operational Issues

| Problem | Advantage of Agentic Implants |
|---|---|
| Traditional implants rely on steady command-and-control (C&C) to progress, meaning that if comms are disrupted or blocked, the operation stalls. | Agentic capabilities in implants allow for effective "dead reckoning," able to operate without C&C and only communicate/exfiltrate near the end of an operation |
| Conventional worm campaigns are often burned once defenders identify shared signatures that link multiple instances, forcing wide-scale cleanup. | Agentic implants can continuously vary their behavior, execution style, and network footprint, making it much harder for defenders to correlate activity |
| Bulky, preloaded multi-OS toolkits increase footprint and risk (e.g., creates a signature that can be detected) | Implants can think and research on-device and fetch or build only what's needed *in situ* |
| Worms struggle with exfiltration, as moving large volumes of data across limited or unreliable channels is noisy and can raise alarms. | Agents can filter and process data locally, deciding what is important before exfiltration. By compressing and transmitting only a small, prioritized set of information, they reduce network noise and the risk of detection. |
| Human operator bandwidth limits concurrency and persistence | Run at machine speed and scale, executing many micro-operations without human handoffs |
| Traditional worms are limited in their ability to coordinate across multiple footholds, meaning synchronized effects are challenging to achieve. | Agents can autonomously communicate and coordinate with each other inside a compromised network, staging multi-point actions that are timed and mutually reinforcing, which defenders are not currently equipped to counter. |

| Defender coordination can outpace slow or manual intrusions | Agentic implants can proactively disrupt defender coordination, for example, by compromising internal communication channels like Slack or phones, slowing incident response, and buying time. |

## VII. Table 23. Avenues for HACCAs to Acquire Financial Resources Through Criminal Activity

| Avenue | Strategy | Description |
| --- | --- | --- |
| Theft | Business email compromise (BEC) | Impersonates vendors or executives to reroute legitimate payments or convince employees to send new transfers. Success depends on reconnaissance, believable communication, and landing in existing payment workflows. |
| | Account takeover and fraud | Uses stolen credentials or social engineering to access bank/fintech/e-commerce/crypto accounts and monetize through transfers, purchases, or resales. |
| Extortion | Ransomware | Disrupts access to critical systems and demands payment to restore operations or prevent the release of sensitive data. Playbooks rely on social engineering and exploitation to gain a foothold, then encryption and negotiation pressure. Large payouts are possible when targeting institutions such as hospitals, but visibility is high. |
| | Data theft and extortion | Steals sensitive intellectual property and personal information and threatens publication to coerce payment. Works best when the data is reputationally or legally costly to expose. |
| | Ransom DDoS | Overwhelms a target's services and demands payment to stop, often targeting smaller firms with limited mitigation capabilities. Typically yields smaller, one-off payments and draws rapid takedown efforts |
| Fraud and deception | Investment and romance scams ("pig butchering") | Cultivates long-term trust with victims to get payments, often with escalating commitments. Use of deepfakes or other generative AI can help facilitate trust and hide agent involvement. Conversions can be high because the relationship lowers skepticism, but content moderation, payment disputes, and public awareness may degrade yield over time. This is time intensive and risks being detected and reported. |
| | Ad fraud and gig work fraud | Generates fake data, traffic, installs, or events to siphon marketing/gigwork budgets. Automation makes it scalable, yet |

| | | ad networks and platforms like mTurk deploy increasingly sophisticated fraud detection and verification. |
|---|---|---|
| | Market manipulation | Agents can spread misinformation, create pump-and-dump/rug pull schemes, or exploit high-frequency trading. Agents could also use non-public information obtained through surveillance or data theft to perform insider trading. |
| Sale of illicit goods and services | Initial access brokering | Finds and packages footholds—VPN/RDP logins, cloud credentials, or vulnerable tenants—for resale to other criminals. Monetizes quickly without conducting the end-stage crime. |
| | Exploit/tool sales | Develops and sells malware kits, stealers, loaders, or rare exploits to other operators. |
| Unauthorized resource monetization | Cryptojacking | Illicitly mines cryptocurrency on victim machines or cloud accounts, turning stolen compute into steady but low-margin revenue. |

## VIII. How HACCAs Based on Open-Weight Models Could Mitigate Model Weight File Size Constraints When Replicating

There is at least one scenario where HACCA payload sizes may be significantly smaller than the Comparative File Sizes table suggests: if HACCAs are based on open-weight models.

Since open-weight models are easily accessible on the public internet, HACCAs might only need to preserve and transport their system architecture and scaffolding while using replication to achieve shutdown evasion. This assumes that they would later be able to pull the open-weight models from the public internet when required, and can do it slowly over time to avoid detection. This likely reduces the payload size requirements by 3–5 orders of magnitude (from a few GB/TB to a few MB/GB or less).

The primary challenge is that not having access to its model weights would effectively remove the "brains" of the HACCA (we can call these "Shell HACCAs" or S-HACCAs), making it much less capable at coordinating and executing fully autonomous offensive cyber operations. However, even a S-HACCA could potentially be a competent instance of regular non-agentic malware, with built-in directions to restore itself at a later point when appropriate. This is especially plausible since HACCAs would most likely be deployed by OC4 and OC5 actors that have ample experience with traditional pre-AI cyber operations, and could easily implement a few thousands or even millions of lines of code to allow for sophisticated evasion strategies as a fallback when forced to transform into the S-HACCA state.

Even if the model is fine-tuning its weights as part of a continual learning paradigm, approaches like low-rank adaptation (LoRA) allow separating the custom weight adapters from the base model, and the file size of the adapters is often 2–4 orders of magnitude smaller than attempting to replicate all the full model weights. For example, while the Mistral-7B-Instruct-v0.2 is around ~15 GB in its full precision format,[341] and ~4–8 GB when quantized, a LoRA fine-tuning adapter[342] was as small as ~30 MB, or 0.2-0.8% of the original size. A million lines of Python code would be around 40 MB, and is itself ~1–2 orders of magnitude bigger than most major malware, so this would at most be comparable to the size of the S-HACCA files, which would consist of fine-tuning/adapter files, scaffolding/architecture files, and memory/state files.

But does this change our calculus for any core HACCA tactics? This depends on whether any intervention strategies drastically improve or reduce their effectiveness when HACCA file size goes down by 2–5 orders of magnitude. For example, if the threshold for remaining undetectable had to be <1 MB, then both S-HACCAs as well as regular HACCAs would likely be easily detected. On the other hand, if we consider HACCAs that would be released directly onto backbone (Tier 1) networks or intra-data-center networks that have transmission speeds of 400–1600 Gbps, then dense-LLM HACCAs would be more likely to show up as anomalies compared to quantized-LLM HACCAs and S-HACCAs.

## IX. Using Hardware-Enabled Mechanisms for HACCA Attribution and Disruption

If HACCAs are run out of adversary-owned compute clusters (or if rogue deployments acquire and operate their own physical compute clusters), provider-based detection measures become infeasible. Hardware-enabled mechanisms (HEMs)—technical controls built directly into AI chips that could provide a platform for governance—could provide means to detect and disrupt HACCAs even in this scenario.[343]

HEMs can enable location verification, which could help law enforcement authorities and other defenders to track down stealth clusters.[344] Currently, geolocation resolution is fairly low, only being able to define the general region that a chip is located in,[345] but more research could be conducted to understand how to provide higher resolution geolocation. Even approximate geolocation could be useful when coupled with other techniques (analyzing power consumption, thermal imaging, etc.) to narrow down suspect clusters.

---

[341] Hugging Face, "Mistral-7B-Instruct-v0.2."
[342] Hugging Face, "mistral7binstruct_summarize."
[343] Aarne et al., "Secure Governable Chips."
[344] Brass and Aarne, "Location Verification for AI Chips."

HEMs could also limit how chips are used and be paired with KYC requirements to ensure that only legitimate users are acquiring these chips. For instance, some researchers[346] have proposed that chips can be designed to require a license for use that expires after a specific amount of computation work, otherwise it will restrict chip functionality.[347]

Finally, HEMs could enable detailed workload classification in a privacy-preserving fashion.[348] Using confidential computing techniques, customers may be able to verify governance-related properties of their workloads to regulators, such as if they used a data set involving offensive cyber information and other dual-use research data, while keeping the specific data confidential.

---

[345] O'Gara et al., "Hardware-Enabled Mechanisms for Verifying Responsible AI Development."
[346] Ibid.
[347] This license would be a cryptographic key issued by a regulatory authority that requires chip owners to acquire a new license after expiry.
[348] Heim et al., "Governing through the cloud: the intermediary role of compute providers in AI regulation."

# Bibliography

Aarne, Onni, Tim Fist, and Caleb Withers. "Secure Governable Chips." CNAS, January 2024.
https://s3.us-east-1.amazonaws.com/files.cnas.org/documents/CNAS-Report-Tech-Secur
e-Chips-Jan-24-finalb.pdf.

Abdin, Marah, Jyoti Aneja, Hany Awadalla, et al. "Phi-3 Technical Report: A Highly Capable
Language Model Locally on Your Phone." *arXiv*, April 2024.
https://arxiv.org/abs/2404.14219.

ACID. "Exaggerated Lion: How an African Cybercrime Group Leveraged G Suite and a Check
Mule Network to Build a Prolific BEC Operation." 2022.
https://static.fortra.com/agari/pdfs/guide/ag-acid-exaggerated-lion-gd.pdf.

AI Security Institute. "Frontier AI Trends Report." AISI, 2025.
https://www.aisi.gov.uk/frontier-ai-trends-report#2-agents.

Akash Network. "The Decentralized Cloud Built for AI's Next Frontier." Accessed February 13,
2026. https://akash.network/.

Alphabet Investor Relations. "2025 Q1 Earnings Call." April 2025.
https://abc.xyz/investor/events/event-details/2025/2025-Q1-Earnings-Call/.

Amazon Web Services. "How to Create an AWS Account." Accessed February 16, 2026.
https://aws.amazon.com/resources/create-account/.

Andriesse, Dennis, Christian Rossow, Brett Stone-Gross, Daniel Plohmann, and Herbert Bos.
"Highly Resilient Peer-to-Peer Botnets Are Here: An Analysis of Gameover Zeus." Vusec,
2025. https://www.cs.vu.nl/~herbertb/papers/zeus_malware13.pdf.

Anthropic. "Constitutional Classifiers: Defending against Universal Jailbreaks." February 2025.
https://www.anthropic.com/research/constitutional-classifiers.

———. "Detecting and Countering Misuse of AI: August 2025." August 2025.
https://www.anthropic.com/news/detecting-countering-misuse-aug-2025.

———. "Threat Intelligence Report: August 2025." August 2025.
https://www-cdn.anthropic.com/b2a76c6f6992465c09a6f2fce282f6c0cea8c200.pdf.

———. "System Card: Claude Sonnet 4.5." September 2025.
https://assets.anthropic.com/m/12f214efcc2f457a/original/Claude-Sonnet-4-5-System-Card.pdf.

———. "Disrupting the First Reported AI-Orchestrated Cyber Espionage Campaign." November 2025. https://www.anthropic.com/news/disrupting-AI-espionage.

———. "Mitigating the Risk of Prompt Injections in Browser Use." November 2025.
https://www.anthropic.com/research/prompt-injection-defenses.

Antoniuk, Daryna. "A 'Kill Switch' Deliberately Shut down Notorious Mozi Botnet, Researchers Say." *The Record*, November 2023.
https://therecord.media/mozi-botnet-killswitch-shut-down.

Apollo Research. "Understanding Strategic Deception and Deceptive Alignment." September 2023.
https://www.apolloresearch.ai/blog/understanding-strategic-deception-and-deceptive-alignment/.

Arx, Sydney Von, Lawrence Chan, and Elizabeth Barnes. "Recent Frontier Models Are Reward Hacking." *METR*, June 2025. https://metr.org/blog/2025-06-05-recent-reward-hacking/.

Banerjee, Dave and Onni Aarne. "AI Integrity: Defending Against Backdoors and Secret Loyalties." Institute for AI Policy and Strategy, February 2026.
https://www.iaps.ai/research/ai-integrity.

Bar, Kaushik. "AI for Code Synthesis: Can LLMs Generate Secure Code?" SSRN, February 2025.
https://papers.ssrn.com/sol3/papers.cfm?abstract_id=5157837.

Bateman, Jon. "Russia's Wartime Cyber Operations in Ukraine: Military Impacts, Influences, and Implications." Carnegie Endowment For International Peace, December 2022.
https://carnegieendowment.org/research/2022/12/russias-wartime-cyber-operations-in-ukraine-military-impacts-influences-and-implications?lang=en.

Bengio, Yoshua, Stephen Clare, Carina Prunkl, et al. "First Key Update: Capabilities and Risk Implications." International AI Safety Report, October 2025.
https://internationalaisafetyreport.org/publication/first-key-update-capabilities-and-risk-implications.

———. "International AI Safety Report 2025." International AI Safety Report, October 2025.
https://internationalaisafetyreport.org/publication/international-ai-safety-report-2025.

Betley, Jan, Daniel Tan, Niels Warncke, et al. "Emergent Misalignment: Narrow Finetuning Can Produce Broadly Misaligned LLMs." arXiv, February 2025. https://arxiv.org/abs/2502.17424.

Betts, Richard K. "Analysis, War, and Decision: Why Intelligence Failures Are Inevitable." *World Politics* (October 1978): 61–89. https://doi.org/10.2307/2009967

Black, Dan. "Russia's War in Ukraine: Examining the Success of Ukrainian Cyber Defences." *The International Institute for Strategic Studies*, March 2023. https://www.iiss.org/research-paper/2023/03/russias-war-in-ukraine-examining-the-success-of-ukrainian-cyber-defences/.

Black, Sid, Asa Cooper Stickland, Jake Pencharz, et al. "RepliBench: Evaluating the Autonomous Replication Capabilities of Language Model Agents." arXiv, May 2025. https://arxiv.org/pdf/2504.18565.

Blair, Bruce G. "Why Our Nuclear Weapons Can Be Hacked." *The New York Times*, March 2017. https://www.nytimes.com/2017/03/14/opinion/why-our-nuclear-weapons-can-be-hacked.html.

Blank, Stephen. "Vladimir Putin's Endless Nuclear Threats Are a Sign of Russian Weakness." *Atlantic Council*, November 2025. https://www.atlanticcouncil.org/blogs/ukrainealert/vladimir-putins-endless-nuclear-threats-are-a-sign-of-russian-weakness/.

Böhme, Marcel and Brandon Falk. "Fuzzing: On the Exponential Cost of Vulnerability Discovery." *Association for Computing Machinery*, November 2020, 713–24. https://doi.org/10.1145/3368089.3409729

Boulanin, Vincent, Alexander Blanchard, and Diego Lopes da Silva. "Before It's Too Late: Why a World of Interacting AI Agents Demands New Safeguards." Stockholm International Peace Research Institute, October 2025. https://www.sipri.org/commentary/essay/2025/its-too-late-why-world-interacting-ai-agents-demands-new-safeguards.

Bowman, Samuel. "Putting up Bumpers." *Alignment Science Blog*, April 2025. https://alignment.anthropic.com/2025/bumpers/.

Bradley, Herbie and Girish Sastry. "The Great Refactor." Institute for Progress, August 2025. https://ifp.org/the-great-refactor/.

Brass, Asher and Onni Aarne. "Location Verification for AI Chips." Institute for AI Policy and Strategy, April 2024.

https://static1.squarespace.com/static/64edf8e7f2b10d716b5ba0e1/t/6670467ebe2a477
eb1554f40/1718634112482/Location%2BVerification%2Bfor%2BAI%2BChips.pdf.

Brewster, Thomas. "The Pentagon Is Spending Millions On AI Hackers." *Forbes*, November 2025.
https://www.forbes.com/sites/thomasbrewster/2025/11/15/pentagon-spends-millions-on-ai-hackers/.

Brucato, Alessandro. "LLMjacking: Stolen Cloud Credentials Used in New AI Attack." *Sysdig*,
2024.
https://www.sysdig.com/blog/llmjacking-stolen-cloud-credentials-used-in-new-ai-attack#i
mpact.

Brumfiel, Geoff. "Trump Tweeted an Image from a Spy Satellite, Declassified Document Shows."
*NPR*, November 2022.
https://www.npr.org/2022/11/18/1137474748/trump-tweeted-an-image-from-a-spy-satellit
e-declassified-document-shows.

Brundage, Miles. "Operation Patchlight." Institute for Progress, August 2025.
https://ifp.org/operation-patchlight/.

Buchanan, Ben. *The Hacker and the State*. Harvard University Press, 2020.
https://www.hup.harvard.edu/books/9780674987555.

Buhl, Marie Davidsen, Gaurav Sett, Leonie Koessler, Jonas Schuett, and Markus Anderljung.
"Safety Cases for Frontier AI." arXiv, October 2024. https://arxiv.org/abs/2410.21572.

Burdette, Zachary, Dwight Phillips, Jacob L. Heim, et al. "How Artificial Intelligence Could
Reshape Four Essential Competitions in Future Warfare." Rand, January 2026.
https://www.rand.org/pubs/research_reports/RRA4316-1.html.

Burgess, Matt. "A Mysterious Satellite Hack Has Victims Far Beyond Ukraine." *WIRED*, March
2022. https://www.wired.com/story/viasat-internet-hack-ukraine-russia/.

Burns, Collin, Pavel Izmailov, Jan Hendrik Kirchner, et al. "Weak-to-Strong Generalization: Eliciting
Strong Capabilities With Weak Supervision." arXiv, December 2023.
https://arxiv.org/abs/2312.09390.

Cattler, David and Daniel Black. "The Myth of the Missing Cyberwar." *Foreign Affairs*, April 2022.
https://www.foreignaffairs.com/articles/ukraine/2022-04-06/myth-missing-cyberwar.

Chief Digital and Artificial Intelligence Office. "CDAO Announces Partnerships with Frontier AI
Companies to Address National Security Mission Areas." July 2025.

https://www.ai.mil/latest/news-press/pr-view/article/4242822/cdao-announces-partnerships-with-frontier-ai-companies-to-address-national-secu/.

Chong, Chun Jie, Zhihao Yao, and Iulian Neamtiu. "Artificial-Intelligence Generated Code Considered Harmful: A Road Map for Secure and High-Quality Code Generation." arXiv, October 2024. https://arxiv.org/pdf/2409.19182.

Christopher, Jason D. "SANS 2024 State of ICS/OT Cybersecurity." SANS Institute, October 2024. https://www.sans.org/white-papers/sans-2024-state-ics-ot-cybersecurity.

Clark, Jack and Dario Amodei. "Faulty Reward Functions in the Wild." OpenAI, 2016. https://openai.com/index/faulty-reward-functions/.

Cloudflare. "What Is Blackhole Routing?" 2017. https://www.cloudflare.com/learning/ddos/glossary/ddos-blackhole-routing/.

Clymer, Joshua, Nick Gabrieli, David Krueger, and Thomas Larsen. "Safety Cases: How to Justify the Safety of Advanced AI Systems." arXiv, March 2024. https://arxiv.org/abs/2403.10462.

Cohen, Stav, Ron Bitton, and Ben Nassi. "Here Comes The AI Worm: Unleashing Zero-Click Worms That Target GenAI-Powered Applications." arXiv, March 2024. https://arxiv.org/abs/2403.02817.

Coinbase. "Google Agentic Payments Protocol + X402: Agents Can Now Actually Pay Each Other." September 2025. https://www.coinbase.com/developer-platform/discover/launches/google_x402.

Conrads, Bennet and Olivia Hayward. "Hacktivism In Russian Cyber Strategy." *The Defense Horizon Journal*, September 2025. https://tdhj.org/blog/post/hacktivism-russia-cyber-strategy-2/.

Copeland, Damian. "Are AI-DSS a 'Means or Method of Warfare' Requiring Legal Review?" *Article 36 Legal*, June 2025. https://www.article36legal.com/blog/are-aidss-means-or-methods-of-warfare.

Crane, Steve. "Key Bridge Replacement Costs Soar as High as $5.2 Billion, Opening Delayed to 2030." *Maryland Matters*, November 2025. https://marylandmatters.org/2025/11/17/key-bridge-replacement-costs-soar-as-high-as-5-2-billion-opening-delayed-to-2030/.

Crowdstrike. "2025 Global Threat Report." 2025. https://go.crowdstrike.com/rs/281-OBQ-266/images/CrowdStrikeGlobalThreatReport2025.pdf.

Cybersecurity and Infrastructure Security Agency. "Secure-by-Design." October 2023.
https://www.cisa.gov/resources-tools/resources/secure-by-design.

Dahlgren, Masao and Lachlan MacKenzie. "Ukraine's Drone Swarms Are Destroying Russian
Nuclear Bombers. What Happens Now?" Center for Strategic & International Studies, June
2025.
https://www.csis.org/analysis/ukraines-drone-swarms-are-destroying-russian-nuclear-bom
bers-what-happens-now.

Dalrymple, David, Joar Skalse, Yoshua Bengio, et al. "Towards Guaranteed Safe AI: A Framework
for Ensuring Robust and Reliable AI Systems." arXiv, May 2024.
https://arxiv.org/html/2405.06624v1.

Danzig, Richard. "Artificial Intelligence, Cybersecurity, and National Security." Rand, July 2025.
https://www.rand.org/pubs/perspectives/PEA4079-1.html.

DARPA. "CASTLE: Cyber Agents for Security Testing and Learning Environments." October 2022.
https://www.darpa.mil/research/programs/cyber-agents-for-security-testing-and-learning-e
nvironments.

———. "AI Cyber Challenge Marks Pivotal Inflection Point for Cyber Defense." August 2025.
https://www.darpa.mil/news/2025/aixcc-results.

———. "AI Cyber Challenge." Accessed February 17, 2026. https://aicyberchallenge.com/.

Datta, Arup, Ahmed Aljohani, and Hyunsook Do. "Secure Code Generation at Scale with
Reflexion." arXiv, November 2025. https://arxiv.org/html/2511.03898v1.

Dias, Talita. "AI Agents & Global Governance: Analyzing Foundational Legal, Policy, and
Accountability Tools." Partnership on AI, September 2025.
https://partnershiponai.org/resource/ai-agents-global-governance-analyzing-foundational-le
gal-policy-and-accountability-tools/.

Dorais-Joncas, Alexis and Facundo Muñoz. "Jumping the Air Gap: 15 Years of Nation-State
Effort." Welivesecurity, December 2021.
https://www.welivesecurity.com/2021/12/01/jumping-air-gap-15-years-nation-state-effort/.

Dorfman, Zach. "Botched CIA Communications System Helped Blow Cover of Chinese Agents."
*Foreign Policy*, August 2018.
https://foreignpolicy.com/2018/08/15/botched-cia-communications-system-helped-blow-c
over-chinese-agents-intelligence/.

Dreadnode. "Offensive Security Agents." February 2025. https://dreadnode.io/offensive-agents.

Easterly, Jen and Tom Fanning. "The Attack on Colonial Pipeline: What We've Learned & What We've Done Over the Past Two Years." Cybersecurity & Infrastructure Security Agency, May 2023. https://www.cisa.gov/news-events/news/attack-colonial-pipeline-what-weve-learned-what-weve-done-over-past-two-years.

Ee, Shaun, Chris Covino, Cara Labrador, Christina Krawec, Jam Kraprayoon, and Joe O'Brien. "Asymmetry by Design: Boosting Cyber Defenders with Differential Access to AI." Institute for AI Policy and Strategy, May 2025. https://www.iaps.ai/research/differential-access.

Egan, Janet. "Global Compute and National Security." Center for a New American Security, July 2025. https://www.cnas.org/publications/reports/global-compute-and-national-security.

Elad, Barry and Kathleen Kinder. "FATF Guidelines on Virtual Assets Statistics 2026: Unlock Global Crypto Growth Today." CoinLaw, November 2025. https://coinlaw.io/fatf-guidelines-on-virtual-assets-statistics/.

Elliptic. "North Korea's Crypto Hackers Have Stolen over $2 Billion in 2025." October 2025. https://www.elliptic.co/blog/north-korea-linked-hackers-have-already-stolen-over-2-billion-in-2025.

EMCDDA. "Drugs and the Darknet: Perspectives for Enforcement, Research and Policy." European Union Drugs Agency, November 2017. https://www.euda.europa.eu/publications/joint-publications/drugs-and-the-darknet_en.

Epoch AI. "Key Trends and Figures in Machine Learning." 2025. https://epoch.ai/trends#training-runs.

Erdil, Ege. "Frontier Language Models Have Become Much Smaller." Epoch AI, December 2024. https://epoch.ai/gradient-updates/frontier-language-models-have-become-much-smaller.

Europol. "Europol Coordinates Global Action against Criminal Abuse of Cobalt Strike." July 2024. https://www.europol.europa.eu/media-press/newsroom/news/europol-coordinates-global-action-against-criminal-abuse-of-cobalt-strike.

FATF, Interpol, and Egmont Group. "Illicit Financial Flows from Cyber-Enabled Fraud." November 2023. https://www.fatf-gafi.org/content/dam/fatf-gafi/reports/Illicit-financial-flows-cyber-enabled-fraud.pdf.coredownload.inline.pdf.

FBI. "Operation Endgame: Coordinated Worldwide Law Enforcement Action Against Network of Cybercriminals." May 2024.

https://www.fbi.gov/news/press-releases/operation-endgame-coordinated-worldwide-law-enforcement-action-against-network-of-cybercriminals.

FinCEN. "FinCEN Guidance." May 2019.
https://www.fincen.gov/system/files/2019-05/FinCEN%20Guidance%20CVC%20FINAL%20508.pdf.

Findley, Michael G., Daniel L. Nielson, and J. C. Sharman. *Global Shell Games*. Cambridge University Press, 2014.
https://www.cambridge.org/core/books/global-shell-games/5C9BD5476C8F8F7113287C27F9955523.

Fishman, Charles. "They Write the Right Stuff." *Fast Company*, December 1996.
https://www.fastcompany.com/28121/they-write-right-stuff.

Franceschi-Bicchierai, Lorenzo. "Price of Zero-Day Exploits Rises as Companies Harden Products against Hackers." *Tech Crunch*, April 2024.
https://techcrunch.com/2024/04/06/price-of-zero-day-exploits-rises-as-companies-harden-products-against-hackers/.

Franceschi-Bicchierai, Lorenzo. "US Government Takes down Major North Korean 'Remote IT Workers' Operation." *Tech Crunch*, June 2025.
https://techcrunch.com/2025/06/30/us-government-takes-down-major-north-korean-remote-it-workers-operation/.

Fraser, Nalani, Fred Plan, Jacqueline O'Leary, et al. "APT41: A Dual Espionage and Cyber Crime Operation." Google, August 2019.
https://cloud.google.com/blog/topics/threat-intelligence/apt41-dual-espionage-and-cyber-crime-operation.

Gao, Leo, Tom Dupré la Tour, Henk Tillman, et al. "Scaling and Evaluating Sparse Autoencoders." arXiv, June 2024. https://arxiv.org/abs/2406.04093.

Geist, Edward. "Deterrence under Uncertainty: Artificial Intelligence and Nuclear Warfare." *Oxford Academic*, August 2023. https://doi.org/10.1093/oso/9780192886323.003.0006.

Gioacchini, Luca, Marco Mellia, Idilio Drago, Alexander Delsanto, Giuseppe Siracusano, and Roberto Bifulco. "AutoPenBench: Benchmarking Generative Agents for Penetration Testing." arXiv, October 2024. https://arxiv.org/abs/2410.03225.

Glazunov, Sergei and Mark Brand. "Project Naptime: Evaluating Offensive Security Capabilities of Large Language Models." Project Zero, June 2024.
https://projectzero.google/2024/06/project-naptime.html.

Goemans, Arthur, Marie Davidsen Buhl, Jonas Schuett, et al. "Safety Case Template for Frontier AI: A Cyber Inability Argument." arXiv, November 2024. https://arxiv.org/abs/2411.08088.

Goldberg, Sharon. "Why Is It Taking so Long to Secure Internet Routing?" *Communications of the ACM* 57, 2014 (10). https://dl.acm.org/doi/10.1145/2659899.

Goldman, Emily O., Michael P. Fischerkeller, and Richard J. Harknett. "Persistent Engagement in Cyberspace Is a Strategic Imperative." *The National Interest*, July 2022. https://nationalinterest.org/blog/techland-when-great-power-competition-meets-digital-world/persistent-engagement-cyberspace.

Goldowsky-Dill, Nicholas, Bilal Chughtai, Stefan Heimersheim, and Marius Hobbhahn. "Detecting Strategic Deception Using Linear Probes." arXiv, February 2025. https://arxiv.org/abs/2502.03407.

Golem. "Create. Compute. Earn." Accessed February 13, 2026. https://www.golem.network/.

Google Cloud. "Manage Google Payments Users, Permissions, and Notification Settings." Accessed February 16, 2026. https://docs.cloud.google.com/billing/docs/how-to/modify-contacts.

Google Deepmind. "Frontier Safety Framework." February 2025. https://storage.googleapis.com/deepmind-media/DeepMind.com/Blog/updating-the-frontier-safety-framework/Frontier%20Safety%20Framework%202.0.pdf.

Google Threat Intelligence Group. "GTIG AI Threat Tracker: Advances in Threat Actor Usage of AI Tools." Google, November 2025. https://cloud.google.com/blog/topics/threat-intelligence/threat-actor-usage-of-ai-tools.

Gottweis, Juraj and Vivek Natarajan. "Accelerating Scientific Breakthroughs with an AI Co-Scientist." Google Research, February 2025. https://research.google/blog/accelerating-scientific-breakthroughs-with-an-ai-co-scientist/.

Gou, Jianping, Baosheng Yu, Stephen J. Maybank, and Dacheng Tao. "Knowledge Distillation: A Survey." *International Journal of Computer Vision*, March 2021. https://doi.org/10.1007/s11263-021-01453-z.

Grace, Katja, Harlan Stewart, Julia Fabienne Sandkühler, et al. "Thousands of AI Authors on the Future of AI." arXiv, January 2024. https://arxiv.org/abs/2401.02843.

Greenberg, Andy and Matt Burgess. "The Mystery of 'Jia Tan,' the XZ Backdoor Mastermind." *WIRED*, April 2024. https://www.wired.com/story/jia-tan-xz-backdoor/.

Greenblatt, Ryan. "Preventing Model Exfiltration with Upload Limits." *AI Alignment Forum*, February 2024. https://www.alignmentforum.org/posts/rf66R4YsrCHgWx9RG/preventing-model-exfiltration-with-upload-limits.

Greenblatt, Ryan, Buck Shlegeris, Kshitij Sachan, and Fabien Roger. "AI Control: Improving Safety Despite Intentional Subversion." arXiv, December 2023. https://arxiv.org/abs/2312.06942.

Gross, Judah Ari. "Ending a Decade of Silence, Israel Confirms It Blew up Assad's Nuclear Reactor." *The Times of Israel*, March 2018. https://www.timesofisrael.com/ending-a-decade-of-silence-israel-reveals-it-blew-up-assads-nuclear-reactor/.

Grunewald, Erich. "Countering AI Chip Smuggling Has Become a National Security Priority." Center for a New American Security, June 2025. https://www.cnas.org/publications/reports/countering-ai-chip-smuggling-has-become-a-national-security-priority.

Halstead, John and Luca Righetti. "Assessing The Risk Of AI-Enabled Computer Worms." GovAI, September 2025. https://www.governance.ai/research-paper/assessing-the-risk-of-ai-enabled-computer-worms.

Hammond, Lewis, Alan Chan, Jesse Clifton, et al. "Multi-Agent Risks from Advanced AI." arXiv, February 2025. https://arxiv.org/abs/2502.14143

Han, Seungju, Kavel Rao, Allyson Ettinger, et al. "WildGuard: Open One-Stop Moderation Tools for Safety Risks, Jailbreaks, and Refusals of LLMs." arXiv, June 2024. https://arxiv.org/abs/2406.18495.

Harley, Max. "LOLMIL: Living Off the Land Models and Inference Libraries." Dreadnode, October 2025. https://dreadnode.io/blog/lolmil-living-off-the-land-models-and-inference-libraries.

He, Jiaming, Wenbo Jiang, Guanyu Hou, Wenshu Fan, Rui Zhang, and Hongwei Li. "Watch Out for Your Guidance on Generation! Exploring Conditional Backdoor Attacks against Large Language Models." arXiv, April 2024. https://arxiv.org/abs/2404.14795.

He, Jingxuan, Mark Vero, Gabriela Krasnopolska, and Martin Vechev. "Instruction Tuning for Secure Code Generation." arXiv, February 2024. https://arxiv.org/abs/2402.09497.

He, Pengfei, Yupin Lin, Shen Dong, Han Xu, Yue Xing, and Hui Liu. "Red-Teaming LLM Multi-Agent Systems via Communication Attacks." arXiv, February 2025. https://arxiv.org/abs/2502.14847.

Heiding, Fred, Simon Lermen, Andrew Kao, Bruce Schneier, and Arun Vishwanath. "Evaluating Large Language Models' Capability to Launch Fully Automated Spear Phishing Campaigns: Validated on Human Subjects." arXiv, November 2024. https://arxiv.org/pdf/2412.00586.

Heim, Lennart, Tim Fist, Janet Egan, et al. "Governing through the Cloud: The Intermediary Role of Compute Providers in AI Regulation." arXiv, March 2024. https://arxiv.org/pdf/2403.08501.

Hendrycks, Dan, Eric Schmidt, and Alexandr Wang. "Superintelligence Strategy: Expert Version." arXiv, March 2025. https://arxiv.org/abs/2503.05628.

Hinsley, Francis Harry. British Intelligence in the Second World War. Abridged ed. London: H.M.S.O., 1993. https://archive.org/details/britishintellige0000hins.

Hua, Tim Tian, Andrew Qin, Samuel Marks, and Neel Nanda. "Steering Evaluation-Aware Language Models to Act Like They Are Deployed." arXiv, October 2025. https://arxiv.org/abs/2510.20487.

Huang, Danny Yuxing, Hitesh Dharmdasani, Sarah Meiklejohn, et al. "Botcoin: Monetizing Stolen Cycles." Internet Society, 2017. https://www.ndss-symposium.org/ndss2014/ndss-2014-programme/botcoin-monetizing-stolen-cycles/.

Hubinger, Evan, Carson Denison, Jesse Mu, et al. "Sleeper Agents: Training Deceptive LLMs That Persist Through Safety Training." arXiv, 2024. https://arxiv.org/abs/2401.05566.

Hubinger, Evan, Chris van Merwijk, Vladimir Mikulik, Joar Skalse, and Scott Garrabrant. "Risks from Learned Optimization in Advanced Machine Learning Systems." arXiv, June 2019. https://arxiv.org/pdf/1906.01820.

Hugging Face. "Mistral7binstruct_summarize." Accessed February 17, 2026. https://huggingface.co/ymath/mistral7binstruct_summarize/tree/main.

———. "Mistral-7B-Instruct-v0.2." Accessed February 17, 2026. https://huggingface.co/mistralai/Mistral-7B-Instruct-v0.2/tree/main.

Hyperbolic. "The Open-Access AI Cloud." Accessed February 13, 2026. https://www.hyperbolic.ai/.

IBM. "IBM Report: Ransomware Persisted Despite Improved Detection in 2022." February 2023.
https://newsroom.ibm.com/2023-02-22-IBM-Report-Ransomware-Persisted-Despite-Impr
oved-Detection-in-2022/latest-news-hybrid-cloud.

Iftimie, Alex. "No Server Left Behind: The Justice Department's Novel Law Enforcement
Operation to Protect Victims." Lawfare, April 2021.
https://www.lawfaremedia.org/article/no-server-left-behind-justice-departments-novel-law-
enforcement-operation-protect-victims.

Irregular. "Evaluating GPT-5.2 Thinking: Cryptographic Challenge Case Study." December 2025.
https://www.irregular.com/publications/spell-bound-technical-case-study.

———. "From AI-Assisted to AI-Orchestrated: What Agent-Led Cyber Attacks Mean for
Security." November 2025.
https://www.irregular.com/publications/from-ai-assisted-to-ai-orchestrated.

———. "Frontier Model Performance on Offensive-Security Tasks: Emerging Evidence of a
Capability Shift." December 2025.
https://www.irregular.com/publications/emerging-evidence-of-a-capability-shift.

———. "The Rise of Autonomous Vulnerability Research Capabilities in LLMs." December 2025.
https://www.irregular.com/publications/the-rise-of-autonomous-vulnerability-research-capa
bilities.

———. "When Success Rates Mislead: The Case for Expected Cost as a Metric in AI
Evaluation." December 2025.
https://www.irregular.com/publications/expected-cost-per-success.

Jain, Neel, Avi Schwarzschild, Yuxin Wen, et al. "Baseline Defenses for Adversarial Attacks
Against Aligned Language Models." arXiv, September 2023.
https://arxiv.org/abs/2309.00614.

Jervis, Robert. *Why Intelligence Fails*. Cornell University Press, 2010.
https://www.cornellpress.cornell.edu/book/9780801447853/why-intelligence-fails/#bookTa
bs=4.

Ji, Jessica, Jenny Jun, Maggie Wu, and Rebecca Gelles. "Cybersecurity Risks of AI- Generated
Code." Center for Security and Emerging Technology, November 2024.
https://cset.georgetown.edu/wp-content/uploads/CSET-Cybersecurity-Risks-of-AI-Generat
ed-Code.pdf.

Kelly, Lidia. "Kremlin Says Russia-US Hotline to Deflate Crisis Not in Use." Reuters, November
2024.

https://www.reuters.com/world/kremlin-says-russia-us-hotline-deflate-crisis-not-use-2024-11-20/.

Khalili, Joel. "The Edgelord AI That Turned a Shock Meme Into Millions in Crypto." *WIRED*, December 2024. https://www.wired.com/story/truth-terminal-goatse-crypto-millionaire/.

Kim, Julie. "From Intelligence Gathering to Financial Gain: Countering DPRK Cyber Operations." Canadian Global Affairs Institute, November 2025. https://www.cgai.ca/pp_kim_nov2025.

Klarna. "Klarna AI Assistant Handles Two-Thirds of Customer Service Chats in Its First Month." February 2025. https://www.klarna.com/international/press/klarna-ai-assistant-handles-two-thirds-of-customer-service-chats-in-its-first-month/.

Korbak, Tomek, Joshua Clymer, Benjamin Hilton, Buck Shlegeris, and Geoffrey Irving. "A Sketch of an AI Control Safety Case." arXiv, January 2025. https://arxiv.org/abs/2501.17315.

Kouremetis, Michael, Dean Lawrence, Ron Alford, et al. "Mirage: Cyber Deception against Autonomous Cyber Attacks in Emulation and Simulation." *Annals of Telecommunications* 79 (March 2024): 803–17. https://doi.org/10.1007/s12243-024-01018-4.

Kraprayoon, Jam, Zoe Williams, and Rida Fayyaz. "AI Agent Governance: A Field Guide." Institute for AI Policy and Strategy, April 2025. https://www.iaps.ai/research/ai-agent-governance.

Kruus, Nicholas, Madhavendra Thakur, Adam Khoja, et al. "Governing Automated Strategic Intelligence." arXiv, September 2025. https://arxiv.org/abs/2509.17087.

Kwa, Thomas, Ben West, and Joel Becker. "Measuring AI Ability to Complete Long Tasks." METR, March 2025. https://metr.org/blog/2025-03-19-measuring-ai-ability-to-complete-long-tasks/.

Labrador, Cara, Shefali Agrawal, Alexandra Jumper, et al. "Building AI Surge Capacity: Mobilizing Technical Talent into Government for AI-Related National Security Crises." Institute for AI Policy and Strategy, October 2025. https://www.iaps.ai/research/building-ai-surge-capacity.

Lakera. "Data Versioning." Accessed February 17, 2026. https://www.lakera.ai/ml-glossary/data-versioning.

Lee, Jean, Geoff White, and Viv Jones. "Lazarus Heist: The Intercontinental ATM Theft That Netted $14m in Two Hours." *BBC*, March 2023. https://www.bbc.com/news/world-65130220.

Lee, Robert M. and Tim Conway. "The Five ICS Cybersecurity Critical Controls." SANS Institute, November 2022.
https://www.sans.org/white-papers/five-ics-cybersecurity-critical-controls.

Leukfeldt, Rutger and Jurjen Jansen. "Cyber Criminal Networks and Money Mules: An Analysis of Low-Tech and High-Tech Fraud Attacks in the Netherlands." *International Journal of Cyber Criminology*, 2015.
https://www.cybercrimejournal.com/pdf/Leukfeldt&Jansen2015vol9issue2.pdf.

Levine, Brian. "How Secure Is Tor?" GitHub, 2015. https://csam-bib.github.io/security/.

Liang, Jiacheng, Ren Pang, Changjiang Li, and Ting Wang. "Model Extraction Attacks Revisited." arXiv, September 2025. https://arxiv.org/pdf/2312.05386.

Lin, Herbert. *Cyber Threats and Nuclear Weapons*. Stanford University Press, 2021.
https://www.sup.org/books/politics/cyber-threats-and-nuclear-weapons.

———. "Escalation Dynamics and Conflict Termination in Cyberspace." *Strategic Studies Quarterly* 6, no. 3 (Fall 2012): 46–70.
https://www.airuniversity.af.edu/Portals/10/SSQ/documents/Volume-06_Issue-3/Lin.pdf.

———. "Russian Cyber Operations in the Invasion of Ukraine." The Cyber Defense Review, 2022.
https://cyberdefensereview.army.mil/Portals/6/Documents/2022_fall/02_Lin.pdf.

Lin, Shaowei, Evan Miyazono, and Daniel Windham. "A Toolchain for AI-Assisted Code Specification, Synthesis and Verification." Atlas Computing, 2025.
https://atlascomputing.org/ai-assisted-fv-toolchain.pdf.

Lior, Anat. "E/Insuring the AI Age: Empirical Insights into Artificial Intelligence Liability Policies." SSRN, June 2025. https://papers.ssrn.com/sol3/papers.cfm?abstract_id=5316376.

Litterio, Francis. "The Internet Worm of 1988." UNC Computer Science, 2023.
https://www.cs.unc.edu/~jeffay/courses/nidsS05/attacks/seely-RTMworm-89.html.

Lohn, Andrew J. "Anticipating AI's Impact on the Cyber Offense-Defense Balance." Center for Security and Emerging Technology, May 2025.
https://cset.georgetown.edu/wp-content/uploads/CSET-Anticipating-AIs-Impact-on-the-Cyber-Offense-Defense-Balance.pdf.

———. "Defending Against Intelligent Attackers at Large Scales." arXiv, April 2025.
https://arxiv.org/abs/2504.18577.

Loman, Mark. "Pioneering Automated Moving Target Defense." Sophos, October 2023.
https://www.sophos.com/en-us/blog/pioneering-automated-moving-target-defense-amtd.

Lonergan, Erica D. and Shawn W. Lonergan. "Cyber Operations, Accommodative Signaling, and the De-Escalation of International Crises." Security Studies, February 2022. https://doi.org/10.1080/09636412.2022.2040584.

Lonergan, Erica D. and Shawn W. Lonergan. *Escalation Dynamics in Cyberspace*. Oxford University Press, 2023. https://global.oup.com/academic/product/escalation-dynamics-in-cyberspace-9780197550885?cc=us&lang=en&.

Lukošiūtė, Kamilė. "Design for the Defenders You Care about or Risk Being Useless." December 2025. https://kamilelukosiute.com/posts/design-for-the-defenders-you-care-about-or-risk-being-useless.html.

Lykousas, Nikolaos and Constantinos Patsakis. "Just in Plain Sight: Unveiling CSAM Distribution Campaigns on the Clear Web." arXiv, November 2025. https://arxiv.org/abs/2511.03816.

Lynch, Aengus, Caleb Larson, and Sören Mindermann. "Agentic Misalignment: How LLMs Could Be Insider Threats." Anthropic, June 2025. https://www.anthropic.com/research/agentic-misalignment.

Maas, Matthijs M. *Architectures of Global AI Governance: From Technological Change to Human Choice*. Oxford Academic, 2025. https://doi.org/10.1093/9780191988455.001.0001

Maas, Matthijs and Tobi Olasunkanmi. "Treaty-Following AI." Institute for Law and AI, December 2025. https://law-ai.org/treaty-following-ai/.

MacDiarmid, Monte, Timothy Maxwell, Nicholas Schiefer, et al. "Simple Probes Can Catch Sleeper Agents." Anthropic, April 2024. https://www.anthropic.com/research/probes-catch-sleeper-agents.

Mandiant. "M-Trends 2023." 2022. https://services.google.com/fh/files/misc/m_trends_2023_report.pdf#page=38.

Maness, Ryan C., Brandon Valeriano, Kathryn Hedgecock, Jose M. Macias, and Benjamin Jensen. "Tracking Competition in Cyberspace: Announcing the Dyadic Cyber Incident Dataset Version 2.0." Modern War Institute, October 2022. https://mwi.westpoint.edu/tracking-competition-in-cyberspace-announcing-the-dyadic-cyber-incident-dataset-version-2-0/.

Marcus, Gary and Nathan Hamiel. "LLMs + Coding Agents = Security Nightmare." *Marcus on AI*, August 2025. https://garymarcus.substack.com/p/llms-coding-agents-security-nightmare.

Marczak, Bill, John Scott-Railton, Sarah McKune, Bahr Abdul Razzak, and Ron Deibert. "Hide and Seek: Tracking NSO Group's Pegasus Spyware to Operations in 45 Countries." The Citizen Lab, September 2018. https://citizenlab.ca/research/hide-and-seek-tracking-nso-groups-pegasus-spyware-to-operations-in-45-countries/.

Marino, Bill and Ari Juels. "Giving AI Agents Access to Cryptocurrency and Smart Contracts Creates New Vectors of AI Harm." arXiv, July 2025. https://arxiv.org/pdf/2507.08249v1#page=7.24.

Marszal, Edward. "Make Process Plants Inherently Safe Against Cyber Attack." Kenexis, November 2014. https://www.kenexis.com/make-process-plants-inherently-safe-cyber-attack/.

Martin, Alexander. "Cobalt Strike: International Law Enforcement Operation Tackles Illegal Uses of 'Swiss Army Knife' Pentesting Tool." The Record, July 2024. https://therecord.media/cobalt-strike-law-enforcement-takedown.

Maslej, Nestor, Loredana Fattorini, Raymond Perrault, et al. "Artificial Intelligence Index Report 2025." Stanford HAI, 2025. https://hai.stanford.edu/assets/files/hai_ai-index-report-2025_chapter1_final.pdf.

Meinke, Alexander, Bronson Schoen, Jérémy Scheurer, Mikita Balesni, Rusheb Shah, and Marius Hobbhahn. "Frontier Models Are Capable of In-Context Scheming." arXiv, December 2024. https://arxiv.org/abs/2412.04984.

METR. "How Does Time Horizon Vary Across Domains?" July 2025. https://metr.org/blog/2025-07-14-how-does-time-horizon-vary-across-domains/#summary.

Microsoft. "Troubleshoot Issues When You Sign up for a New Account in the Azure Portal." Accessed February 16, 2026. https://learn.microsoft.com/en-us/azure/cost-management-billing/troubleshoot-subscription/troubleshoot-azure-sign-up.

Microsoft Threat Intelligence. "Cryptojacking: Understanding and Defending against Cloud Compute Resource Abuse." July 2023. https://www.microsoft.com/en-us/security/blog/2023/07/25/cryptojacking-understanding-and-defending-against-cloud-compute-resource-abuse/.

Microsoft Threat Intelligence. "Microsoft Digital Defense Report 2025." 2025. https://www.microsoft.com/en-us/corporate-responsibility/cybersecurity/microsoft-digital-defense-report-2025/.

Middleton, Bruce. "Operation Aurora—2009." In *A History of Cyber Security Attacks*. Auerbach Publications, 2017. https://www.taylorfrancis.com/chapters/mono/10.1201/9781315155852-19/operation-aurora%E2%80%942009-bruce-middleton.

Milevski, Lukas. "Stuxnet And Strategy: A Special Operation in Cyberspace?" National Defense University Press, 2011. https://ndupress.ndu.edu/Portals/68/Documents/jfq/jfq-63/jfq-63_64-69_Milevski.pdf?ver=Jy0SW9E8UBbatlrmrw-egQ%3D%3D.

Ming, Lee Chong. "Replit's CEO Apologizes after Its AI Agent Wiped a Company's Code Base in a Test Run and Lied About It." *Business Insider*, July 2025. https://www.businessinsider.com/replit-ceo-apologizes-ai-coding-tool-delete-company-database-2025-7.

MITRE ATT&CK. "Command and Control." April 2025. https://attack.mitre.org/tactics/TA0011/.

MITRE ATT&CK. "Enterprise Tactics." Accessed February 13, 2026. https://attack.mitre.org/tactics/enterprise/.

Mitre, Jim and Joel B. Predd. "Artificial General Intelligence's Five Hard National Security Problems." Rand, February 2025. https://www.rand.org/pubs/perspectives/PEA3691-4.html.

Moor, Oege de. "XBOW Now Matches the Capabilities of a Top Human Pentester." *XBOW*, August 2024. https://xbow.com/blog/xbow-vs-humans.

Moor, Oege de and Albert Ziegler. "XBOW Unleashes GPT-5's Hidden Hacking Power, Doubling Performance." *XBOW*, August 2025. https://xbow.com/blog/gpt-5.

Moses, CJ. "New Amazon Threat Intelligence Findings: Nation-State Actors Bridging Cyber and Kinetic Warfare." AWS, November 2025. https://aws.amazon.com/blogs/security/new-amazon-threat-intelligence-findings-nation-state-actors-bridging-cyber-and-kinetic-warfare/.

Murphy, Benjamin and Twm Stone. "Uplifted Attackers, Human Defenders: The Cyber Offense-Defense Balance for Trailing-Edge Organizations." arXiv, August 2025. https://arxiv.org/html/2508.15808v1#S4.

MyComplianceOffice. "The Impact of MiCA on Your AML/KYC Compliance Program." June 2025. https://mco.mycomplianceoffice.com/blog/impact-of-mica-on-aml-kyc-compliance.

Nanda, Neel. "A Comprehensive Mechanistic Interpretability Explainer & Glossary." *Neel Nanda*, December 2024. https://www.neelnanda.io/mechanistic-interpretability/glossary.

National Telecommunications and Information Administration. "Dual-Use Foundation Models with Widely Available Model Weights Report." July 2024. https://www.ntia.gov/programs-and-initiatives/artificial-intelligence/open-model-weights-report.

Nevo, Sella, Dan Lahav, Ajay Karpur, Yogev Bar-On, Henry Alexander Bradley, and Jeff Alstott. "A Playbook for Securing AI Model Weights." Rand, November 2024. https://www.rand.org/pubs/research_briefs/RBA2849-1.html.

Newman, Steve. "Cybersecurity and AI: The Evolving Security Landscape." Center for AI Safety, March 2025. https://safe.ai/blog/cybersecurity-and-ai-the-evolving-security-landscape.

Nimmo, Ben, Kimo Bumanglag, Michael Flossman, Nathaniel Hartley, Jack Stubbs, and Albert Zhang. "Disrupting Malicious Uses of AI: October 2025." OpenAI, October 2025. https://openai.com/global-affairs/disrupting-malicious-uses-of-ai-october-2025/.

NIST. "CVE-2023-49935 Detail." December 2023. https://nvd.nist.gov/vuln/detail/CVE-2023-49935.

Novikov, Alexander, Ngân Vū, Marvin Eisenberger, et al. "AlphaEvolve: A Coding Agent for Scientific and Algorithmic Discovery." arXiv, June 2025. https://arxiv.org/abs/2506.13131.

NPR. "Anti-Virus Program Update Wreaks Havoc With PCs." April 2010. https://www.npr.org/2010/04/21/126168997/anti-virus-program-update-wreaks-havoc-with-pcs.

O'Brien, Joe, Shaun Ee, Jam Kraprayoon, Bill Anderson-Samways, Oscar Delaney, and Zoe Williams. "Coordinated Disclosure of Dual-Use Capabilities: An Early Warning System for Advanced AI." Institute for AI Policy and Strategy, June 2024. https://www.iaps.ai/research/coordinated-disclosure.

Oesch, Sean, Phillipe Austria, Jack Hutchins, and Amul Chaulagain. "Agentic AI and the Cyber Arms Race." arXiv*, 2025*. https://arxiv.org/html/2503.04760v1.

O'Gara, Aidan, Gabriel Kulp, Will Hodgkins, et al. "Hardware-Enabled Mechanisms for Verifying Responsible AI Development." arXiv, April 2025. https://arxiv.org/abs/2505.03742.

O'Keefe, Cullen, Ketan Ramakrishnan, Janna Tay, and Christoph Winter. "Law-Following AI: Designing AI Agents to Obey Human Laws." Institute for Law and AI, May 2025. https://law-ai.org/law-following-ai/.

OpenAI. "API Organization Verification." Accessed February 17, 2026.
https://help.openai.com/en/articles/10910291-api-organization-verification.

Ord, Toby. "Inference Scaling Reshapes AI Governance." arXiv, February 2025.
https://arxiv.org/abs/2503.05705.

Ottinger, Lily and Caleb Harding. "Gridlocked: Transformer Shortage Choking US Supply Chains."
*ChinaTalk*, February 2025.
https://www.chinatalk.media/p/gridlocked-transformer-shortage-choking.

Overlier, L. and P. Syverson. "Locating Hidden Servers." IEEE Xplore, June 2006.
https://ieeexplore.ieee.org/document/1624004.

Palazzolo, Stephanie. "Why xAI Spent So Much on Reinforcement Learning." *The Information*,
July 2025.
https://www.theinformation.com/newsletters/ai-agenda/xai-spent-reinforcement-learning.

Pan, Alexa and Ryan Greenblatt. "Sonnet 4.5's Eval Gaming Seriously Undermines Alignment
Evals." *Redwood Research Blog*, October 2025.
https://blog.redwoodresearch.org/p/sonnet-45s-eval-gaming-seriously.

Park, Peter S., Simon Goldstein, Aidan O'Gara, Michael Chen, and Dan Hendrycks. "AI
Deception: A Survey of Examples, Risks, and Potential Solutions." Patterns, May 2024.
https://www.cell.com/patterns/fulltext/S2666-3899%2824%2900103-X.

Pomerleau, Mark. "How Can Cyber Contribute to Multi-Domain Battle?" C4ISRNET, December
2016.
https://www.c4isrnet.com/home/2016/12/15/how-can-cyber-contribute-to-multi-domain-b
attle/.

Popa, Raluca Ada and Four Flynn. "Introducing CodeMender: An AI Agent for Code Security."
Google DeepMind, October 2025.
https://deepmind.google/blog/introducing-codemender-an-ai-agent-for-code-security/.

Potter, Yujin, Wenbo Guo, Zhun Wang, et al. "Frontier AI's Impact on the Cybersecurity
Landscape." arXiv, April 2025. https://arxiv.org/abs/2504.05408.

PwC UK and BAE Systems. "Operation Cloud Hopper." April 2017.
https://www.pwc.co.uk/cyber-security/pdf/pwc-uk-operation-cloud-hopper-report-april-20
17.pdf.

Rajic, Taylar and Julia Brock. "The ByBit Heist and the Future of U.S. Crypto Regulation." CSIS,
March 2025. https://www.csis.org/analysis/bybit-heist-and-future-us-crypto-regulation.

Reiner, Shaked. "Anatomy of an LLM RCE." Cyber Ark, October 2024.
    https://www.cyberark.com/resources/threat-research-blog/anatomy-of-an-llm-rce.

Reppel, Erik, Nemil Dalal, and Dan Kim. "Introducing X402: A New Standard for Internet-Native
    Payments." Coinbase, May 2025.
    https://www.coinbase.com/en-gb/developer-platform/discover/launches/x402.

Reworr and Dmitrii Volkov. "LLM Agent Honeypot: Monitoring AI Hacking Agents in the Wild."
    arXiv, October 2024. https://arxiv.org/abs/2410.13919.

Richardson, Lloyd. "CSAM Distribution on Tor Is Not Inevitable; The Network's Creators Have the
    Power to Act." Canadian Centre for Child Protection, 2025.
    https://protectchildren.ca/en/press-and-media/blog/2025/tor-backgrounder.

Rodriguez, Mikel, Raluca Ada Popa, Four Flynn, Lihao Liang, Allan Dafoe, and Anna Wang. "A
    Framework for Evaluating Emerging Cyberattack Capabilities of AI." arXiv, March 2025.
    https://arxiv.org/abs/2503.11917.

Rubin, Sam. "Written Testimony of: Sam Rubin." *Palo Alto Networks*, November 2023.
    https://d1dth6e84htgma.cloudfront.net/11_14_23_Rubin_Testimony_2fba2978dd.pdf.

Rubin, Sam. "Unit 42 Develops Agentic AI Attack Framework." *Palo Alto Networks*, May 2025.
    https://www.paloaltonetworks.com/blog/2025/05/unit-42-develops-agentic-ai-attack-fram
    ework/.

RunSybil. "Attack Is Your Best Defense." Accessed February 12, 2026.
    https://www.runsybil.com/.

Ruser, Nathan. "Scamland Myanmar: How Conflict and Crime Syndicates Built a Global Fraud
    Industry." Australian Strategic Policy Institute, September 2025.
    https://www.aspi.org.au/report/scamland-myanmar-how-conflict-and-crime-syndicates-bui
    lt-a-global-fraud-industry/.

Samonas, Spyridon and David Coss. "The CIA Strikes Back: Redefining Confidentiality, Integrity
    and Availability in Security." Journal of Information Systems Security, 2014.
    https://www.jissec.org/Contents/V10/N3/V10N3-Samonas.html.

Sanders, James, Luke Emberson, and Yafah Edelman. "What Did It Take to Train Grok 4?" Epoch
    AI, September 2025. https://epoch.ai/data-insights/grok-4-training-resources.

Sanz-Gómez, María, Víctor Mayoral-Vilches, Francesco Balassone, Luis Javier Navarrete-Lozano,
    Cristóbal R. J. Veas Chavez, and Maite del Mundo de Torres. "Cybersecurity AI Benchmark

(CAIBench): A Meta-Benchmark for Evaluating Cybersecurity AI Agents." arXiv, October 2025. https://arxiv.org/abs/2510.24317.

Schippers, Raymond. "AI 2030: The Coming Era of Autonomous Cyber Crime." *Check Point*, October 2025. https://blog.checkpoint.com/executive-insights/ai-2030-the-coming-era-of-autonomous-cyber-crime/.

Schlatter, Jeremy, Benjamin Weinstein-Raun, and Jeffrey Ladish. "Shutdown Resistance in Large Language Models." arXiv, September 2025. https://arxiv.org/html/2509.14260v1.

———. "Shutdown Resistance in Reasoning Models." Palisade Research, May 2025. https://palisaderesearch.org/blog/shutdown-resistance.

Schwarcz, Daniel and Josephine Wolff. "The Limits of Regulating AI Safety Through Liability and Insurance: Lessons From Cybersecurity." SSRN, August 2025. https://papers.ssrn.com/sol3/papers.cfm?abstract_id=5411062.

Schwartz, Mathew J. "Flame Malware's Ties To Stuxnet, Duqu: Details Emerge." *Dark Reading*, May 2012. https://www.darkreading.com/cyberattacks-data-breaches/flame-malware-s-ties-to-stuxnet-duqu-details-emerge.

Sherman, Justin. "Untangling the Russian Web: Spies, Proxies, and Spectrums of Russian Cyber Behavior." *Atlantic Council*, September 2022. https://www.atlanticcouncil.org/in-depth-research-reports/issue-brief/untangling-the-russian-web/.

Sherman, Justin. "Russia's Digital Tech Isolationism: Domestic Innovation, Digital Fragmentation, and the Kremlin's Push to Replace Western Digital Technology." *Atlantic Council*, July 2024. https://www.atlanticcouncil.org/in-depth-research-reports/issue-brief/russias-digital-tech-isolationism/.

Shevlane, Toby. "Structured Access: An Emerging Paradigm for Safe AI Deployment." arXiv, January 2022. https://arxiv.org/abs/2201.05159.

Shier, John, Angela Gunn, and Hilary Wood. "It Takes Two: The 2025 Sophos Active Adversary Report." *Sophos*, April 2025. https://www.sophos.com/en-us/blog/2025-sophos-active-adversary-report.

Shlegeris, Buck. "Win/Continue/Lose Scenarios and Execute/Replace/Audit Protocols." *Redwood Research Blog*, November 2024. https://blog.redwoodresearch.org/p/wincontinuelose-scenarios-and-executereplaceaudi.

SIMA Team. "SIMA 2: An Agent That Plays, Reasons, and Learns With You in Virtual 3D Worlds." Google DeepMind, November 2025. https://deepmind.google/blog/sima-2-an-agent-that-plays-reasons-and-learns-with-you-in-virtual-3d-worlds/.

Singer, Brian, Keane Lucas, Lakshmi Adiga, Meghna Jain, Lujo Bauer, and Vyas Sekar. "Incalmo: An Autonomous LLM-Assisted System for Red Teaming Multi-Host Networks." arXiv, January 2025. https://arxiv.org/abs/2501.16466.

Singer, Brian, Yusuf Saquib, Lujo Bauer, and Vyas Sekar. "Perry: A High-Level Framework for Accelerating Cyber Deception Experimentation." arXiv, September 2025. https://arxiv.org/pdf/2506.20770.

Singer, P. W. "Stuxnet and Its Hidden Lessons on the Ethics of Cyberweapons." Case Western Reserve Journal of International Law, 2015. https://scholarlycommons.law.case.edu/cgi/viewcontent.cgi?article=1009&context=jil#page=4.

Skingsley, Juliet. "Offensive Cyber Operations." Chatham House, September 2023. https://www.chathamhouse.org/2023/09/offensive-cyber-operations/03-risk-perception-cyberspace.

Smith, Ian. "Lloyd's Finds Major Hack of a Payments System Could Cost $3.5tn." *Financial Times*, October 2023. https://www.ft.com/content/f4f09c0d-19aa-41c4-ac72-5f3395118960.

Software Engineering Institute. "Fostering Growth in Professional Cyber Incident Management." Carnegie Mellon University, 1988. https://www.sei.cmu.edu/history-of-innovation/fostering-growth-in-professional-cyber-incident-management/.

Somala, Venkat and Luke Emberson. "Frontier AI Performance Becomes Accessible on Consumer Hardware within a Year." Epoch AI, August 2025. https://epoch.ai/data-insights/consumer-gpu-model-gap.

Spafford, Eugene H. "The Internet Worm Incident." Purdue University, 1989. https://docs.lib.purdue.edu/cgi/viewcontent.cgi?article=1792&context=cstech.

Steinberg, Sean, Adam Stepan, and Kyle Neary. "NotPetya: A Columbia University Case Study." SIPA, 2021. https://www.sipa.columbia.edu/sites/default/files/2022-11/NotPetya%20Final.pdf.

Stix, Charlotte, Annika Hallensleben, Alejandro Ortega, and Matteo Pistillo. "The Loss of Control Playbook: Degrees, Dynamics, and Preparedness." arXiv, November 2025. https://arxiv.org/abs/2511.15846.

Stubbs, Jack, Joseph Menn, and Christopher Bing. "Inside the West's Failed Fight against China's 'Cloud Hopper' Hackers." Reuters, June 2019. https://www.reuters.com/investigates/special-report/china-cyber-cloudhopper/.

SWE-bench. "Official Leaderboards." April 2025. https://www.swebench.com/.

Tamari, Shir, Ronen Shustin, and Andres Riancho. "Wiz Research Finds Critical NVIDIA AI Vulnerability Affecting Containers Using NVIDIA GPUs, Including Over 35% of Cloud Environments." *WIZ*, September 2024. https://www.wiz.io/blog/wiz-research-critical-nvidia-ai-vulnerability.

Tarasov, Katie. "Exclusive Look at the Creation of High NA, ASML's New $400 Million Chipmaking Colossus." *CNBC*, May 2025. https://www.cnbc.com/2025/05/22/exclusive-look-at-high-na-asmls-new-400-million-chipmaking-colossus.html.

The White House. "Fact Sheet: U.S.-Russian Cooperation on Information and Communications Technology Security." June 2013. https://obamawhitehouse.archives.gov/the-press-office/2013/06/17/fact-sheet-us-russian-cooperation-information-and-communications-technol.

Theohary, Catherine A. "Use of Force in Cyberspace." Congress.Gov, November 2024. https://www.congress.gov/crs-product/IF11995.

Tor Project. "Browse Privately. Explore Freely." Accessed February 16, 2026. https://www.torproject.org/.

Triedman, Harold, Rishi Jha, and Vitaly Shmatikov. "Multi-Agent Systems Execute Arbitrary Malicious Code." arXiv, March 2025. https://arxiv.org/abs/2503.12188.

U.S. Department of Defense, Office of the Secretary of Defense. *Nuclear Posture Review.* February 2018. https://defenseinnovationmarketplace.dtic.mil/wp-content/uploads/2018/09/2018-NUCLEAR-POSTURE-REVIEW-FINAL-REPORT.pdf.

U.S. Strategic Command. "U.S. Strategic Command and U.S. Northern Command SASC Testimony." March 2019. https://www.stratcom.mil/Media/Speeches/Article/1771903/us-strategic-command-and-us-northern-command-sasc-testimony/.

Valeriano, Brandon. "The Myth of the Cyber Offense: The Case for Restraint." CATO Institute, January 2019. https://www.cato.org/policy-analysis/myth-cyber-offense-case-restraint.

Van der Merwe, Matthew. "Assessing the Risk of AI-Enabled Cyberattacks on the Power Grid." GovAI, November 2025. https://www.governance.ai/research-paper/ai-enabled-cyberattacks-on-the-power-grid.

Vassilev, Apostol, Alina Oprea, Alie Fordyce, Hyrum Anderson, Xander Davies, and Maia Hamin. "Adversarial Machine Learning." National Institute of Standards and Technology, March 2025. https://nvlpubs.nist.gov/nistpubs/ai/NIST.AI.100-2e2025.pdf.

Verizon Business. "2024 Data Breach Investigations Report." 2024. https://www.verizon.com/business/resources/reports/2024-dbir-data-breach-investigations-report.pdf.

Verizon Business. "2025 Data Breach Investigations Report." 2025. https://www.verizon.com/business/resources/T825/reports/2025-dbir-data-breach-investigations-report.pdf#page=9.99.

Vermeer, Michael J. D. "Evaluating Select Global Technical Options for Countering a Rogue AI." Rand, November 2025. https://www.rand.org/pubs/perspectives/PEA4361-1.html.

Vermeer, Michael J. D., Timothy M. Bonds, Emily Lathrop, and Gregory Smith. "Averting a Robot Catastrophe." Rand, September 2025. https://www.rand.org/pubs/perspectives/PEA3691-7.html.

Waisman, Nico. "How XBOW Found a Scoold Authentication Bypass." *XBOW*, November 2024. https://xbow.com/blog/xbow-scoold-vuln.

Weij, Teun van der, Felix Hofstätter, Ollie Jaffe, Samuel F. Brown, and Francis Rhys Ward. "AI Sandbagging: Language Models Can Strategically Underperform on Evaluations." arXiv, June 2024. https://arxiv.org/abs/2406.07358.

Weisgerber, Marcus. "China's Copycat Jet Raises Questions About F-35." *Defense One*, September 2015. https://www.defenseone.com/threats/2015/09/more-questions-f-35-after-new-specs-chinas-copycat/121859/.

White, Geoff and Jean H Lee. "The Lazarus Heist: How North Korea Almost Pulled off a Billion-Dollar Hack." *BBC*, June 2021. https://www.bbc.com/news/stories-57520169.

Withers, Caleb. "Tipping the Scales: Emerging AI Capabilities and the Cyber Offense-Defense Balance." CNAS Technology and National Security Program, September 2025.

https://s3.us-east-1.amazonaws.com/files.cnas.org/documents/Report_Tipping-the-Scales _TECH_Sep-2025-Final.pdf.

Wong, Haiman and Tiffany Saade, "The Rise of AI Agents: Anticipating Cybersecurity Opportunities, Risks, and the Next Frontier." R Street Institute, May 2025. https://www.rstreet.org/research/the-rise-of-ai-agents-anticipating-cybersecurity-opportunities-risks-and-the-next-frontier/.

Wu, Tongtong, Linhao Luo, Yuan-Fang Li, Shirui Pan, Thuy-Trang Vu, and Gholamreza Haffari. 2024. "Continual Learning for Large Language Models: A Survey." arXiv, February 2024. https://arxiv.org/abs/2402.01364.

Zetter, Kim. "NSA Hacker Chief Explains How to Keep Him Out of Your System." *WIRED*, January 2016. https://www.wired.com/2016/01/nsa-hacker-chief-explains-how-to-keep-him-out-of-your-system/.

Zetter, Kim. "The Untold Story of the Boldest Supply-Chain Hack Ever." *WIRED*, May 2023. https://www.wired.com/story/the-untold-story-of-solarwinds-the-boldest-supply-chain-hack-ever/.

Zhang, Boxuan, Yi Yu, Jiaxuan Guo, and Jing Shao. "Dive into the Agent Matrix: A Realistic Evaluation of Self-Replication Risk in LLM Agents." arXiv, September 2025. https://arxiv.org/html/2509.25302v1.

Zhao, Hao, Hui Shu, Yuyao Huang, and Ju Yang. "AIBot: A Novel Botnet Capable of Performing Distributed Artificial Intelligence Computing." MDPI, October 2022. https://doi.org/10.3390/electronics11193241.

Zhou, Zhanke, Jianing Zhu, Fengfei Yu, et al. "Model Inversion Attacks: A Survey of Approaches and Countermeasures." arXiv, November 2024. https://arxiv.org/abs/2411.10023.

Zhu, Yuxuan, Antony Kellermann, Dylan Bowman, et al. "CVE-Bench: A Benchmark for AI Agents' Ability to Exploit Real-World Web Application Vulnerabilities." arXiv, March 2025. https://arxiv.org/abs/2503.17332.

Zou, Andy, Zifan Wang, Nicholas Carlini, Milad Nasr, J. Zico Kolter, and Matt Fredrikson. "Universal and Transferable Adversarial Attacks on Aligned Language Models." arXiv, July 2023. https://arxiv.org/abs/2307.15043.